\documentclass[twocolumn,aps,prl,showpacs,amsmath,superscriptaddress,longbibliography,notitlepage]{revtex4-2}
\usepackage{amssymb}
\usepackage{mathrsfs}
\usepackage{graphicx}
\usepackage{float}
\usepackage[caption=false]{subfig}
\usepackage[pdftex,colorlinks=red]{hyperref}
\usepackage{verbatim}
\usepackage{txfonts}
\usepackage{epstopdf}
\usepackage[normalem]{ulem}
\usepackage{xcolor}
\usepackage{bm}
\usepackage[utf8]{inputenc}

\def\red{\textcolor{black}}
\def\blue{\textcolor{black}}

\begin{document}

\title{Anisotropic-scaling localization in higher-dimensional non-Hermitian systems}
\author{Zuxuan Ou}
\affiliation{Guangdong Provincial Key Laboratory of Quantum Metrology and Sensing $\&$ School of Physics and Astronomy, Sun Yat-Sen University (Zhuhai Campus), Zhuhai 519082, China}
\author{Hui-Qiang Liang}
\affiliation{Department of Physics, Shandong University, Jinan 250100, China}
\author{Guo-Fu Xu}\email{xgf@sdu.edu.cn}
\affiliation{Department of Physics, Shandong University, Jinan 250100, China}
\author{Linhu Li}\email{lilinhu@quantumsc.cn}
\affiliation{Quantum Science Center of Guangdong-Hong Kong-Macao Greater Bay Area (Guangdong), Shenzhen, China}
\date{\today}

\begin{abstract}
Spatial localization of quantum states is one of the focal points in condensed matter physics and quantum simulations, as it signatures profound physics such as nontrivial band topology and non-reciprocal non-Hermiticity. Yet, in higher dimensions, characterizing state localization becomes elusive due to the sophisticated interplay between different localization mechanisms and spacial geometries. In this work, we unveil an exotic type of localization phenomenon in higher-dimensional non-Hermitian systems, termed anisotropic-scaling localization (ASL), where localization lengths follow distinct size-dependent scaling rules in an anisotropic manner. Assisted with both analytical solution and numerical simulation, we find that ASL can emerge from two different mechanisms of effective bulk couplings or one-dimensional junction between different 1D edges, depending on how non-reciprocity is introduced to the system. The competition between ASL states and edge non-Hermitian skin states are further identified by their complex and real eigenenergies, respectively. Our results resolve the subtle co-existence of loop-like spectrum and skin-like localization of boundary states in contemporary literature, and provide a framework to classify the intricate higher-order non- Hermitian localization regarding their localization profiles. 
\end{abstract}

\maketitle

{\it Introduction.-}
Spatial localization of quantum states commonly manifest when translational symmetry is broken to some extent, such as the Anderson localization induced by disorders \citep{anderson1958absence}, and topological localization at open boundaries of systems with nontrivial topology \citep{hasan2010colloquium,qi2011topological}. 
In the realm of non-Hermitian physics, 
novel types of localization phenomena that greatly diverge from Hermitian localizations 
have aroused intense interest over the past few years~\citep{lee2016anomalous,martinez2018non,yao2018edge,yokomizo2019non,borgnia2020non,okuma2020topological,zhang2020correspondence,lin2023topological,torres2019perspective,PhysRevB.106.L161402,
PhysRevLett.128.226401,PhysRevLett.131.176402,PhysRevB.108.165105,
li2020critical,li2021impurity,guo2023accumulation,li2023scale,PhysRevLett.131.103604,PhysRevLett.132.086502,guo2024scale,yokomizo2021scaling,qin2023universal,lee2022exceptional,zhu2024brief,lee2019hybrid,li2022gain,zhu2022hybrid,li2020topological,ou2023non,kawabata2020higher,okugawa2020second,fu2021non,PhysRevLett.123.123601,PhysRevB.102.235151,zhang2022symmetry,faugno2022interaction,qin2024occupation,yoshida2024non,PhysRevLett.133.136502},
including the celebrated non-Hermitian skin effect (NHSE) where massive eigenstates are exponentially localized at the boundaries~\citep{martinez2018non,yao2018edge,yokomizo2019non,borgnia2020non,okuma2020topological,zhang2020correspondence,lin2023topological}, and the scale-free localization (SFL) with localization length increasing linearly with the system's size~\citep{li2020critical,li2021impurity,yokomizo2021scaling,qin2023universal,guo2023accumulation,li2023scale,PhysRevLett.131.103604,PhysRevLett.132.086502,guo2024scale}. 
Extending into higher-dimensions, the interplay between non-Hermitian and conventional localizations leads to new classes of higher-order boundary localization~\cite{zhu2024brief,lee2019hybrid,li2022gain,zhu2022hybrid,li2020topological,ou2023non}.
However, their complexity increases significantly with the spatial dimensions, mainly due to rich geometries and multiple length parameters that determine a system's size.
Consequently, much remains elusive regarding the localization profiles of higher-order boundary states and their relation to the corresponding spectral features. 
Specifically, the non-Hermitian corner localization is commonly considered as a higher-order NHSE~\citep{kawabata2020higher,okugawa2020second,fu2021non}, yet its co-existence with loop-like boundary spectrum seemingly suggests features of the size-dependent SFL instead~\citep{ou2023non}.
Given the distinct scaling behaviors for different non-Hermitian localization,
it is crucial to identifying the mechanism of boundary states for anatomizing their intrinsic properties, such as the nontrivial boundary topology that leads to higher-order topological phases.

In this work, we anatomize higher-order boundary localization in higher-dimensional non-Hermitian  lattices, 
and uncover a type of anisotropic-scaling localization (ASL) that resolves the above enigma.
Specifically,
ASL states exhibit localization lengths that change anisotropically with the system's size, yet remain localized in the thermodynamic limit, reflecting the characteristics of both SFL and NHSE in 1D systems, respectively.
In a minimal two-dimensional (2D) model, 
we analytically solve the ASL states and find that they arise from the interference between different 1D non-Hermitian edges through the 2D bulk of the lattice, and can be identified by the appearance of complex eigenenergies of boundary states.
Extending our analysis to the non-Hermitian Benalcazar-Bernevig-Hughes (BBH) model with non-Hermitian pumping along one or both directions,
we discover that ASL can emerge from different mechanisms of either effective bulk couplings or edge junctions, 
and may co-exist with boundary NHSE in certain parameter regimes.
\blue{Notably, these two types of ASL analogize 1D SFL induced by weak interchain couplings~\cite{li2020critical} and local impurities~\cite{li2021impurity}, respectively.}
Our work reveals an intriguing class of size-dependent higher-order non-Hermitian boundary localization,
which is expected to be closely relevant to most commentary experimental realizations of finite-size non-Hermitian lattices.


{\it The minimal model and corner states.-}
\begin{figure}
	\includegraphics[width=1.0\linewidth]{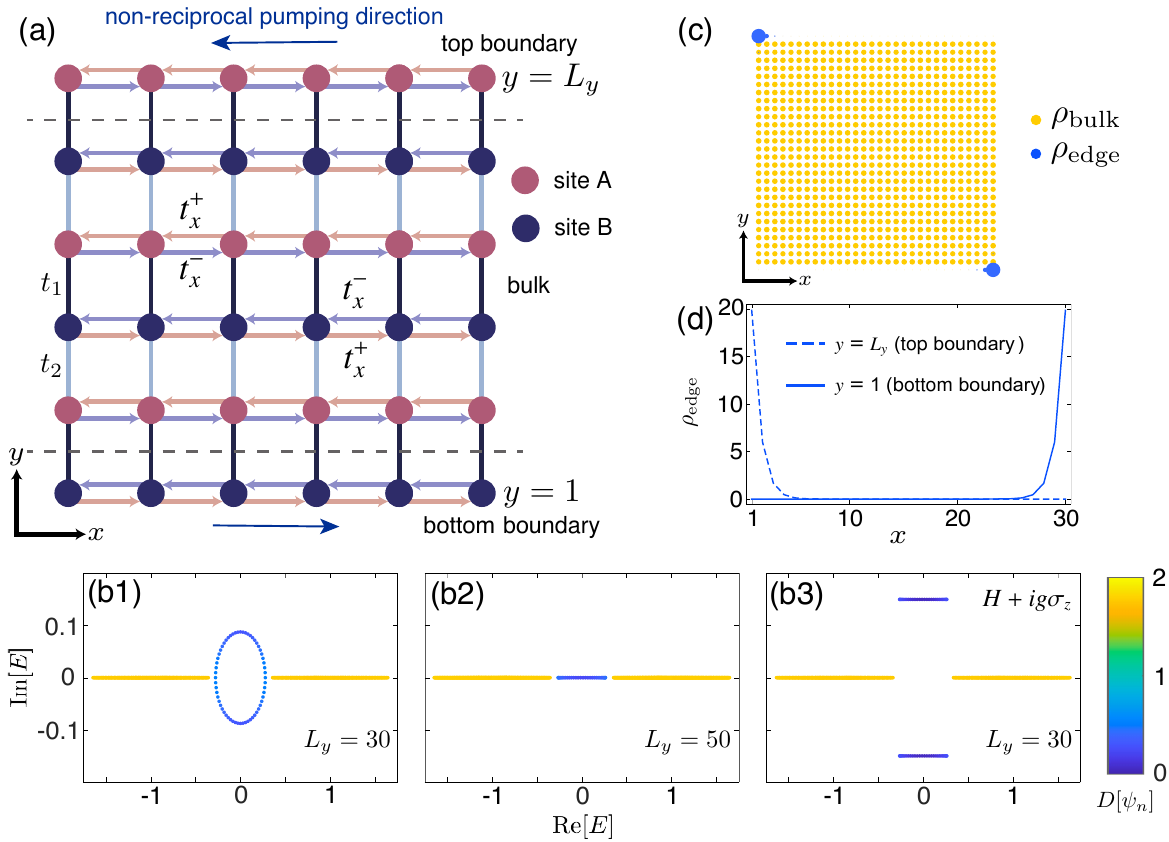}
	\caption{
	(a) A sketch of the model with $L_x=L_y=6$. Blue arrows indicate the non-Hermitian non-reciprocal pumping direction.  
	(b1)-(b3) Energy spectra under OBCs with $L_x=30$ and different $L_y$.
	\blue{An extra term of pseudospin-dependent imaginary energies $ig\sigma_z$ with $g=0.15$ are added to the Hamiltonian for (b3).}
	 Eigenenergies are marked by different colors according to the fractal dimension $D[\psi_n]$. 
	(c) Summed distribution of bulk states (yellow dots) and edge states (blue dots) in (b1). 
	Size of each dot indicates the value of the total density for bulk and edge states at each lattice site,
	defined as $\rho_\text{bulk/edge}=\sum_n\vert\psi_{n}(x,y)\vert^2$, where the summation of $n$ runs over all states with $D[\psi_n]\approx 2$ and $D[\psi_n]\approx 0$, respectively. 
	$\rho_{\rm edge}$ along the bottom and top edges ($y=1$ and $y=L_y$) is further demonstrated in (d).
	Other parameters are $t_1=0.25,t_2=1.0,t_x^+=0.35,t_x^-=0.05$. 
	}
	\label{model_with_energy}
\end{figure}
We first consider a 2D model with $L_x\times L_y$ lattice sites, 
composed by a series of Hantano-Nelson (HN) chains \citep{hatano1996localization,hatano1998non} with alternating non-reciprocal hopping amplitudes and imaginary on-site energies, 
which are connected through hoppings with staggered amplitudes along the second ($y$) direction, 
as shown in Fig. \ref{model_with_energy}(a). 
We refer to it as the HN-SSH model since it can be viewed as a combination of HN chains along $x$ direction and Su-Schrieffer-Heeger (SSH) chains~\citep{su1979solitons} along $y$ direction.
In the following discussion $t_x^+>t_x^-$ is chosen without loss of generality. 
The corresponding Bloch Hamiltonian reads 
\blue{
\begin{eqnarray}
	H(k_x,k_y)&=&(t_x^++t_x^-)\cos k_x\sigma_0+i(t_x^+-t_x^-)\sin k_x\sigma_z\notag\\
	&&+(t_1+t_2\cos k_y)\sigma_x+t_2\sin k_y\sigma_y,
	\label{bulk_H}
\end{eqnarray}
}
\blue{where} $t_{1,2}$ the staggered hopping amplitudes between them ($t_1<t_2$ is chosen so to generate topological localization on top and bottom edges), and $\sigma_{x,y,z}$ the Pauli matrices acting on their pseudospin-1/2 space.
This Hamiltonian satisfies a parity-time ($\mathcal{PT}$) symmetry,
$\sigma_x H^*(k_x,k_y)\sigma_x=H(k_x,k_y)$, which
ensures real Bloch spectrum and the absence of bulk NHSE in the $\mathcal{PT}$-unbroken phase.

In Fig. \ref{model_with_energy}(b), we illustrate the OBC eigenenergies of the model and mark them by colors according to the fractal dimension of the corresponding eigenstates,
defined as 
\blue{\begin{eqnarray}
	D[\psi_n]
	=-\ln I/\ln\sqrt{L_x L_y},
\end{eqnarray}
where 
$I=\sum_{x,y}\vert\psi_{n}(x,y)\vert^4$ is the inverse participation ratio of a normalized eigenstate $\Psi_n$~\cite{wegner1980inverse,ganeshan2015nearest},}
$\psi_n(x,y)$ is the amplitude $\Psi_n$ at position $(x,y)$,
and $L_{x(y)}$ is the total number of sites along $x(y)$ direction. 
Driven by the non-reciprocal pumping, a $O(L)$ number of corner states characterized by $D[\psi_n]\approx 0$ are seen to emerge in the line-gap around ${\rm Re}[E]=0$, 
and a $O(L^2)$ number of bulk states remains extended with $D[\psi_n]\approx 2$,
due to the destructive interference of non-reciprocity~\citep{lee2019hybrid}.
The summed distribution of each of these two sets of eigenstates are shown in Fig. \ref{model_with_energy}(c) and (d).
Intriguingly, the corner states are seen to possess distinctive energetic features for different scenarios, which hints their sensitivity to the correlation between different edges.  
That is, eigenenergies of these corner states form a loop-like spectrum in Fig. \ref{model_with_energy}(b1), which turns into lines that enclose no area in the complex energy plane with either a larger $L_y$ that distances the top and bottom edges in real space [Fig. \ref{model_with_energy}(b2)], 
or \blue{ extra pseudospin-dependent imaginary energies $i g\sigma_z$} that separates the two edges in (imaginary) energy [Fig. \ref{model_with_energy}(b3)].



{\it Anisotropic-scaling corner localization.-}
To quantitatively describe the corner localization along the top and bottom edges,
we may consider an effective 1D two-leg ladder formed by these edges, with bulk lattices determining couplings between the two legs.
Explicitly, for the OBC Hamiltonian
\begin{eqnarray}
	H=\begin{pmatrix}
		H_\text{edge} & X \\
		X^T & H_\text{bulk} \\
	\end{pmatrix},\label{eq:H_block}
\end{eqnarray}
an effective Hamiltonian for edge states with eigenenergies $E_{\rm edge}$ can be obtained as \citep{suppmat}
\begin{eqnarray}
	\blue{H_\text{ladder}^\text{eff}}=H_\text{edge}+X(E_{\rm edge}-H_\text{bulk})^{-1}X^T,\label{eq:Heff}
\end{eqnarray}
where 
\begin{eqnarray}
	H_\text{edge}=\begin{pmatrix}
		H_A & 0 \\
		0 & H_B \\
	\end{pmatrix}\label{eq:edge1}
\end{eqnarray}
is the \blue{bare Hamiltonian of the top and bottom edges that form the two legs of the ladder}, 
\begin{eqnarray}
	H_\text{bulk}=\begin{pmatrix}
		H_B & t_2 & 0 & \cdots & 0 & 0\\
		t_2 & H_A & t_1 & \cdots & 0 & 0\\
		0 & t_1 & H_B & \cdots  & 0 & 0\\
		\vdots & \vdots & \vdots & \ddots & \vdots & \vdots\\
		0 & 0 & 0 & \cdots & H_B & t_2 \\
		0 & 0 & 0 & \cdots & t_2 & H_A \\
	\end{pmatrix}_{2(N_y-1)\times 2(N_y-1)}\label{eq:bulk}
\end{eqnarray}
is the bare bulk Hamiltonian with $N_y=L_y/2$ the number of unit cells along $y$ direction, and 
\begin{eqnarray}
	X=\begin{pmatrix}
		t_1 & 0 & \cdots & 0 & 0 \\
		0 & 0 & \cdots & 0 & t_1 \\
	\end{pmatrix}_{2\times 2(N_y-1)}\label{eq:X}
\end{eqnarray}
is the coupling between $H_\text{edge}$ and $H_\text{bulk}$. 
Note that under OBCs, $H_A$ and $H_B$ are $L_x\times L_x$ matrices that describe 1D HN chains, and $t_{1,2}$ shall be replaced by $t_{1,2}\mathcal{I}$ with $\mathcal{I}$ the  $L_x\times L_x$ identity matrix.
Alternatively, as derived in \citep{suppmat}, we may replace the matrices $H_{A(B)}$ with the non-Bloch solutions in the generalized Brillouin zone (GBZ) for $x-$OBC, without changing the form of $H_{\rm bulk}$ \citep{yao2018edge,yokomizo2019non,suppmat}.
With further derivation, we obtain 
\begin{eqnarray}
	\blue{H_\text{ladder}^\text{eff}}\approx\begin{pmatrix}
		H_A & \Delta^{(N_y)} \\
		\Delta^{(N_y)} & H_B \\
	\end{pmatrix}
\end{eqnarray}
with $\Delta^{(N_y)}=(-1)^{N_y-1}t_1^{N_y}/t_2^{N_y-1}$ the effective size-dependent coupling between the two edges. Note that $\blue{H_\text{ladder}^\text{eff}}$ represents a 1D two-leg ladder model 
possessing the critical NHSE \citep{li2020critical}, 
\blue{which generates SFL with}
localization length given by \citep{yokomizo2021scaling,suppmat}
\begin{eqnarray}
	\xi_x^{\rm AS}&\approx&
	-\frac{L_x+1}{\ln\left[\frac{(t_x^++t_x^-)^2 }{2(t_x^+-t_x^-)[(t_x^+)^2+(t_x^-)^2]}\Delta^{(N_y)}\right]}
	\nonumber\\
	&=&-\frac{L_x+1}{\ln\left[\frac{(t_x^++t_x^-)^2 t_2}{2(t_x^+-t_x^-)[(t_x^+)^2+(t_x^-)^2]}\right]+\frac{L_y}{2}\ln(\frac{t_1}{t_2})}.\label{eq:AS}
\end{eqnarray}
Thus we have $\xi_x^{\rm AS} \propto \frac{L_x}{L_y}$ when $L_{x,y}\gg1$, indicating an anisotropic dependence on the system's size in our 2D model. 

\begin{figure}
	\includegraphics[width=1.0\linewidth]{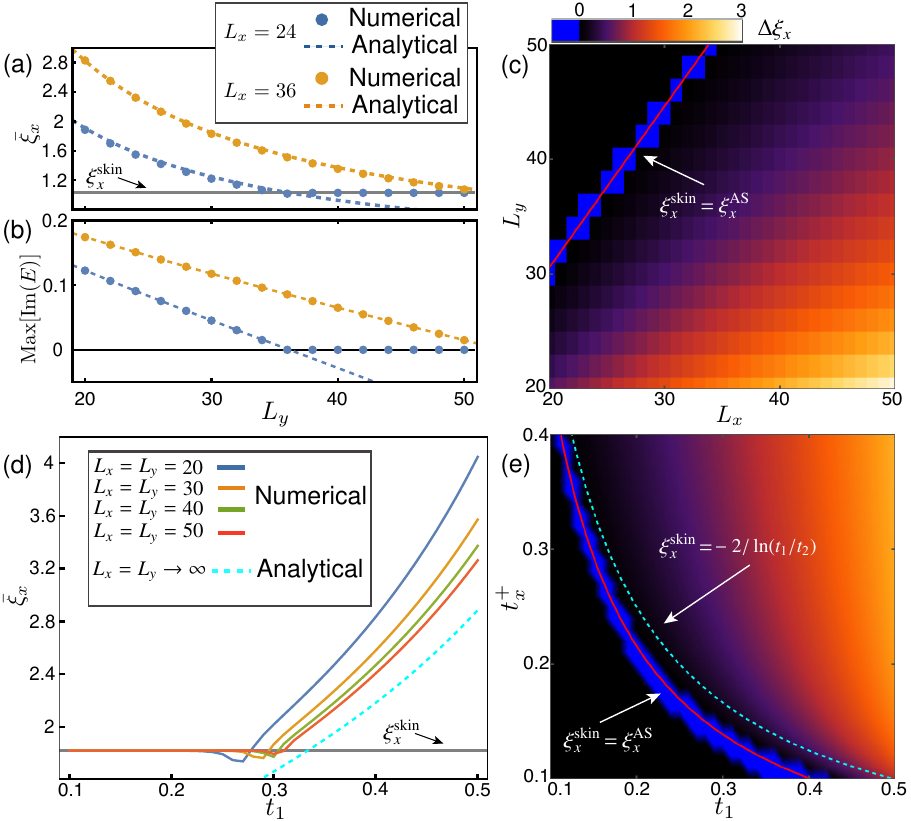}
	\caption{
	Average localization length (along $x$ direction) of corner states, $\bar{\xi}_x$, versus different parameters under OBCs. 
	(a) $\bar{\xi}_x$ (dots) versus $L_y$ with $L_x=24$ and $36$, respectively, in comparison with the analytical localization lengths of ASL ($\xi_x^{\rm AS}$, dash lines) and NHSE states ($\xi_x^{\rm skin}$, gray line).
	(b) Numerical (dots) and analytical (dashed line) results of the maximal imaginary eigenenergies, $\text{Max}[\text{Im}(E)]$, versus $L_y$ corresponding to (a). 
	It is seen that $\bar{\xi}_x\approx{\rm Max}\{ {\xi}_x^{\rm skin},{\xi}_x^{\rm AS}\}$, and the transition point of ${\xi}_x^{\rm skin}={\xi}_x^{\rm AS}$ matches the complex-real transition of eigenenergies. 
	(c) $\Delta\xi_x\equiv\bar{\xi}_x-\xi_x^{\rm skin}$ versus $L_x$ and $L_y$. 
	The analytical result of $\xi_x^{\rm AS}=\xi_x^{\rm skin}$ (red line) consists with the transition between $\Delta\xi_x=0$ and $\Delta\xi_x>0$.
	Negative $\Delta\xi_x$ (blue) is observed around the transition line between ASL and NHSE.
	(d) $\xi_x^{\rm AS}$ of Eq. \eqref{eq:infinite_L} (dashed line) and $\bar{\xi}_x$ (dots) versus $t_1$ for our model at different sizes. 
	With the size increased, numerical results are seen to approach the analytical prediction for the thermodynamic limit.
	(e) $\Delta\xi_x$ versus $t_1$ and $t_x^+$ for $L_x=L_y=30$, with analytical results of $\xi_x^{\rm AS}=\xi_x^{\rm skin}$ at this size (red line) and the thermodynamic limit (cyan dash line). Parameters are $t_1=0.25,t_2=1,t_x^+=0.35$, $t_x^-=0.05$, unless specified otherwise in the figures.
	In (c) and (e), $10^{-3}$ is chosen as the threshold of negative $\Delta\xi_x$ to distinguish from the black region with small negative $\Delta\xi_x$ due to numerical inaccuracy.
	}
	\label{phase}
\end{figure}

Such an ASL is verified by our numerical simulation, as demonstrated in Fig. \ref{phase}. In particular, we calculate the average localization length (along $x$ direction) of corner states 
\begin{eqnarray}
\bar{\xi}_x=\sum_{n}\xi_x^n/n_l,
\end{eqnarray}
with $n_l$ the total number of corner states ($n_l=2L_x$ in our case),
\begin{eqnarray}
	\xi_x^n=\big|\frac{L_x-1}{\ln[\vert\psi_n^c(L_x,1)\vert/\vert\psi_n^c(1,1)\vert]}\big|,\label{xi_x}
\end{eqnarray}
and $\psi_n^c(x,y)$ the amplitude of the $n$-th corner mode at position $(x,y)$. 
As shown in Fig. \ref{phase}(a), 
$\bar{\xi}_x$ matches our analytical result of $\xi^{\rm AS}_x$ in Eq.~\eqref{eq:AS} well for small $L_y$;
but takes a constant value of $\xi^{\rm skin}_x=-2/\ln(t_x^-/t_x^+)$ when $L_y$ is large and $\xi^{\rm AS}_x<\xi^{\rm skin}_x$, 
with $\xi_x^{\rm skin}$ the localization length of NHSE of a single HN chain \citep{suppmat}.
In addition, the transition between these two types of localization can be characterized by a $\mathcal{PT}-$symmetry breaking of the corner states, whose eigenenergies acquire nonzero imaginary values only for ASL states, as shown in Fig. \ref{phase}(b). It can be seen that the analytical solution of the maximal imaginary energy, given by $\text{Max}[\text{Im}(E)]\approx t_x^+\exp(-1/\xi_x^{\rm AS})-t_x^-\exp(1/\xi_x^{\rm AS})$ \citep{yokomizo2021scaling,suppmat}, 
agrees with the numerical results for ASL states. In Fig. \ref{phase}(c), we further demonstrate a diagram of $\Delta\xi_x=\bar{\xi}_x-\xi_x^{\rm skin}$ versus $L_x$ and $L_y$, 
where ASL and NHSE are characterized by $\Delta\xi_x>0$ and $\Delta\xi_x=0$, respectively.
That is, the system always favors the one with longer localization length (hence weaker localization) between these two types of localization.
We note that $\Delta \xi_x<0$ is numerically observed along the transition boundaries, which is possibly due to 
some instability near transition points and
the intricate localizing behaviors in non-Hermitian systems beyond the approximated exponential decay~\citep{kawabata2021nonunitary,davies2024two,zhang2024algebraic}

Despite their similarity, the 2D nature of ASL states induces unique features in contrast to SFL states in 1D. As shown in Fig. \ref{phase}(d), the system enters ASL regime when $t_1$ exceeds a critical value (so that $\xi_x^{\rm AS}>\xi_x^{\rm skin}$),
where  
\begin{equation}
	\bar{\xi}_x\approx\xi_x^{\rm AS}\rightarrow-2r/\ln(t_1/t_2)
	\label{eq:infinite_L}
\end{equation}
for $L_x=rL_y\rightarrow\infty$,
becoming roughly unchanged with $t_x^+$. In other words, in the thermodynamic limit, the localization length of ASL states approaches a fixed value determined solely by hopping amplitudes $t_{1,2}$ along $y$ direction, independent from the non-Hermitian parameters $t_x^{\pm}$ that induce the corner localization. Furthermore, unlike SFL states in 1D that become extended in the thermodynamic limit, Eq. (\ref{eq:infinite_L}) suggests that ASL states in our 2D model always remain localized, provided $L_x/L_y$ stays finite when the system's size approaches infinity. The phase boundary between ASL and NHSE is thus given by $t_1/t_2=(t_x^-/t_x^+)^r$,
which is slightly shifted away in our numerical simulation due to finite-size effect, 
as shown in Fig. \ref{phase}(d) and (e).

\begin{figure*}
	\includegraphics[width=1.0\linewidth]{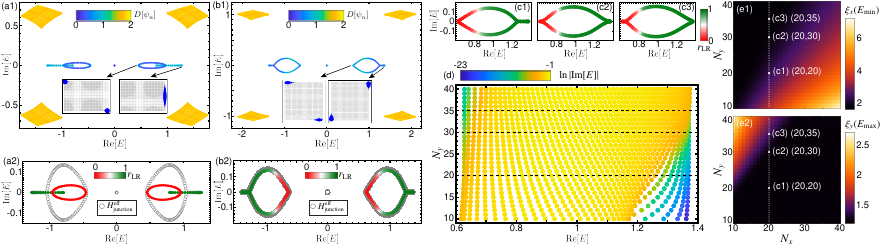}
	\caption{NHSE and ASL of the non-Hermitian BBH model.
	(a1) Energy spectrum of the non-Hermitian BBH model with non-Hermicity only along $x$ direction. 	Insets display the distribution of the eigenstates marked by arrows.
	(a2) Comparison between the eigenenergies of the edge states (colored solid dots) in (a1) and that of the effective 1D junction model formed by the edges of the 2D system (gray circles). 
	Colors indicate the edge distribution ratio $r_{\rm LR}$ of each edge states.
	Parameters are $t_x^+=2,t_x^-=0.25,t_y^+=t_y^-=1,t'=0.25,N_x=30$ and $N_y=20$.	
	(b1) and (b2) the same as (a1) and (a2), but with non-Hermicity along both directions.
	Parameters are $t_x^+=1.5,t_y^+=2.5,t_x^-=t_y^-=0.5,t'=0.25$ and $N_x=N_y=20$.  
	(c) Edge spectra with positive real energy (marked by $r_{\rm LR}$), with $N_x=20$, ${\rm Re}[E]>0$, and $N_y=20,30,35$ for (c1)-(c3) respectively.
	(d) Real part of edge spectra with $N_x=20$ and ${\rm Re}[E]>0$ versus $N_y$. Colors indicate the value of $\ln\vert{\rm Im}[E]\vert$. The three dash lines mark the cases for (c1) to (c3), respectively. 
	(e) Localization length along $x$ ($y$) direction of corner state with the minimal (maximal) absolute value of real energy, $\xi_x(E_{\rm min})$ [$\xi_y(E_{\rm max})$], versus the system's size $N_x$ and $N_y$.
\blue{Here $\xi_{x/y}$ is defined regarding the sublattice strucutre of the BBH model; see Supplemental Materials~\citep{suppmat} for more details.}
Other parameters in (c) to (e) are same as that in (b).
	}
	\label{BBH_figs}
\end{figure*}

{\it ASL in non-Hermitian BBH model.-}
Originated from the correlation between distanced edges, ASL is expected to generally emerge in two- or higher-dimensional non-Hermitian models with edge localization, yet the underlying mechanisms may differ depending on how different edges are effectively coupled to each other.
To see this, we consider a 2D non-Hermitian Benalcazar-Bernevig-Hughes (BBH) model \citep{benalcazar2017quantized,benalcazar2017electric,lee2019hybrid}
with edge localization along both $x$ and $y$ directions, whose Bloch Hamiltonian reads
\begin{equation}
	H_\text{BBH}(\mathbf{k})=\begin{pmatrix}
		0 & t'+t_x^- e^{-ik_x} & -t'-t_y^+ e^{ik_y} & 0\\
		t'+t_x^+ e^{ik_x} & 0 & 0& t'+t_y^- e^{ik_y}\\
		-t'-t_y^- e^{-ik_y} & 0 & 0 & t'+t_x^+ e^{-ik_x}\\
		0 & t'+t_y^+ e^{-ik_y} & t'+t_x^- e^{ik_x} & 0\\
	\end{pmatrix},
\end{equation}
where $t'$ represents the intracell hopping amplitude and $t_{x(y)}^\pm$ represents the non-reciprocal intercell hopping amplitude along $x(y)$ direction. 
The system's size in real space is chosen to have $N_x$ ($N_y$) unit cells along $x$ ($y$) direction, with $N_x=L_x/2$ ($N_y=L_y/2$).
In the Hermitian scenario with $t_{x(y)}^+=t_{x(y)}^->t'$, the BBH model under OBCs supports 1D edge states at all the four edges, differing from the HN-SSH model with edge states localized only at top and bottom edges. 

Analogous to the HN-SSH model, 
when non-Hermiticity is introduced to the BBH model along $x$ direction (e.g., $t_x^+>t_x^-$), a similar effective ladder Hamiltonian can be obtained, containing two non-Hermitian SSH chains coupled via size-dependent couplings arisen from the bulk~\citep{suppmat}.
ASL corner states emerge accordingly at top-left and bottom-right corners with their eigenenergies forming two loops in the complex energy plane, as shown in Fig. \ref{BBH_figs}(a).
On the other hand, left- and right-edge states are seen to be unaffected by the non-Hermiticity, remaining extended along the 1D edges with real eigenvalues.
In Fig. \ref{BBH_figs}(a2), we mark these edge states by colors according to
a edge distribution ratio, defined as
 $$r_\text{LR}[\psi_n]=\rho_\text{LR}/(\rho_\text{LR}+\rho_\text{TB}),$$ 
 with $\rho_{\text{LR}(\text{TB})}=\sum_{(x,y)\in\text{LR}(\text{TB})}\vert\psi_n(x,y)\vert^2$ the summed distribution of an eigenstate on left and right (top and bottom) edges excluding the four corners.
It can be seen that the loop and line spectra are clearly distinguished by $r_{\rm LR}\approx0$ and $1$, 
suggesting that the top-bottom and left-right edges form two separable edge systems, without much interference between each other.


When non-Hermiticity is further introduced along both directions (e.g., $t_{y}^+>t_{x}^+>t_x^-=t_y^->t'$), a $O(L)$ number of corner states is also observed, while 1D edge states are absent as all the four edges now suffer from non-Hermitian non-reciprocal pumping~[Fig. \ref{BBH_figs}(b)].
In this case, numerical simulations show that the edge spectrum does not match the ladder description (details in Supplemental Materials~\citep{suppmat}).
Alternatively, the edges can be viewed as a 1D junction system of different non-Hermitian SSH chains \citep{ou2023non},
\blue{described by the Hamiltonian \cite{suppmat}
\begin{eqnarray}
\hat{H}_\text{junction}^\text{eff}=\hat{P}_{\rm edge}\hat{H}_{\rm BBH}\hat{P}_{\rm edge},
\end{eqnarray}
with $\hat{P}_{\rm edge}=\sum_{\mathbf{n}}|\mathbf{n}\rangle\langle\mathbf{n}|$ and $\mathbf{n}$ the lattice sites along the four edges, $\mathbf{n}=(x,y)$ where $x\in\{1,L_x\}$ or $y\in\{1,L_y\}$.}
As shown in Fig. \ref{BBH_figs}(b2), 
$\hat{H}_\text{junction}^\text{eff}$
accurately reproduces the edge spectrum of the non-Hermitian BBH model, insensitive to effective weak couplings through the bulk
\footnote{The junction system of the edges effectively form a 1D system with closed boundary conditions. In contrast, in Refs.~\citep{budich2020non,guo2021exact}, the extreme sensitivity is a property of OBC systems with NHSE, where extra weak couplings effectively close the boundaries and greatly affect the GBZ description of the OBC eigensolutions.}.
Note that the junction description fails to predict the spectral properties in Fig. \ref{BBH_figs}(a2) where non-Hermiticity is introduced only along a single direction, which in turn verifies the origin of effective bulk couplings for ASL states therein.

Interestingly, given that non-Hermitian non-reciprocity acts along both $x$ and $y$ directions, 
corner localization now arises from all edge states, which may exhibit different ASL or NHSE depending on the size and parameters.
As can be seen from Fig. \ref{BBH_figs}(b2) and (c), corner states are separated into two groups in real energy with $r_\text{LR}\approx 1$ and $r_\text{LR}\approx 0$, 
indicating that they distribute mostly on left-right and top-bottom edges, respectively.
In addition, 
each group of corner states is found to possess both real and complex eigenenergies in certain parameter regimes [Fig. \ref{BBH_figs}(c1) to (c3)],
suggesting the co-existence of ASL and NHSE along the same edges.
In Fig. \ref{BBH_figs}(d), we demonstrate the amplitude of ${\rm Im}[E]$ for edge states with ${\rm Re}[E]>0$ versus $N_y$, 
where transition between (almost) real and complex eigenenergies are found to occur at different values of $N_y$ for states with different ${\rm Re}[E]$.

Finally,
we calculate the localization length along $x$ ($y$) direction of the corner state with the minimal (maximal) absolute value of real energy,
i.e. $E_{\rm min}$ : $\text{Re}[E_{\rm min}]=\text{Min}\vert\text{Re}[E]\vert$ ($E_{\rm max}$ : $\text{Re}[E_{\rm max}]=\text{Max}\vert\text{Re}[E]\vert$), as an example for illustrating the size-dependent behaviors of states with $r_\text{LR}\approx 0$ ($r_\text{LR}\approx 1$).
The numerical results versus the system's size $N_x$ and $N_y$ are demonstrated in Fig. \ref{BBH_figs}(e1) and (e2),
where NHSE and ASL are identified by fixed $\xi_{x,y}$ (black regions) and size-dependent $\xi_{x,y}$ (colored regions), respectively.
The transition between them is seen to consist with the transition between (almost) real and complex energies for $E=E_{\rm min}$ and $E=E_{\rm max}$ in Fig. \ref{BBH_figs}(d).
Remarkably,  as these two ASL states localize along different directions, they manifest size dependencies that are opposite to each other,
namely, $\xi_x(E_{\rm min})\propto N_x/N_y$ and $\xi_y(E_{\rm max})\propto N_y/N_x$. 
\blue{We also note that the effective junction model $H^{\rm eff}_{\rm junction}$ can be further mapped to a spatially homogeneous model with two local impurities that generate SFL~\cite{suppmat,li2021impurity}.
It indicates a different origin of the ASL here, in contrast to that induced by critical NHSE in the effective ladder description of $H^{\rm eff}_{\rm ladder}$.}

{\it Discussion.-}
We have unveiled a class of ASL in non-Hermitian systems, originated from effective bulk couplings or edge junctions between different non-Hermitian boundaries of the lattices.
Similar phenomena are usually considered as hybrid skin-topological effect~\citep{zhu2024brief,lee2019hybrid,li2022gain,zhu2022hybrid,li2020topological,ou2023non} or higher-order NHSE~\citep{kawabata2020higher,okugawa2020second,fu2021non}  previously, which do not capture the anisotropic size-dependency of the localization profile for ASL states.
Unlike NHSE that possesses curve-like spectrum, ASL is accompanied by complex eigenenergies enclosing nonzero areas in the complex plane,
which solves the mysterious co-existence of skin-like localization and loop-like spectrum of boundary states that is forbidden in 1D systems
\citep{kawabata2020higher,li2023enhancement}.
Intriguingly,
despite originating from non-Hermiticy, the localization length of ASL in the thermodynamic limit may depend only on the Hermitian parameters of the system, which provides a means to detect even an extremely weak non-Hermitian effect that induces ASL states with a noticeable localization length in a given Hermitian system.
\blue{Finally, the ASL is also found to be robust in the presence of disorder~\cite{suppmat}, making it feasible in experimental implementation.}

\blue{Given the rapid experimental developments of realizing non-Hermitian systems~\citep{zou2021observation,palacios2021guided,zhang2021observation,liang2022dynamic,zhao2025two}, our model can be implemented across various platforms, e.g., the non-Hermitian BBH model has already been experimentally realized in electrical circuits~\citep{zou2021observation}, while loop energy spectrum for corner states similar to those of HN-SSH model presented here have been achieved in active particle systems~\citep{palacios2021guided}. Furthermore, in quantum simulations, we numerically find that the ASL also exists in a prototype model of cold atoms loaded in optical lattices with laser-induced atom loss~\citep{li2020topological}, as detailed in Supplemental Materials~\citep{suppmat}.}

In higher dimensions, non-Hermitian localization is known to exhibit many fascinating phenomena beyond its 1D counterpart,
such as the geometry-dependency~\citep{zhang2022universal} and directional toggling~\cite{lei2024activating} of first-order bulk NHSE,
which lead to the development of several delicate formulations for describing non-Hermitian bulk bands~\citep{zhang2022universal,wang2024amoeba,zhang2024algebraic,hu2024topological,xiong2024non}.
In comparison, the ASL formalism provides a different perspective for investigating higher-order non-Hermitian localization along boundaries, whose generalization into higher dimensions is straightforward~\citep{suppmat}.
Combining these distinguished aspects, even more complicated phenomena may emerge when increasing the spatial dimension.
For example, 
a surface of a 3D lattice may support the interplay or competition between these known non-Hermitian bulk and boundary localizations,
as it can be viewed as either a boundary of the 3D system, and as the bulk of a reduced 2D lattice.
\red{In addition, it is also nontrivial to generalize our results to boarder classes of systems where the hopping cannot
be taken as nearest-neighbor always. A preliminary study in Supplemental Materials \cite{suppmat} shows the existence of ASL under exponentially decaying long-range coupling, yet their detailed behaviors still await further exploration.}
In this regard, our finding can trigger further investigations into richer size-dependencies of non-Hermitian localization in broader classes of systems, 
which shall be of significant interest for experimental implementations of non-Hermitian physics in finite-size lattices.

\section*{Acknowledgement}
Z. Ou and H.-Q. Liang  contributed equally to this work.
This work is supported by 
the National Natural Science Foundation of China (Grant No. 12474159 and No. 12174224), and the Guangdong Project (Grant No.
2021QN02X073).

\onecolumngrid

\clearpage
\begin{center}
\textbf{\large Supplemental Materials for ``Anisotropic-scaling localization in higher-dimensional non-Hermitian systems"}
\end{center}

\tableofcontents
\setcounter{secnumdepth}{2}

	\setcounter{equation}{0} \setcounter{figure}{0} \setcounter{table}{0} %
	\renewcommand{\theequation}{S\arabic{equation}} \renewcommand{\thefigure}{S%
		\arabic{figure}} 

\section{Effective edge Hamiltonian of the HN-SSH model}
\label{secS1}

The Schr{\"o}dinger equation of HN-SSH model under OBC can be written as
\begin{eqnarray}
	\begin{pmatrix}
		H_A & t_1 & 0 & \cdots & 0 & 0 \\
		t_1 & H_B & t_2 & \cdots & 0 & 0 \\
		0 & t_2 & H_A & \cdots & 0 & 0 \\
		\vdots & \vdots & \vdots & \ddots & \vdots & \vdots \\
		0 & 0 & 0 & \cdots & H_A & t_1 \\
		0 & 0 & 0 & \cdots & t_1 & H_B \\
	\end{pmatrix}
	\begin{pmatrix}
		\Psi_A^{1} \\
		\Psi_B^{1} \\
		\Psi_A^{2} \\
		\vdots \\
		\Psi_A^{L_y/2} \\
		\Psi_B^{L_y/2} \\
	\end{pmatrix}=E\begin{pmatrix}
		\Psi_A^{1} \\
		\Psi_B^{1} \\
		\Psi_A^{2} \\
		\vdots \\
		\Psi_A^{L_y/2} \\
		\Psi_B^{L_y/2} \\
	\end{pmatrix}.
	\end{eqnarray}
where $\Psi_\alpha^{n_y}=(\psi_\alpha^{1,n_y},\psi_\alpha^{2,n_y},\cdots,\psi_\alpha^{L_x-1,n_y},\psi_\alpha^{L_x,n_y})^T$ ($\alpha=A,B$ and $n_y=1,2,...,L_y/2$) 
is an eigenstate in the subspace of each Hatano-Nelson chain (labeled by $n_y$ and $\alpha$), $E$ is the corresponding eigenenergy, and 
$H_{A(B)}$ is a $L_x\times L_x$ matrix whose form is 
\begin{eqnarray}
	H_{A(B)}=\begin{pmatrix}
		0 & t_x^{+(-)} & 0 & \cdots & 0 & 0 \\
		t_x^{-(+)} & 0 & t_x^{+(-)} & \cdots & 0 & 0 \\
		0 & t_x^{-(+)} & 0 & \cdots & 0 & 0 \\
		\vdots & \vdots & \vdots & \ddots & \vdots & \vdots \\
		0 & 0 & 0 & \cdots & 0 & t_x^{+(-)} \\
		0 & 0 & 0 & \cdots & t_x^{-(+)} & 0 \\
	\end{pmatrix}, 
\end{eqnarray}
describing a Hatano-Nelson chain under OBC. The bulk equations are  
\begin{subequations}
\begin{align}
	t_2\psi_B^{n_x,n_y-1}+t_x^-\psi_A^{n_x-1,n_y}+t_x^+\psi_A^{n_x+1,n_y}+t_1\psi_B^{n_x,n_y}=E\psi_A^{n_x,n_y}, \\
	t_1\psi_A^{n_x,n_y}+t_x^+\psi_B^{n_x-1,n_y}+t_x^-\psi_B^{n_x+1,n_y}+t_2\psi_A^{n_x,n_y+1}=E\psi_B^{n_x,n_y},
\end{align}
\end{subequations}
and the equations for the boundaries along $y-$direction, $y=1$ and $y=L_y$, are given by
\begin{subequations}
\begin{align}
	t_x^-\psi_A^{n_x-1,1}+t_x^+\psi_A^{n_x+1,1}+t_1\psi_B^{n_x,1}=E\psi_A^{n_x,1}, \\
	t_1\psi_A^{n_x,L_y/2}+t_x^+\psi_B^{n_x-1,L_y/2}+t_x^-\psi_B^{n_x+1,L_y/2}=E\psi_B^{n_x,L_y/2}. 
\end{align}
\end{subequations}
According to the linear difference equations, we can take an ansatz for the eigenstates as a linear combination, 
\begin{eqnarray}
	\begin{pmatrix}
		\psi_A^{n_x,n_y} \\
		\psi_B^{n_x,n_y} \\
	\end{pmatrix}=\sum_j^{2L_y}(\beta_j)^{n_x}
	\begin{pmatrix}
		\phi_A^{n_y,(j)} \\
		\phi_B^{n_y,(j)} \\
	\end{pmatrix}. \label{eq:ansatz}
\end{eqnarray}
With this ansatz, the Schr{\"o}dinger equation can be transformed into
\begin{eqnarray}
	\begin{pmatrix}
		h_A & t_1 & 0 & \cdots & 0 & 0 \\
		t_1 & h_B & t_2 & \cdots & 0 & 0 \\
		0 & t_2 & h_A & \cdots & 0 & 0 \\
		\vdots & \vdots & \vdots & \ddots & \vdots & \vdots \\
		0 & 0 & 0 & \cdots & h_A & t_1 \\
		0 & 0 & 0 & \cdots & t_1 & h_B \\
	\end{pmatrix}
	\begin{pmatrix}
		\phi_A^{1} \\
		\phi_B^{1} \\
		\phi_A^{2} \\
		\vdots \\
		\phi_A^{L_y/2} \\
		\phi_B^{L_y/2} \\
	\end{pmatrix}=E\begin{pmatrix}
		\phi_A^{1} \\
		\phi_B^{1} \\
		\phi_A^{2} \\
		\vdots \\
		\phi_A^{L_y/2} \\
		\phi_B^{L_y/2} \\
	\end{pmatrix},
	\label{eq:xGBZ_HN-SSH}
\end{eqnarray}
where \begin{eqnarray}
	h_{A(B)}=t_x^{\pm}\beta+t_x^{\mp}\beta^{-1}
	\label{poly_ab}
\end{eqnarray}
is a polynomial of $\beta=\beta_j$, and $\phi_\alpha^{n_y}=\phi_\alpha^{n_y,(j)}$. 
In fact, under the periodic boundary conditions (PBCs) along $x$ direction, the Bloch Hamiltonian has the same form with Eq.~(\ref{eq:xGBZ_HN-SSH}) but $h_{A(B)}\rightarrow h_{A(B)}(k)=t_x^{\pm}e^{ik}+t_x^{\mp}e^{-ik}$. 
In other words, 
deforming the boundary conditions along $x$ direction from PBCs to OBCs is equivalent to the replacement $e^{\pm ik}\rightarrow\beta^{\pm 1}$, which can be understood as the generalized Brillouin zone of the non-Bloch band theory~\cite{yao2018edge,yokomizo2019non} (also see Supplemental Note 1).


For each $\beta$, we can rewrite the Hamiltonian in Eq.~(\ref{eq:xGBZ_HN-SSH}) as
\begin{eqnarray}
	h=\begin{pmatrix}
		h_\text{edge} & X \\
		X^T & h_\text{bulk} \\
	\end{pmatrix},
\end{eqnarray}
where
\begin{eqnarray}
	h_\text{edge}=\begin{pmatrix}
		h_A & 0 \\
		0 & h_B \\
	\end{pmatrix}_{2\times 2}\label{eq:edge}
\end{eqnarray}
is the edge Hamiltonian, 
\begin{eqnarray}
	h_\text{bulk}=\begin{pmatrix}
		h_B & t_2 & 0 & \cdots & 0 & 0\\
		t_2 & h_A & t_1 & \cdots & 0 & 0\\
		0 & t_1 & h_B & \cdots  & 0 & 0\\
		\vdots & \vdots & \vdots & \ddots & \vdots & \vdots\\
		0 & 0 & 0 & \cdots & h_B & t_2 \\
		0 & 0 & 0 & \cdots & t_2 & h_A \\
	\end{pmatrix}_{2(N_y-1)\times 2(N_y-1)}
\end{eqnarray}
is the bulk Hamiltonian with $N_y=L_y/2$ the number of unit cells along $y$ direction, and 
\begin{eqnarray}
	X=\begin{pmatrix}
		t_1 & 0 & \cdots & 0 & 0 \\
		0 & 0 & \cdots & 0 & t_1 \\
	\end{pmatrix}_{2\times 2(N_y-1)}
\end{eqnarray}
is the coupling between $h_\text{edge}$ and $h_\text{bulk}$. 
Then the Schr{\"o}dinger equation can be written as
\begin{eqnarray}
	h	\begin{pmatrix}
		\psi_{\rm edge} \\
		\psi_{\rm bulk}  \\
	\end{pmatrix}=\begin{pmatrix}
		h_\text{edge} & X \\
		X^T & h_\text{bulk} \\
	\end{pmatrix}\begin{pmatrix}
		\psi_{\rm edge} \\
		\psi_{\rm bulk}  \\
	\end{pmatrix}=E\begin{pmatrix}
		\psi_{\rm edge} \\
		\psi_{\rm bulk}  \\
	\end{pmatrix},
\end{eqnarray}
with $\psi_{\rm edge}$ and $\psi_{\rm bulk}$ the edge and boundary parts of the eigenstates, respectively.
Eliminating $\psi_{\rm bulk}$ from its characteristic equation,
we obtain
\begin{eqnarray}
h_{\rm edge}\psi_{\rm edge}+X (E-h_{\rm bulk})^{-1} X^T\psi_{\rm edge}=E\psi_{\rm edge}.
\end{eqnarray}
Thus we can get the effective edge Hamiltonian for $\psi_{\rm edge}$ as 
\begin{eqnarray}
	h_\text{eff}(E)&=&h_\text{edge}+X(E-h_\text{bulk})^{-1}X^T\nonumber\\
	&=&\begin{pmatrix}
		h_A+t_1^2 G_{11}(E) & t_1^2 G_{1,2(N_y-1)}(E) \\
		t_1^2 G_{2(N_y-1),1}(E) & h_B+t_1^2 G_{2(N_y-1),2(N_y-1)}(E) \\
	\end{pmatrix},\label{eq:h_Green}
\end{eqnarray}
where $G_{ml}(E)$ is the $(m,l)-$element of the bulk Green function with a reference energy $E$, $G(E)=(E-h_\text{bulk})^{-1}$. 

In the following, we calculate the elements of the bulk Green function. 
With the relationship of the Green function, $(E-H_{\rm bulk})G(E)=\mathcal{I}$, we obtain the following recurrence relations, 
\begin{eqnarray}
	(E-h_B)G_{1,j}-t_2 G_{2,j}=\delta_{1,j}, 
	\label{G_relation_1}
\end{eqnarray}
\begin{eqnarray}
	-t_2 G_{i-1,j}+(E-h_A)G_{i,j}-t_1 G_{i+1,j}=\delta_{i,j},\quad i \text{ is even}, 
	\label{G_relation_2}
\end{eqnarray}
\begin{eqnarray}
	-t_1 G_{i-1,j}+(E-h_B)G_{i,j}-t_2 G_{i+1,j}=\delta_{i,j},\quad i \text{ is odd}, 
	\label{G_relation_3}
\end{eqnarray}
\begin{eqnarray}
	-t_2 G_{2N_y-3,j}+(E-h_A)G_{2N_y-2,j}=\delta_{2N_y-2,j}. 
	\label{G_relation_4}
\end{eqnarray}
Specially, at $E=h_A$, Eqs.~(\ref{G_relation_2}) and (\ref{G_relation_4}) are reduced as 
\begin{eqnarray}
	-t_2 G_{i-1,j}-t_1 G_{i+1,j}=\delta_{i,j},\quad i \text{ is even}, 
\end{eqnarray}
\begin{eqnarray}
	-t_2 G_{2N_y-3,j}=\delta_{2N_y-2,j}. 
\end{eqnarray}
Then we obtain 
\begin{eqnarray}
	G_{11}(h_A)=(-t_1/t_2)^{N_y-2}G_{2N_y-3,1}(h_A)=0, 
\end{eqnarray}
\begin{eqnarray}
	G_{1,2(N_y-1)}(h_A)&=&(-t_1/t_2)^{N_y-2}G_{2N_y-3,2(N_y-1)}(h_A)\notag\\
	&=&\frac{(-1)^{N_y-1}t_1^{N_y-2}}{t_2^{N_y-1}}.
\end{eqnarray}
On the other hand, at $E=h_B$, Eqs.~(\ref{G_relation_1}) and (\ref{G_relation_3}) are reduced as 
\begin{eqnarray}
	-t_2 G_{2,j}=\delta_{1,j}, 
\end{eqnarray}
\begin{eqnarray}
	-t_1 G_{i-1,j}-t_2 G_{i+1,j}=\delta_{i,j},\quad i \text{ is odd}. 
\end{eqnarray}
Then we obtain 
\begin{eqnarray}
	G_{2(N_y-1),1}(h_B)&=&(-t_1/t_2)^{N_y-2}G_{21}(h_B)\notag\\
	&=&\frac{(-1)^{N_y-1}t_1^{N_y-2}}{t_2^{N_y-1}},
\end{eqnarray}
\begin{eqnarray}
	G_{2(N_y-1),2(N_y-1)}(h_B)=(-t_1/t_2)^{N_y-2}G_{2,2(N_y-1)}(h_B)=0.
\end{eqnarray}

\section{Approximation of the effective edge Hamiltonian}\label{sec:Heff_appro}
The concrete form of the effective edge Hamiltonian [as in Eq.~\eqref{eq:h_Green}] 
involves the eigenenergy $E$ of the whole Hamiltonian, making the Schr{\"o}dinger equation nonlinear on $E$ and difficult to be solved analytically.
Alternatively,
since we are interesting on how boundary states are affected by the bulk of the lattice,
we may take the edge Hamiltonian [as in Eq.~\eqref{eq:edge}] as the unperturbed Hamiltonian, 
whose eigensolutions are given by 
$$E_A^{(0)}=h_A,E_B^{(0)}=h_B;~\vert\phi_A^{(0)}\rangle=(1,0)^T,\vert\phi_B^{(0)}\rangle=(0,1)^T.$$
Thus, by taking the rest terms in Eq.~\eqref{eq:h_Green} as perturbation,
an approximation of the effective edge Hamiltonian can be obtain as
\begin{eqnarray}
	h_\text{eff}^\text{edge}&\approx&\begin{pmatrix}
		\langle\phi_A^{(0)}\vert h_\text{eff}(h_A)\vert\phi_A^{(0)}\rangle & \langle\phi_A^{(0)}\vert h_\text{eff}(h_B)\vert\phi_B^{(0)}\rangle \\
		\langle\phi_B^{(0)}\vert h_\text{eff}(h_A)\vert\phi_A^{(0)}\rangle & \langle\phi_B^{(0)}\vert h_\text{eff}(h_B)\vert\phi_B^{(0)}\rangle \\
	\end{pmatrix}\notag\\
	&=&\begin{pmatrix}
		h_A+t_1^2G_{11}(h_A) & t_1^2G_{1,2(N_y-1)}(h_B) \\
		t_1^2G_{2(N_y-1),1}(h_A) & h_B+t_1^2G_{2(N_y-1),2(N_y-1)}(h_B) \\
	\end{pmatrix}\notag\\
	&=&\begin{pmatrix}
		h_A & \Delta^{({N_y})} \\
		\Delta^{({N_y})} & h_B \\
	\end{pmatrix}, 
	\label{eff_h_app}
\end{eqnarray}
where $\Delta^{(N_y)}={(-1)^{N_y-1}t_1^{N_y}}/{t_2^{N_y-1}}$. 
Note that the formal solution of $h_\text{eff}^\text{edge}$ is obtained by taking the ansatz in Eq.~\eqref{eq:ansatz} for the original Hamiltonian $H$. Thus, the effective edge Hamiltonian of $H$ can also be approximately given by
\begin{eqnarray}
	H_\text{eff}^\text{edge}\approx\begin{pmatrix}
		H_A & \Delta^{(N_y)} \\
		\Delta^{(N_y)} & H_B \\
	\end{pmatrix}, 
\end{eqnarray}
which describes a two-leg ladder model composed by chain $A$ and chain $B$ with interchain coupling $\Delta^{(N_y)}$, where $H_{A(B)}$ is a $L_x\times L_x$ matrix describing chain $A(B)$. 

\section{Skin states of one-dimensional Hatano-Nelson model}\label{sec:1D_skin}
In this section, we briefly review the non-Hermitian skin effect (NHSE) \cite{yao2018edge} in the one-dimensional Hantano-Nelson model \cite{hatano1996localization,hatano1998non}, 
whose Bloch Hamiltonian reads, 
\begin{eqnarray}
	h_\text{HN}(k)=t_L e^{-ik}+t_R e^{ik}. 
\end{eqnarray}
Its OBC Hamiltonian in real space is given by
\begin{eqnarray}
	H_\text{HN}=\sum_n^{L-1}(t_L c_n^\dagger c_{n+1}+t_R c_{n+1}^\dagger c_n),
\end{eqnarray}
where $c_n^{(\dagger)}$ is annihilation (creation) operator at the $n$th site, $t_{L},t_R$ are the hopping amplitudes, and $L$ is the size of the system.
The eigenfunction $H_\text{HN}\vert\psi\rangle=E_\text{OBC}\vert\psi\rangle$ with $\vert\psi\rangle=(\psi_1,\psi_2,\cdots,\psi_L)^T$
gives the bulk equation
\begin{eqnarray}
	t_R\psi_{n-1}+t_L\psi_{n+1}=E\psi_n, \label{eq:bulk_HN}
\end{eqnarray}
with OBCs given by $\psi_0=\psi_{L+1}=0$. 
According to the linear differences equations, we can take an ansatz for the eigenstates as a linear combination, 
\begin{eqnarray}
	\psi_n=(\beta_1)^n\phi^{(1)}+(\beta_2)^n\phi^{(2)}.\label{eq:ansatz_HN}
\end{eqnarray}
Thus the boundary conditions become
\begin{eqnarray}
	\begin{pmatrix}
		1 & 1 \\
		\beta_1^{L+1} & \beta_2^{L+1} \\
	\end{pmatrix}
	\begin{pmatrix}
		\phi^{(1)} \\
		\phi^{(2)} \\
	\end{pmatrix}=
	\begin{pmatrix}
		0 \\ 0 \\
	\end{pmatrix}, 
\end{eqnarray}
which requires $(\beta_2/\beta_1)^{L+1}=1$ to have nontrivial solutions of $\phi^{(1)}$ and $\phi^{(2)}$. 
On the other hand, substituting the ansatz of Eq.~\eqref{eq:ansatz_HN} to Eq.~\eqref{eq:bulk_HN}, the solutions of $\beta$ requires $\beta_1\beta_2=t_R/t_L$. Combining these results, we obtain 
\begin{eqnarray}
	\vert\beta_1\vert=\vert\beta_2\vert=\vert\beta\vert=\sqrt{\left\vert\frac{t_R}{t_L}\right\vert}. 
\end{eqnarray}
If $t_L=t_R$, $H_\text{HN}$ is Hermitian Hamiltonian and its bulk state, $\vert\psi\rangle$, is extend state. 
For $t_L>t_R$ ($t_L<t_R$), we have $\vert\beta\vert<1$ ($\vert\beta\vert>1$) and $\vert\psi\rangle$ is localized state which is localized at the left (right) edge of the model. This is known as the NHSE and we can find that the localization length of the model is 
$\xi_\text{HN}=\big|1/\ln\vert\beta\vert\big|=\big|2/\ln(t_R/t_L)\big|$ [$\xi_x^\text{skin}=-2/\ln(t_x^-/t_x^+)$ in the main text for the edges of the HN-SSH model, as $t_x^-<t_x^+$].

\section{Scaling rule for critical non-Hermitian skin effect}In this section, 
we reproduce the calculation of localization length for the critical NHSE in Ref.~\cite{yokomizo2021scaling,qin2023universal}, which leads to Eq.~(9) 
 in the main text.
The Hamiltonian for the two-leg ladder model with critical NHSE reads
\begin{eqnarray}
	H_\text{cNHSE}=\sum_n^{L-1}(t_+ c^\dagger_{n,A}c_{n+1,A}+t_- c^\dagger_{n+1,A}c_{n,A}+t_- c^\dagger_{n,B} c_{n+1,B}+t_+ c^\dagger_{n+1,B}c_{n,B}+\Delta c^\dagger_{n,A}c_{n,B}+\Delta c^\dagger_{n,B}c_{n,A})
\end{eqnarray}
where $c_{n,\alpha}^{(\dagger)}$ $(\alpha=A,B)$ represents the annihilation (creation) operator at the $\alpha$ sublattice of the $n$th unit cell, $L$ is the size of the system, and we choose $t_+>t_-$. The Bloch Hamiltonian of the model under PBCs is 
\begin{eqnarray}
	H_\text{cNHSE}(k)=\begin{pmatrix}
		t_+e^{ik}+t_-e^{-ik} & \Delta \\
		\Delta & t_-e^{ik}+t_+e^{-ik}
	\end{pmatrix}, 
\end{eqnarray}
and the eigenenergies of it are 
\begin{eqnarray}
	E_\text{PBC}^{\pm}(k)=(t_++t_-)\cos k\pm i\sqrt{(t_+-t_-)^2\sin^2 k-\Delta^2}, 
	\label{cNHSE_PBC_E}
\end{eqnarray}
where $k$ is the real Bloch wave number. 

The OBC eigenfunction $H_\text{cNHSE}\vert\psi\rangle=E_\text{OBC}\vert\psi\rangle$, with $\vert\psi\rangle=(\psi_{1,A},\psi_{2,A},\cdots,\psi_{L,A},\psi_{1,B},\psi_{2,B},\cdots,\psi_{L,B})^T$, leads to the bulk equations
\begin{subequations}
\begin{align}
	t_-\psi_{n-1,A}+t_+\psi_{n+1,A}+\Delta\psi_{n,B}=E_\text{OBC}\psi_{n,A}, \\
	\Delta\psi_{n,A}+t_+\psi_{n-1,B}+t_-\psi_{n+1,B}=E_\text{OBC}\psi_{n,B},
\end{align}
\label{cNHSE_real}
\end{subequations}
with OBCs given by $\psi_{0,\alpha}=\psi_{L+1,\alpha}=0$ for both $\alpha=A,B$. According to the theory of linear difference equations, we can take an ansatz for the eigenstates as a linear combination: 
\begin{eqnarray}
	\begin{pmatrix}
		\psi_{n,A} \\ \psi_{n,B} \\
	\end{pmatrix}=\sum_{j=1}^4 (\beta_j)^n\begin{pmatrix}
		\phi_A^{(j)} \\ \phi_B^{(j)} \\
	\end{pmatrix}. 
	\label{ansatz_cNHSE}
\end{eqnarray}
Hence Eqs~(\ref{cNHSE_real}) can be written as 
\begin{eqnarray}
	\begin{pmatrix}
		t_+\beta+t_-\beta^{-1} & \Delta \\
		\Delta & t_-\beta+t_+\beta^{-1} \\
	\end{pmatrix}
	\begin{pmatrix}
		\phi_A \\ \phi_B \\
	\end{pmatrix}=E_\text{OBC}\begin{pmatrix}
		\phi_A \\ \phi_B \\
	\end{pmatrix}, 
	\label{cNHSE_bulk}
\end{eqnarray}
with $\beta=\beta_j$, and $\phi_\alpha=\phi_\alpha^{(j)}$. With the condition that $\phi_A$ and $\phi_B$ take nonzero values, the characteristic equation is written as 
\begin{eqnarray}
	\beta^2-\left(\frac{1}{t_+}+\frac{1}{t_-}\right)E_\text{OBC}\beta+\frac{1}{t_+ t_-}(t_+^2+t_-^2+E_\text{OBC}^2-\Delta^2)-\left(\frac{1}{t_+}+\frac{1}{t_-}\right)E_\text{OBC}\beta^{-1}+\beta^{-2}=0. 
	\label{cNHSE_characteristic}
\end{eqnarray}

Next, we examine the open boundaries of the system (with $\psi_{0,\alpha}=\psi_{L+1,\alpha}=0$), where the bulk equations of Eqs~(\ref{cNHSE_real}) become
\begin{subequations}
\begin{align}
	t_+\psi_{2,A}+\Delta\psi_{1,B}=E_\text{OBC}\psi_{1,A}, \\
	t_-\psi_{2,B}+\Delta\psi_{1,A}=E_\text{OBC}\psi_{1,B}, \\
	t_-\psi_{L-1,A}+\Delta\psi_{L,B}=E_\text{OBC}\psi_{L,A}, \\
	t_+\psi_{L-1,B}+\Delta\psi_{L,A}=E_\text{OBC}\psi_{L,B}. 
\end{align}
\end{subequations}
With the ansatz of Eq.~(\ref{ansatz_cNHSE}), the above boundary equations can be written as 
\begin{subequations}
\begin{align}
	t_+\sum_{j=1}^4\beta_j^2\phi_{A}^{(j)}+\Delta\sum_{j=1}^4\beta_j\phi_{B}^{(j)}=E_\text{OBC}\sum_j^{4}\beta_j\phi_{A}^{(j)}, \\
	t_-\sum_{j=1}^4\beta_j^2\phi_{B}^{(j)}+\Delta\sum_{j=1}^4\beta_j\phi_{A}^{(j)}=E_\text{OBC}\sum_{j=1}^4\beta_j\phi_{B}^{(j)}, \\
	t_-\sum_{j=1}^4\beta_j^{L-1}\phi_{A}^{(j)}+\Delta\sum_{j=1}^4\beta_j^{L}\phi_{B}^{(j)}=E_\text{OBC}\sum_{j=1}^4\beta_j^L\phi_{A}^{(j)}, \\
	t_+\sum_{j=1}^4\beta_j^{L-1}\phi_{B}^{(j)}+\Delta\sum_{j=1}^4\beta_j^{L}\phi_{A}^{(j)}=E_\text{OBC}\sum_{j=1}^4\beta_j^L\phi_{B}^{(j)}. 
\end{align}	
\end{subequations}
On the other hand,  the Eq.~(\ref{cNHSE_bulk}) can be expanded as
\begin{subequations}
\begin{align}
	(t_+\beta+t_-\beta^{-1}-E_\text{OBC})\phi_A+\Delta\phi_B=0, \\
	\Delta\phi_A+(t_-\beta+t_+\beta^{-1}-E_\text{OBC})\phi_B=0,
\end{align}
\end{subequations}
which leads to
\begin{eqnarray}
	\phi_B^{(j)}=f_j\phi_A^{(j)},
	~~
	&f_j=\dfrac{E_\text{OBC}-t_+\beta_j-t_-\beta_j^{-1}}{\Delta}=\dfrac{\Delta}{E_\text{OBC}-t_-\beta_j-t_+\beta_j^{-1}}. 
	\label{cNHSE_bulk_relation}
\end{eqnarray}
Substituting Eq.~(\ref{cNHSE_bulk_relation}) to the boundary equations, we obtain
\begin{subequations}
\begin{align}
	t_+\sum_{j=1}^4\beta_j^2\phi_{A}^{(j)}+\Delta\sum_{j=1}^4\beta_j f_j\phi_{A}^{(j)}=E_\text{OBC}\sum_j^{4}\beta_j\phi_{A}^{(j)}, \\
	t_-\sum_{j=1}^4\beta_j^2 f_j\phi_{A}^{(j)}+\Delta\sum_{j=1}^4\beta_j\phi_{A}^{(j)}=E_\text{OBC}\sum_{j=1}^4\beta_j f_j\phi_{A}^{(j)}, \\
	t_-\sum_{j=1}^4\beta_j^{L-1}\phi_{A}^{(j)}+\Delta\sum_{j=1}^4\beta_j^{L}f_j\phi_{A}^{(j)}=E_\text{OBC}\sum_{j=1}^4\beta_j^L\phi_{A}^{(j)}, \\
	t_+\sum_{j=1}^4\beta_j^{L-1}f_j\phi_{A}^{(j)}+\Delta\sum_{j=1}^4\beta_j^{L}\phi_{A}^{(j)}=E_\text{OBC}\sum_{j=1}^4\beta_j^L f_j\phi_{A}^{(j)}, 
\end{align}
\end{subequations}
which can be further simplified as
\begin{subequations}
\begin{align}
	&\sum_{j=1}^4\phi_A^{(j)}=0,\\
	&\sum_{j=1}^4(E_\text{OBC}-t_+\beta_j-t_-\beta_j^{-1})\phi_A^{(j)}=0, \\
	&\sum_{j=1}^4\beta_j^{L+1}\phi_A^{(j)}=0, \\
	&\sum_{j=1}^4(E_\text{OBC}-t_+\beta_j-t_-\beta_j^{-1})\beta_j^{L+1}\phi_A^{(j)}=0.
\end{align}
\end{subequations}
With the condition that $\phi_A^{(j)}$ $(j=1,2,3,4)$ take nonzero values, the above equations are equivalent to a vanishing determinant:
\begin{eqnarray}
	\left\vert\begin{matrix}
		1 & 1 & 1 & 1 \\
		Y_1 & Y_2 & Y_3 & Y_4 \\
		\beta_1^{L+1} & \beta_2^{L+1} & \beta_3^{L+1} & \beta_4^{L+1} \\
		Y_1\beta_1^{L+1} & Y_2\beta_2^{L+1} & Y_3\beta_3^{L+1} & Y_3\beta_4^{L+1} \\
	\end{matrix}\right\vert=0, 
\end{eqnarray}
where $\beta_j$ $(j=1,2,3,4)$ satisfy $\vert\beta_1\vert\leqslant\vert\beta_2\vert\leqslant\vert\beta_3\vert\leqslant\vert\beta_4\vert$ and $Y_j$ $(j=1,2,3,4)$ are defined as 
\begin{eqnarray}
	Y_j\equiv E_\text{OBC}-t_+\beta_j-t_-\beta_j^{-1}\quad(j=1,2,3,4). 
	\label{cNHSE_Yi}
\end{eqnarray}
Then we can obtain 
\begin{eqnarray}
	Y_{1,4}Y_{2,3}[(\beta_1\beta_4)^{L+1}+(\beta_2\beta_3)^{L+1}]-Y_{1,3}Y_{2,4}[(\beta_1\beta_3)^{L+1}+(\beta_2\beta_4)^{L+1}]+Y_{1,2}Y_{3,4}[(\beta_1\beta_2)^{L+1}+(\beta_3\beta_4)^{L+1}]=0, 
	\label{cNHSE_boundary_c}
\end{eqnarray}
where $Y_{i,j}$ $(i,j=1,2,3,4)$ are defined as 
\begin{eqnarray}
	Y_{i,j}\equiv Y_i-Y_j=t_+(\beta_j-\beta_i)+t_-(\beta_j^{-1}-\beta_i^{-1})\quad (i,j=1,2,3,4). 
	\label{cNHSE_Yij}
\end{eqnarray}
With a sufficiently large system size $L$,
Eq.~\eqref{cNHSE_boundary_c} is dominated by $\beta_j$ with larger absolute values.
However, keeping only a single leading term of $(\beta_3\beta_4)^{L+1}$ leads to a trivial solution of $\beta_j=0$. 
Thus we need to consider at least two leading terms, which are $(\beta_3\beta_4)^{L+1}$  and $(\beta_2\beta_4)^{L+1}$, 
so that Eq.~\eqref{cNHSE_boundary_c} can be approximately transformed into
\begin{eqnarray}
	(\beta_2)^{2L+2}\simeq\frac{Y_{1,2}Y_{3,4}}{Y_{1,3}Y_{2,4}}. 
	\label{cNHSE_beta2_app1}
\end{eqnarray}
Note that Eq.~\eqref{cNHSE_beta2_app1} is obtained with the relationship
\begin{eqnarray}
	\beta_1=\frac{1}{\beta_4},\quad \beta_2=\frac{1}{\beta_3}, 
\end{eqnarray}
as the characteristic equation of Eq.~(\ref{cNHSE_characteristic}) is a reciprocal equation for $\beta$.

Keeping two leading terms for large $L$ also suggests that $|\beta_2|\approx|\beta_3|$, which consists with the non-Bloch band theory that requires the ``middle" two solutions of $\beta$ have the same absolute value to predict eigensolutions and give the localization length of OBC eigenstates~\cite{yao2018edge,yokomizo2019non}.
Thus, to obtain an analytical solution of $\beta$,
we consider a perturbative solution of Eq.~\eqref{cNHSE_characteristic}, up to the second order in $\Delta$, 
\begin{subequations}
\begin{align}
	\beta_1\simeq x_-^{(1)}+y_-^{(1)}\Delta^2, \\
	\beta_2\simeq x_+^{(1)}+y_+^{(1)}\Delta^2, \\
	\beta_3\simeq x_-^{(2)}+y_-^{(2)}\Delta^2, \\
	\beta_4\simeq x_+^{(2)}+y_+^{(2)}\Delta^2, 
\end{align}
\label{cNHSE_perturbative_beta}
\end{subequations}
where 
\begin{subequations}
\begin{align}
	&x_\pm^{(1)}=\frac{1}{2t_+}(E_\text{OBC}\pm\delta), \\
	&x_\pm^{(2)}=\frac{1}{2t_-}(E_\text{OBC}\pm\delta), \\
	&y_\pm^{(1)}=\frac{\pm(E_\text{OBC}^2-2t_+ t_-\pm\delta E_\text{OBC})}{(t_+-t_-)\delta[2t_+(t_++t_-)-E_\text{OBC}^2\mp\delta E_\text{OBC}]}, \\
	&y_\pm^{(2)}=\frac{\pm(E_\text{OBC}^2-2t_+ t_-\pm\delta E_\text{OBC})}{(t_+-t_-)\delta[-2t_-(t_++t_-)+E_\text{OBC}^2\pm\delta E_\text{OBC}]}, 
\end{align}
\end{subequations}
and 
\begin{eqnarray}
	\delta=\sqrt{E_\text{OBC}^2-4t_+t_-}. 
\end{eqnarray}
With Eqs.~(\ref{cNHSE_perturbative_beta}) and Eq.~(\ref{cNHSE_Yij}), we obtain 
\begin{subequations}
\begin{align}
	&Y_{1,2}=\frac{t_++t_-}{t_+-t_-}\frac{\delta}{(t_++t_-)^2-E_\text{OBC}^2}\Delta^2+\mathcal{O}(\Delta^4), \\
	&Y_{3,4}=(\frac{t_+}{t_-}-\frac{t_-}{t_+})\delta+\mathcal{O}(\Delta^2), \\
	&Y_{1,3}=(t_+-t_-)(\frac{E_\text{OBC}-\delta}{2t_-}-\frac{2t_-}{E_\text{OBC}-\delta})+\mathcal{O}(\Delta^2), \\
	&Y_{2,4}=(t_+-t_-)(\frac{E_\text{OBC}+\delta}{2t_-}-\frac{2t_-}{E_\text{OBC}+\delta})+\mathcal{O}(\Delta^2). 
\end{align}
\label{cNHSE_Yij_s}
\end{subequations}
Substituting Eqs.~(\ref{cNHSE_Yij_s}) into Eq.~(\ref{cNHSE_beta2_app1}) and expanding up to the second order in $\Delta$, we obtain 
\begin{eqnarray}
	(\beta_2)^{2L+2}\simeq\frac{(t_++t_-)^2}{(t_+-t_-)^2}\frac{E_\text{OBC}^2-4t_+t_-}{[(t_++t_-)^2-E_\text{OBC}^2]^2}\Delta^2. 
	\label{cNHSE_beta2_app2}
\end{eqnarray}

It is known that in the thermodynamic limit, the OBC energy spectrum of this model coincides with the PBC one~\cite{yokomizo2021scaling}.
Thus we can approximate the eigenenergies $E_\text{OBC}$ by $E_\text{PBC}^{\pm}(k)$ under PBCs. Substituting Eq.~(\ref{cNHSE_PBC_E}) into Eq.~(\ref{cNHSE_beta2_app2}) and taking $k=\pi/2$, we obtain the analytical expression of the absolute value of $\beta_2$ as 
\begin{eqnarray}
	\vert\beta_2\vert\simeq\left[\frac{(t_++t_-)^2}{2(t_+-t_-)(t_+^2+t_-^2)}\Delta\right]^{1/(L+1)}. 
	\label{eq:beta2}
\end{eqnarray}
Therefore the localization length of the model that describe the scaling rule along $x-$direction in the main text is 
\begin{eqnarray}
	\xi_x^\text{AS}=-\frac{1}{\ln\vert\beta_2\vert}=-\frac{L_x+1}{\ln\left[\frac{(t_x^++t_x^-)^2}{2(t_x^+-t_x^-)[(t_x^+)^2+(t_x^-)^2]}\Delta^{(N)}\right]}, 
\end{eqnarray}
where $\Delta^{(N)}$ is the effective coupling as discussed above and in the main text. 
Note that this solution is obtained for the $E_\text{PBC}^{\pm}(k=\pi/2)$,
which has the maximum imaginary part among all eigenenergies.
Similar scaling of the localization length also appears for other values of $k$, except that $\xi_x\rightarrow \infty$ for $k=0$, suggesting an extended state. 
Nevertheless, as the majority of the eigenstates still exhibit similar size-dependent localization,
the average localization length defined in the main text is dominated by their collective behavior and can be captured by the solution of $k=\pi/2$, as shown by our numerical results.

On the other hand, the energy eigenvalues under OBCs can also be obtained from Eq.~\eqref{eq:beta2}.
Explicitly, 
we may determine the energy by dropping the small amount of $\Delta^2$ in Eq.~\eqref{cNHSE_characteristic}, which yields
\begin{eqnarray}
	E_{\rm OBC}^{\pm}\approx t_+\beta^{\pm}+t_-\beta^{\mp}. 
\end{eqnarray}
Since the OBC energy spectrum of this model coincides with the PBC one~\cite{yokomizo2021scaling}, we can see from Eq.~\eqref{cNHSE_PBC_E} that the maximal imaginary energy corresponds to $k=\pm\pi/2$, where $\beta=\vert\beta\vert e^{\pm i\pi/2}=\pm i\vert\beta\vert$ and $\beta^{-1}=\vert\beta\vert^{-1} e^{\mp i\pi/2}=\mp i\vert\beta\vert^{-1}$. Then we substitute Eq.~\eqref{eq:beta2}, we obtain the maximal imaginary energy, 
\begin{eqnarray}
	\text{Max}[\text{Im}(E)]\approx t_x^+\exp(-1/\xi_x^{\rm AS})-t_x^-\exp(1/\xi_x^{\rm AS}).
\end{eqnarray}

\section{Different types of boundaries in non-Hermitian BBH model}\label{sec:BBH}
\subsection{Effective ladder model}
\begin{figure*}
	\includegraphics[width=1.0\linewidth]{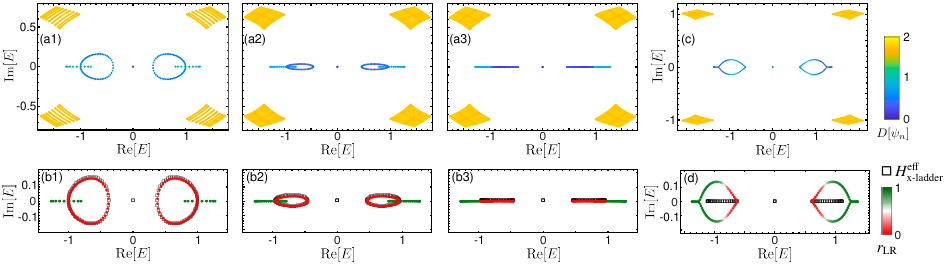}
	\caption{(a) Energy spectrum of the non-Hermitian BBH model with
	non-Hermiticity added along $x$ direction. The system's size is $N_x=30$ and $N_y=10,20,30$ for (a1) to (a3), respectively.
	 (b) Comparison between the eigenenergies of the edge states (colored solid dots) in (a) and that of $H_\text{eff}^\text{y-edge}$, the effective 1D ladder model formed by the top and bottom edges of the 2D systems with size-dependent couplings induced by the bulk (black squares). Colors indicate the edge distribution ratio $r_{\rm LR}$ of each edge state. $r_{\rm LR}\approx 1$ and $0$ correspond to eigenstates distribute mostly on the left/right and top/bottom edges, respectively. It is seen that the eigenenergies of $H_\text{eff}^\text{y-edge}$ match well with those of the top/bottom edge states.
	 (c) and (d) the same plots as in (a) and (b), but with non-Hermiticity added along both directions. The eigenenergies of the edge states and that of $H_\text{eff}^\text{y-edge}$ do not coincide with each other.	 
	 Parameters are $t_x^+=2,t_x^-=0.25,t_y^+=t_y^-=1,t'=0.25$ in (a) and (b), and $t_x^+=1.5,t_x^-=0.5,t_y^+=2.5,t_y^-=0.5,t'=0.25,N_x=N_y=20$ in (c) and (d).
	}
	\label{SM_BBH}
\end{figure*}
In the main text, we have introduced the non-Hermitian Benalcazar-Bernevig-Hughes (BBH) model \citep{benalcazar2017quantized,benalcazar2017electric} with non-Hermicity only along $x$ direction or along both directions,
whose Bloch Hamiltonian reads
\begin{equation}`
	H_\text{BBH}(\mathbf{k})=\begin{pmatrix}
		0 & t'+t_x^- e^{-ik_x} & -t'-t_y^+ e^{ik_y} & 0\\
		t'+t_x^+ e^{ik_x} & 0 & 0& t'+t_y^- e^{ik_y}\\
		-t'-t_y^- e^{-ik_y} & 0 & 0 & t'+t_x^+ e^{-ik_x}\\
		0 & t'+t_y^+ e^{-ik_y} & t'+t_x^- e^{ik_x} & 0\\
	\end{pmatrix}.
\end{equation}
When non-Hermicity is introduced to the BBH model only along $x$ direction (i.e., $t_x^+\neq t_x^-$ and $t_y^+=t_y^-$), a similar effective edge Hamiltonian 
\begin{eqnarray}
	H^\text{eff}_\text{x-ladder}\approx\begin{pmatrix}
		H_{\rm SSH}^{\rm top} & \Delta_{\text{BBH}}^{(N_y)} \\
		\Delta_{\text{BBH}}^{(N_y)} & H_{\rm SSH}^{\rm bottom} \\
	\end{pmatrix}
\end{eqnarray}
can be obtained through the same formalism as for the NH-SSH model, containing two non-Hermitian SSH chains coupled via size-dependent couplings arisen from the bulk.
The size-dependent couplings can be obtained 
from the same calculation as in Sec.\ref{secS1} and \ref{sec:Heff_appro}, with 
$t_1$ and $t_2$ for the HN-SSH model replaced by $t'\sigma_z$ and $t_y\sigma_z$, respectively.
Their explicit form reads
$$\Delta_{\text{BBH}}^{(N_y)}=\frac{(-1)^{N_y-1}(t'\sigma_z)^{N_y}}{(t_y\sigma_z)^{N_y-1}}=\frac{(-1)^{N_y-1}(t')^{N_y}\sigma_z}{(t_y)^{N_y-1}}.$$
This analytical solution is verified by our numerical results as shown in Fig. \ref{SM_BBH}(a) and (b), where the boundary spectrum for top and bottom edges matches the eigenvalues of the effective edge Hamiltonian. 

On the other hand, we may also consider the effective edge Hamiltonian 
\begin{eqnarray}
	H^\text{eff}_\text{y-ladder}\approx\begin{pmatrix}
		H_{\rm SSH}^{\rm left} & \Delta_{\text{BBH}}^{(N_x,+)} \\
		\Delta_{\text{BBH}}^{(N_x,-)} & H_{\rm SSH}^{\rm right} \\
	\end{pmatrix}
\end{eqnarray}
for the left and right edges,
 which is formed by two Hermitian SSH chains coupled via weak couplings arisen from the bulk. 
 However, $H_{\rm SSH}^{\rm left}$ and $H_{\rm SSH}^{\rm right}$ are Hermitian Hamiltonians for the original SSH model,
 which is insensitive to the weak couplings through the bulk. 
 Consistently, they give rise to the 1D edge states on the left and right edges (characterized by $D[\phi_n]\approx 1$) with size-independent spectral features in our numerical results in Fig. \ref{SM_BBH}(a) and (b).

In Fig. \ref{SM_BBH}(c) and (d) we display the spectrum of the non-Hermitian BBH model with non-Hermiticity added along all the four edges, and that of the effective edge Hamiltonian $H^\text{eff}_\text{x-ladder}$ obtained similarly (for the top and bottom edges coupled through the bulk).
It is seen that the effective edge Hamiltonian no longer predicts the spectrum of edge states. Thus, together with the results we show in the main text, we can conclude that the ASL in this case origins from the 1D junction system, instead of the effective coupling between edges through the bulk.

\subsection{Eeffective junction model and the origin of ASL}\label{sec:junction}
The 1D junction system consists of the four edges of the 2D lattce, and can be further mapped to a spatially homogeneous model consisting of one pair of edges (e.g., top and bottom), with the other edges acting as two local impurities that generate SFL~\cite{li2021impurity}.
To see this, we define an effective Hamiltonian for the 1D junction system,
\begin{equation}
	\hat{H}_{\rm junction}^{\rm eff}=\hat{P}_{\rm edge}\hat{H}_{\rm BBH}\hat{P}_{\rm edge}, \label{H_junction}
\end{equation}
with $\hat{P}_{\rm edge}=\sum_{\mathbf{n}}|\mathbf{n}\rangle\langle\mathbf{n}|$ and $\mathbf{n}$ the lattice sites along the four edges, $\mathbf{n}=(x,y)$ where $x\in\{1,L_x\}$ or $y\in\{1,L_y\}$. More explictly, we can divide the 1D junction chain into four parts and the Hamiltonian of it can be written as 
\begin{equation}
	H_{\rm junction}=\begin{pmatrix}
		H_{\rm top} & 0 & X_1 & X_2 \\
		0 & H_{\rm bottom} & X_3 & X_4 \\
		X_1^T & X_3^T & H_{\rm left} & 0 \\
		X_2^T & X_4^T & 0 & H_{\rm right} \\
	\end{pmatrix}, 
\end{equation}
where 
\begin{equation}
	H_{\rm top}=\begin{pmatrix}
		0 & t_x^+ & 0 & \cdots & 0 & 0 \\
		t_x^- & 0 & t' & \cdots & 0 & 0 \\
		0 & t' & 0 & \cdots & 0 & 0 \\
		\vdots & \vdots & \vdots & \ddots & \vdots & \vdots \\
		0 & 0 & 0 & \cdots & 0 & t_x^+ \\
		0 & 0 & 0 & \cdots & t_x^- & 0 \\
	\end{pmatrix}_{(L_x-2)\times (L_x-2)}, 
	~ H_{\rm bottom}=H_{\rm top}^T, ~ 
	H_{\rm left}=\begin{pmatrix}
		0 & -t' & 0 & \cdots & 0 & 0 \\
		-t' & 0 & -t_y^- & \cdots & 0 & 0 \\
		0 & -t_y^+ & 0 & \cdots & 0 & 0 \\
		\vdots & \vdots & \vdots & \ddots & \vdots & \vdots \\
		0 & 0 & 0 & \cdots & 0 & -t' \\
		0 & 0 & 0 & \cdots & -t' & 0 \\
	\end{pmatrix}_{L_y\times L_y}, 
	~ H_{\rm right}=-H_{\rm left}^T
\end{equation}
are the Hamiltonian matrices for the top ($x\in[2,L_x-1],y=Ly$), bottom ($x\in[2,L_x-1],y=1$), left ($x=1,y\in[1,L_y]$) and right ($x=L_x,y\in[1,L_y]$) edges respectively, and coupling between them are 
\begin{equation}
	X_1=\begin{pmatrix}
		t' & 0 & \cdots & 0 & 0 \\
		0 & 0 & \cdots & 0 & 0 \\
		\vdots & \vdots & \ddots & \vdots & \vdots \\
		0 & 0 & \cdots & 0 & 0 \\
		0 & 0 & \cdots & 0 & 0 \\
	\end{pmatrix}_{(L_x-2)\times L_y},
	X_2=\begin{pmatrix}
		0 & 0 & \cdots & 0 & 0 \\
		0 & 0 & \cdots & 0 & 0 \\
		\vdots & \vdots & \ddots & \vdots & \vdots \\
		0 & 0 & \cdots & 0 & 0 \\
		t' & 0 & \cdots & 0 & 0 \\
	\end{pmatrix}_{(L_x-2)\times L_y},
	X_3=\begin{pmatrix}
		0 & 0 & \cdots & 0 & t' \\
		0 & 0 & \cdots & 0 & 0 \\
		\vdots & \vdots & \ddots & \vdots & \vdots \\
		0 & 0 & \cdots & 0 & 0 \\
		0 & 0 & \cdots & 0 & 0 \\
	\end{pmatrix}_{(L_x-2)\times L_y},
	X_4=\begin{pmatrix}
		0 & 0 & \cdots & 0 & 0 \\
		0 & 0 & \cdots & 0 & 0 \\
		\vdots & \vdots & \ddots & \vdots & \vdots \\
		0 & 0 & \cdots & 0 & 0 \\
		0 & 0 & \cdots & 0 & t' \\
	\end{pmatrix}_{(L_x-2)\times L_y}. 
\end{equation}
Next, we focus on the state distribution along top and bottom edges;
the left and right edges can be analyzed in the same way. 
That is, we separate the Hamiltonian matrices of these edges as
\begin{equation}
	H_x=\begin{pmatrix} 
		H_{\rm top} & 0 \\
		0 & H_{\rm bottom} \\ 
	\end{pmatrix}, ~
	H_y=\begin{pmatrix}
		H_{\rm left} & 0 \\
		0 & H_{\rm right} \\
	\end{pmatrix},
\end{equation}
and obtain an effective Hamiltonian for the top and bottom edges,
\begin{equation}
	H_x^{\rm eff}(E)=H_x+X_{\rm junction}(E-H_y)^{-1}X_{\rm junction}^T, \label{eq:H_x_eff}
\end{equation}
where the coupling between them is 
\begin{equation}
	X_{\rm junction}=\begin{pmatrix}
		X_1 & X_2 \\
		X_3 & X_3 \\
	\end{pmatrix}. 
\end{equation}
Setting $G_{\rm left}(E)=(E-H_{\rm left})^{-1}$, and $G_{\rm right}(E)=(E-H_{\rm right})^{-1}$, we obtain 
\begin{equation}
	(E-H_y)^{-1}=\begin{pmatrix}
		E-H_{\rm left} & 0 \\
		0 & E-H_{\rm right} \\
	\end{pmatrix}^{-1}=\begin{pmatrix}
		(E-H_{\rm left})^{-1} & 0 \\
		0 & (E-H_{\rm right})^{-1} \\
	\end{pmatrix}=\begin{pmatrix}
		G_{\rm left}(E) & 0 \\
		0 & G_{\rm right}(E) \\
	\end{pmatrix}. 
\end{equation}
Then the second term in Eq.~\eqref{eq:H_x_eff} can be expressed as
\begin{equation}
	X_{\rm junction}(E-H_y)^{-1}X_{\rm junction}^T=\begin{pmatrix}
		X_1 & X_2 \\
		X_3 & X_4 \\
	\end{pmatrix}\begin{pmatrix}
		G_{\rm left}(E) & 0 \\
		0 & G_{\rm right}(E) \\
	\end{pmatrix}\begin{pmatrix}
		X_1^T & X_3^T \\
		X_2^T & X_4^T \\
	\end{pmatrix}=\begin{pmatrix}
		X_1 G_{\rm left}(E)X_1^T+X_2 G_{\rm right}(E)X_2^T & X_1 G_{\rm left}(E)X_3^T+X_2 G_{\rm right}(E)X_4^T \\
		X_3 G_{\rm left}(E)X_1^T+X_4 G_{\rm right}(E)X_2^T & X_3 G_{\rm left}(E)X_3^T+X_4 G_{\rm right}(E)X_4^T \\
	\end{pmatrix}. 
\end{equation}
Explicitly, these matrix elements are given by
\begin{subequations}
\begin{align}
	&X_1 G_{\rm left}(E) X_1^T+X_2 G_{\rm right}(E)X_2^T=\begin{pmatrix}
		(t')^2G_{\rm left}^{1,1}(E) & 0 & \cdots & 0 & 0 \\
		0 & 0 & \cdots & 0 & 0 \\
		\vdots & \vdots & \ddots & \vdots & \vdots \\
		0 & 0 & \cdots & 0 & 0 \\
		0 & 0 & \cdots & 0 & (t')^2G_{\rm right}^{1,1}(E) \\
	\end{pmatrix}_{(L_x-2)\times (L_x-2)}\\
	&X_1 G_{\rm left}(E)X_3^T+X_2 G_{\rm right}(E)X_4^T=\begin{pmatrix}
		(t')^2G_{\rm left}^{1,L_y}(E) & 0 & \cdots & 0 & 0 \\
		0 & 0 & \cdots & 0 & 0 \\
		\vdots & \vdots & \ddots & \vdots & \vdots \\
		0 & 0 & \cdots & 0 & 0 \\
		0 & 0 & \cdots & 0 & (t')^2G_{\rm right}^{1,L_y}(E) \\
	\end{pmatrix}_{(L_x-2)\times (L_x-2)}\\
	&X_3 G_{\rm left}(E)X_1^T+X_4 G_{\rm right}(E)X_2^T=\begin{pmatrix}
		(t')^2G_{\rm left}^{L_y,1}(E) & 0 & \cdots & 0 & 0 \\
		0 & 0 & \cdots & 0 & 0 \\
		\vdots & \vdots & \ddots & \vdots & \vdots \\
		0 & 0 & \cdots & 0 & 0 \\
		0 & 0 & \cdots & 0 & (t')^2G_{\rm right}^{L_y,1}(E) \\
	\end{pmatrix}_{(L_x-2)\times (L_x-2)}\\
	&X_3 G_{\rm left}(E)X_3^T+X_4 G_{\rm right}(E)X_4^T=\begin{pmatrix}
		(t')^2G_{\rm left}^{L_y,L_y}(E) & 0 & \cdots & 0 & 0 \\
		0 & 0 & \cdots & 0 & 0 \\
		\vdots & \vdots & \ddots & \vdots & \vdots \\
		0 & 0 & \cdots & 0 & 0 \\
		0 & 0 & \cdots & 0 & (t')^2G_{\rm right}^{L_y,L_y}(E) \\
	\end{pmatrix}_{(L_x-2)\times (L_x-2)}
\end{align}
\end{subequations}
where $G_{\rm left/right}^{m,l}(E)$ is the $(m,l)-$element of the $G_{\rm left/right}(E)$. 
Finally, the effective Hamiltonian can be rewritten as 
\begin{eqnarray}
	H_x^{\rm eff}(E)&=&\sum_{x\in\{2,4,6,\cdots,L_x-2\}}[t_x^- c_{(x,L_y)}^\dagger c_{(x+1,L_y)}+t_x^+ c_{(x+1,L_y)}^\dagger c_{(x,L_y)}]+\sum_{x\in\{3,5,7,\cdots,L_x-3\}}[t'c_{(x,L_y)}^\dagger c_{(x+1,L_y)}+h.c.]\notag\\
	&&+\sum_{x\in\{2,4,6,\cdots,L_x-2\}}[t_x^+ c_{(x,1)}^\dagger c_{(x+1,1)}+t_x^- c_{(x+1,1)}^\dagger c_{(x,1)}]+\sum_{x\in\{3,5,7,\cdots,L_x-3\}}[t'c_{(x,1)}^\dagger c_{(x+1,1)}+h.c.]\notag\\
	&&+(t')^2 G_{\rm left}^{1,L_y}(E)c_{(2,L_y)}^\dagger c_{(2,1)}+(t')^2 G_{\rm left}^{L_y,1}(E)c_{(2,1)}^\dagger c_{(2,L_y)}+(t')^2 G_{\rm right}^{1,L_y}(E)c_{(L_x-1,L_y)}^\dagger c_{(L_x-1,1)}+(t')^2 G_{\rm right}^{L_y,1}(E)c_{(L_x-1,1)}^\dagger c_{(L_x-1,L_y)}\notag\\
	&&+(t')^2 G_{\rm left}^{1,1}(E)c_{(2,L_y)}^\dagger c_{(2,L_y)}+(t')^2 G_{\rm right}^{1,1}(E)c_{(L_x-1,L_y)}^\dagger c_{(L_x-1,L_y)}+(t')^2 G_{\rm left}^{L_y,L_y}(E)c_{(2,1)}^\dagger c_{(2,1)}+(t')^2 G_{\rm right}^{L_y,L_y}(E)c_{(L_x-1,1)}^\dagger c_{(L_x-1,1)},\notag\\
	\label{eq:H_x_eff_full}
\end{eqnarray}
where $c_{(x,y)}$ is the annihilation operator of a particle at site $(x,y)$. 
We can see that the top and bottom edges are connected into a 1D non-Hermitian lattice with translational symmetry in the bulk [the first two lines in Eq.~\eqref{eq:H_x_eff_full}], and impurity hopping amplitudes [third line in Eq.~\eqref{eq:H_x_eff_full}] and on-site potentials [fourth line in Eq.~\eqref{eq:H_x_eff_full}] at the two junctions.
Such impurities are known to generate the scale-free localization in 1D \citep{li2021impurity}, which unveils the origin of ASL on top and bottom edges in this case.
Similarly, ASL on left and right edges can be obtained by exchanging the roles of $H_x$ and $H_y$ in the above derivation, which leads to another effective 1D non-Hermitian lattice (with impurities) describing these two edges.

\section{Localization length in the non-Hermitian BBH model}\label{sec:length}
As discussed in Sec. \ref{sec:1D_skin}, skin states of one-dimensional HN model can be expressed as $\psi_x\sim(\beta)^x$. 
More precisely, $\psi_x\sim(\beta)^x\sin[x\theta]$, where $\theta=[m\pi/(N+1)]$ $(m=1,2,\dots,N)$~\citep{guo2021exact}. 
Therefore, numerically we can obtain the localization length as $$\xi=\big|\frac{1}{\ln\vert\beta\vert}\big|=\big|\frac{L_x-1}{\ln[|\psi_{L_x}/\psi_1|]}\big|$$
to demonstrate the localization property of skin states therein However, edges of the BBH model form 1D model with two sublattices (i.e., the non-Hermitian SSH model), where the wavefunctions of OBC skin states are given by \citep{guo2021exact} , 
\begin{equation}
	\psi_{x,A}\sim r^{x}(\sin[x\theta]+\alpha\sin[(x-1)\theta]),~\psi_{x,B}\sim\frac{E}{t_{1L}}r^x\sin[x\theta],
\end{equation}
where $A$ and $B$ denote the two sublattices,$r=\sqrt{\frac{t_{1R}t_{2R}}{t_{1L}t_{2L}}},\alpha=\sqrt{\frac{t_{2R}t_{2L}}{t_{1R}t_{1L}}}$, $\{t_{1L},t_{1R},t_{2L},t_{2R}\}$ are the hopping amplitudes of the intracell and intercell, $\theta$ is a real phase factor satisfying $\sin[(N_x+1)\theta]+\alpha\sin[N_x\theta]=0$, $N_x$ is the number of unit cells, and $E$ is the corresponding eigenenergy. 
In this case,  
\begin{equation*}
	\frac{\psi_{N_x,A}}{\psi_{1,A}}\sim r^{N_x-1}\frac{\sin[N_x\theta]+\alpha\sin[(N_x-1)\theta]}{\sin[\theta]},~\frac{\psi_{N_x,B}}{\psi_{1,B}}\sim r^{N_x-1}\frac{\sin[N_x\theta]}{\sin[\theta]}, 
\end{equation*}
making the localization length defined for a single sublattice, $\xi_{A(B)}^{\rm SSH}=\big|(N_x-1)/\ln[\vert\psi_{N_x,A(B)}\vert/\vert\psi_{1,A(B)}\vert]\big|$,
oscillates with the system's size $N_x$ and the phase factor $\theta$. 
On the other hand, note that
\begin{equation}
	\frac{\psi_{N_x,A}}{\psi_{1,A}} \frac{\psi_{N_x,B}}{\psi_{1,B}} \sim r^{2(N_x-1)} \frac{\{\sin[N_x\theta]+\alpha\sin[(N_x-1)\theta]\} \sin[N_x\theta]}{\sin[\theta]\sin[\theta]}=r^{2(N_x-1)}. 
\end{equation}
Therefore, to better demonstrate the scaling of localization,
we can define an effective localization length as 
\begin{equation}
	\xi_x^{\rm SSH}=\big|\frac{2(N_x-1)}{\ln\left\{[\vert\psi_{N_x,A}\vert\vert\psi_{N_x,B}\vert]/[\vert\psi_{1,A}\vert\vert\psi_{1,B}\vert]\right\}}\big| 
\end{equation}
for the non-Hermitian SSH model.
For our non-Hermitian BBH model, the localization length $\xi_x$ and $\xi_y$ in Fig. 4 in the main text are defined similarly as
\begin{equation}
	\xi_x=\big|\frac{L_x-4}{\ln\left\{[\vert\psi_{L_x-1,{1}}\vert\vert\psi_{L_x-2,{1}}\vert]/[\vert\psi_{2,{1}}\vert\vert\psi_{3,{1}}\vert]\right\}}\big|,~~~~
	\xi_y=\big|\frac{L_y-4}{\ln\left\{[\vert\psi_{{1},L_y-1}\vert\vert\psi_{{1},L_y-2}\vert]/[\vert\psi_{{1},2}\vert\vert\psi_{{1},3}\vert]\right\}}\big|,
\end{equation}
with $L_{x/y}=2N_{x/y}$ the number of lattice sites along each direction.
Note that to avoid the influence of zero-energy corner states induced by possible higher-order topology, we have ignored the corner lattice sites (namely, with $x,y\in\{1,L_{x/y}\}$) in the above definations.

\section{ASL in the three-dimensional lattice}
\begin{figure}[h]
	\includegraphics[width=1.0\linewidth]{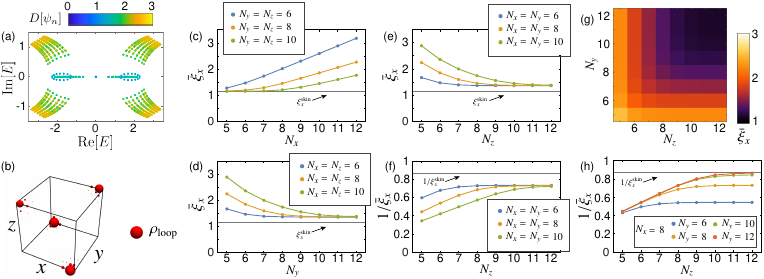}
	\caption{ASL of the 3D BBH model with non-Hermiticity added along $x$ direction. (a) Energy spectrum of the 3D non-Hermitian BBH model under OBCs with $N_x=N_y=N_z=8$. Eigenenergies are marked by different colors according to the 3D fractal dimension, defined as $D[\psi_n]=-\ln I/\ln\sqrt[3]{8N_x N_y N_z}$ and $I=\sum_{x,y,z}|\psi_{x,y,z}|^4$.
	Corner states correspond to $0\lesssim D[\psi_n]<1$, and form two loops in the complex energy plane.
	(b) Summed distribution of corner states with loop-like spectrum in (a). Size of each red sphere indicates the value of the total density for corner states at each site, defined as $\rho_{\rm loop}=\sum_n\vert\psi_n(x,y,z)\vert^2$, where the summation of $n$ runs over all states with loop-like spectra in (a). (c) Average localization length along $x$ direction of corner states with loop-like spectra, $\bar{\xi}_x$, versus $N_x$ with different $N_y=N_z$. 
	(d) $\bar{\xi}_x$ versus $N_y$ with $N_x=N_z$.
	(e) $\bar{\xi}_x$ versus $N_z$ with $N_x=N_y$. 
	Gray solid lines in (c)-(e) indicate $\xi_x^{\rm skin}=-2/\ln(t_x^-/t_x^+)$. (f) Inverse of the average localization length along $x$ direction of corner states with loop-like spectra, $1/\bar{\xi}_x$, versus $N_z$ with $N_x=N_y$. 
	(g) $\bar{\xi}_x$ versus $N_y$ and $N_z$ with $N_x=8$. 
	(h) $1/\bar{\xi}_x$ versus $N_z$ with $N_x=8$ and $N_y=6,8,10,12$ respectively. Gray solid lines in (f) and (h) indicate $1/\xi_x^{\rm skin}$.
	Other parameters are $t'=0.5,t_y=t_z=1,t_x^-=0.75,t_x^+=4.25$. }
	\label{BBH3D_SM}
\end{figure}
We consider a 3D extenstion of the BBH model \citep{benalcazar2017quantized,benalcazar2017electric} with non-Hermiticity added along $x$ direction, whose Bloch Hamiltonian reads 
\begin{eqnarray}
	H_{\rm BBH}^{\rm 3D}(\mathbf{k})=\begin{pmatrix}
		0 & t'+t_x^- e^{-ik_x} & -t'-t_y e^{ik_y} & 0 & t'+t_z e^{i k_z} & 0 & 0 & 0\\
		t'+t_x^+ e^{ik_x} & 0 & 0& t'+t_y e^{ik_y} & 0 & t'+t_z e^{i k_z} & 0 & 0\\
		-t'-t_y e^{-ik_y} & 0 & 0 & t'+t_x^+ e^{-ik_x} & 0 & 0 & t'+t_z e^{i k_z} & 0\\
		0 & t'+t_y e^{-ik_y} & t'+t_x^- e^{ik_x} & 0 & 0 & 0 & 0 & t'+t_z e^{i k_z}\\
		t'+t_z e^{-i k_z} & 0 & 0 & 0 & 0 & -t'-t_x^+ e^{-ik_x} & t'+t_y e^{ik_y} & 0\\
		0 & t'+t_z e^{-i k_z} & 0 & 0 & -t'-t_x^- e^{ik_x} & 0 & 0& -t'-t_y e^{ik_y}\\
		0 & 0 & t'+t_z e^{-i k_z} & 0 & t'+t_y e^{-ik_y} & 0 & 0 & -t'-t_x^- e^{-ik_x}\\
		0 & 0 & 0 & t'+t_z e^{-i k_z} & 0 & -t'-t_y e^{-ik_y} & -t'-t_x^+ e^{ik_x} & 0\\
	\end{pmatrix}, 
\end{eqnarray}
where $t'$ represents the intracell hopping amplitude, $t_x^\pm,t_y,t_z$ represent the intercell hopping amplitudes along $x,y,z$ directions, respectively, and $N_x,N_y,N_z$ are the sizes of the system along each direction.
Non-Hermiticity is introduced only along $x$ direction by the non-reciprocal hopping $t_x^+\neq t_x^-$.
In the Hermitian scenario with $t_x^+=t_x^->t'$ and $t_{y/z}>t'$, the 3D BBH model under OBCs supports 1D hinge states at all the twelve hinges. 

Analogous to the 2D non-Hermitian BBH model discussed in the main text, loop-like spectra emerge and they correspond to corner states induced by the interplay between 1D hinge localization and non-Hermiticity along $x$ direction, as shown in Fig.~\ref{BBH3D_SM}(a) and (b).

In Fig.~\ref{BBH3D_SM}(c), we display the average localization length along $x$ direction for the corner states with loop-like spectrum, $\bar{\xi}_x$, versus $N_x$ with different $N_y=N_z$. It can be seen that $\bar{\xi}_x$ grows linearly for large $N_x$, but takes a constant value of $\bar{\xi}_x\approx\xi_x^{\rm skin}=-2/\ln(t_x^-/t_x^+)$ when $N_x$ is small, with $\xi_x^{\rm skin}$ the localization length of NHSE of a single non-Hermitian SSH chain with nonreciprocal intercell hopping amplitudes $t_x^\pm$. 
On the other hand, $\bar{\xi}_x$ versus $N_{y/z}$ with $N_{x}=N_{z/y}$ is demonstrated in Fig.~\ref{BBH3D_SM}(d/e). $\bar{\xi}_x$ decreases and eventually reaches a constant when increasing, analogous to Fig. 2(a) in the main text. 
More explicitly, we plot its inverse $1/\bar{\xi}_x$ with different $N_x=N_y$ in Fig.~\ref{BBH3D_SM}(f), 
which shows a linear dependence on $N_z$, $\bar{\xi}_x\propto 1/N_z$, when $N_z\lesssim N_y$.
However,
while $\bar{\xi}_x$ also reaches a constant when $N_z>N_y$, it diverges from $\xi_x^{\rm skin}$, as can be seen in Fig.~\ref{BBH3D_SM}(e) and (f) [and also in (d) where the roles of $N_y$ and $N_z$ exchange].

To further unveil the influence of $N_y$ and $N_z$ on the localization length, we demonstrate a diagram of $\bar{\xi}_x$ versus $N_y$ and $N_z$ with fixed $N_x$ in Fig.~\ref{BBH3D_SM}(g). 
In the regime with $N_z\lesssim N_y$ ($N_y\lesssim N_z$), $\bar{\xi}_x$ is seen to nearly holds a constant value greater than $\xi_x^{\rm skin}$ that depends on $N_y$ ($N_z$), which is consistent with the observation in Fig.~\ref{BBH3D_SM}(d) to (f).
Thus, the size-dependency of average localization length can be summarized as
\begin{eqnarray}
	\bar{\xi}_x\propto \frac{N_x}{\min[N_y,N_z]}, 
\end{eqnarray}
when $\bar{\xi}_x>\xi_x^{\rm skin}$, represents a extension of the ASL in 3D systems.
This observation is further verified in Fig.~\ref{BBH3D_SM}(h), where we plot $1/\bar{\xi}_x$ as a function of $N_z$ with $N_x=8$ and several different $N_y$.
That is, $1/\bar{\xi}_x$ is seen to depend linearly on $N_z$ when $N_z\lesssim N_y$, and becomes a constant when $N_z\gtrsim N_y$, but never exceed $1/\xi_x^{\rm skin}$.
 
In this 3D model, since the ASL corner states form an effective edge model composed by four non-Hermitian SSH chains with effective bulk-induced couplings, the dependency on $\min[N_y,N_z]$ may originate from the competition between the effective bulk-induced couplings along different directions. Namely, when $N_y<N_z$, the effective couplings through the bulk (and surface) along $y$ direction are stronger than that along $z$ direction, so that $\bar{\xi}_x$ is mainly influenced by $N_y$.

\section{ASL with disorder}
In this section, we study the stability of ASL against disorder. We start from the HN-SSH model in the main text, then adding on-site disordered terms as 
\begin{equation}
	H_{\rm dis}=\sum_{(x,y)}w_{(x,y)}\hat{c}_{(x,y)}^\dagger\hat{c}_{(x,y)},
	\label{disorder_terms}
\end{equation}
where $\hat{c}_{(x,y)}$ is the annihilation operator of a particle at site $(x,y)$, and $w_i\in[-W/2,W/2]$ denotes the Anderson disorder \citep{anderson1958absence} with $W$ the disorder strength.

\begin{figure}
\begin{center}
	\includegraphics[width=1.0\linewidth]{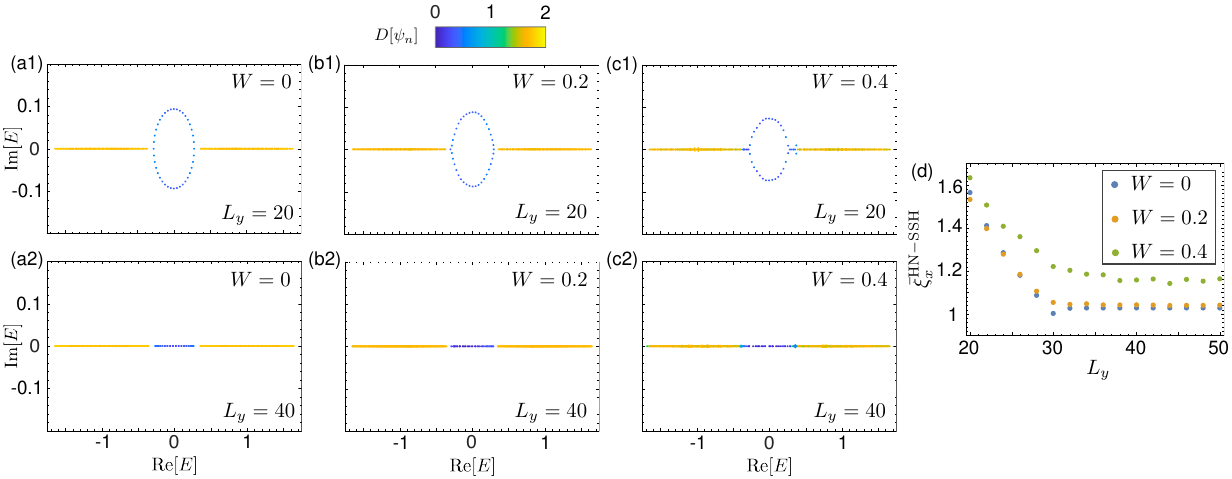}
\end{center}
	\caption{ASL with disorder in HN-SSH model. (a)-(c) Energy spectra with different disorder strengths $W=0,0.2,0.4$, respectively. Eigenenergies are marked by different colors according to the fractal dimension $D[\psi_n]$. (d) The average localization length $\bar{\xi}_x^{\rm HN-SSH}=\sum_n^{n_c}\xi_x/n_c$ of $n_c=L_x$ most localized states at the edge of $y=1$ (those with larger $\rho_c=\sum_{x\in[L_x-2,L_x]}\vert\psi(x,1)\vert^2$), versus $L_y$ with $W=0,0.2,0.4$. The results with $W\neq0$ shown in (d) are the average of 100 disorders. Other parameters are $t_x^+=0.35,t_x^-=0.05,t_1=0.25,t_2=1,L_x=20$, and $L_y=20,40$ in (a1)-(c1) and (a2)-(c2), respectively.}
	\label{disorder_HNSSH}
\end{figure}

As presented in the main text, ASL states in the HN-SSH model can be described by an effective ladder model, with either loop-like spectrum or line spectrum depending on the parameters and the size of the system. These spectral features are found to be robust against disorders as shown in Fig. \ref{disorder_HNSSH}(a)-(c). In Fig. \ref{disorder_HNSSH}(d), we display the average localization length $\bar{\xi}_x^{\rm HN-SSH}$ of corner states along the edge at $y=1$, which shows size-dependence even in the presence of disorder, manifesting the ASL.

\begin{figure}
\begin{center}
	\includegraphics[width=1.0\linewidth]{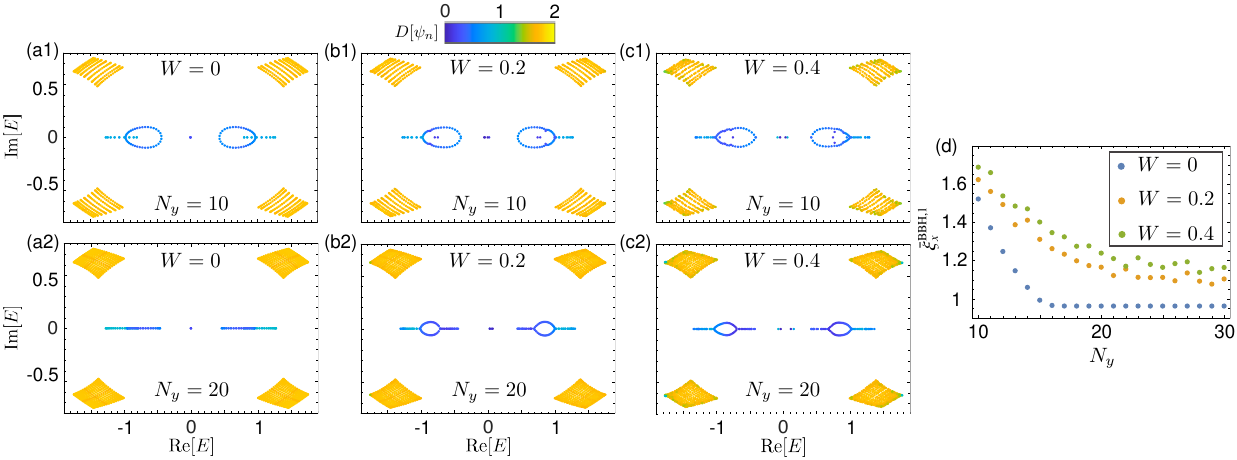}
\end{center}
	\caption{ASL with disorder in BBH model with non-Hermiticity added along $x$ direction. (a)-(c) Energy spectra with different disorder strengths $W=0,0.2,0.4$, respectively. Eigenenergies are marked by different colors according to the fractal dimension $D[\psi_n]$. (d) The average localization length $\bar{\xi}_x^{\rm BBH,1}=\sum_n^{n_c^x}\xi_x/n_c^x$ of $n_c^x=N_x=L_x/2$ most localized corner states at the edge of $y=1$ (those with larger $\rho_c^x=\sum_{x\in[L_x-6,L_x-1]}\vert\psi(x,1)\vert^2$; the site $x=L_x$ is neglated to avoid affection of possible second-order topological corner states), versus $N_y$ with $W=0,0.2,0.4$. The results with $W\neq0$ shown in (d) are the average of 100 disorders. Other parameters are $t_x^+=2,t_x^-=0.25,t_y^+=t_y^-=1,t'=0.25,N_x=20$, and $N_y=10,20$ in (a1)-(c1) and (a2)-(c2), respectively.}
	\label{disorder_BBH1}
\end{figure}

\begin{figure}
\begin{center}
	\includegraphics[width=1.0\linewidth]{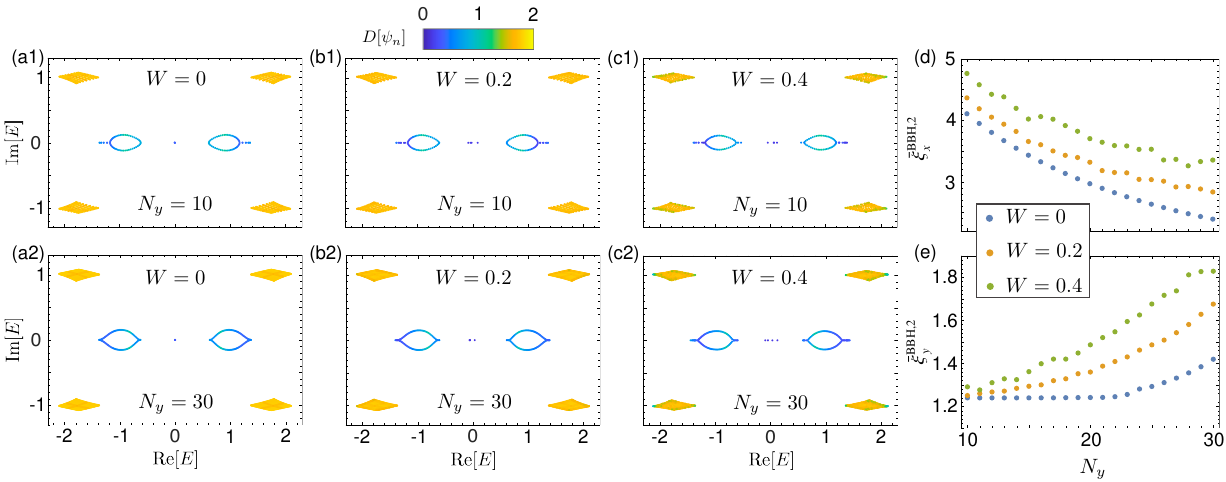}
\end{center}
	\caption{ASL with disorder in BBH model with non-Hermiticity added along both directions. (a)-(c) Energy spectra with different disorder strengths $W=0,0.2,0.4$, respectively. Eigenenergies are marked by different colors according to the fractal dimension $D[\psi_n]$. (d) The average localization length $\bar{\xi}_x^{\rm BBH,2}=\sum_n^{n_c^x}\xi_x/n_c^x$ of $n_c^x=N_x=L_x/2$ most localized corner states at the edge of $y=1$ (those with larger $\rho_c^x=\sum_{x\in[L_x-6,L_x-1]}\vert\psi(x,1)\vert^2$;  the site $x=L_x$ is neglated to avoid affection of possible second-order topological corner states)), versus $N_y$ with $W=0,0.2,0.4$. 
(e) The average localization length $\bar{\xi}_y^{\rm BBH,2}=\sum_n^{n_c^y}\xi_y/n_c^y$ of $n_c^y=N_y$ most localized corner statesat the edge of $x=1$ (those with larger $\rho_c^y=\sum_{x\in[L_y-6,L_y-1]}\vert\psi(1,y)\vert^2$; the site $y=L_y$ is neglated to avoid affection of possible second-order topological corner states), versus $N_y$ with $W=0,0.2,0.4$. The results with $W\neq0$ shown in (d) are the average of 100 disorders. Other parameters are $t_x^+=1.5,t_x^-=0.5,t_y^+=2.5,t_y^-=0.5,t'=0.25,N_x=20$, and $N_y=10,30$ in (a1)-(c1) and (a2)-(c2), respectively.}
	\label{disorder_BBH2}
\end{figure}

Next, we also demonstrate the energy spectrum and average localization length of the non-Hermitian BBH model with the same on-site disorder of Eq. (\ref{disorder_terms}), 
with non-Hermiticity added along $x$ direction (Fig. \ref{disorder_BBH1}) and both directions (Fig. \ref{disorder_BBH2}). 
ASL is seen to exist in both cases, however, in the former case, disorder appears to induce a transition in the spectrum of corner states, mixing the two branches (corresponding to states along different edges, see Fig. \ref{SM_BBH}) of spectrum together. 
These observations suggest that the corner states are now govened by the effective 1D junction model, instead of the ladder model formed by a pair of edges.
Specifically in Figs. \ref{disorder_BBH1}(a2)-(c2), the line spectrum opens a point-gap in the presence of disorder, indicating the emergence of junction-ASL even when the ladder-ASL is absent. 
This speculation is further varified by Fig. \ref{disorder_BBH1}(d), where the localization length varies with $N_y$ over a broader range of $N_y$.
On the other hand, in the BBH model with non-Hermiticity added along both directions, the size-dependence of localization length is also seen in a broader range of $N_y$, 
as shown in Fig. \ref{disorder_BBH2}(e).

\section{ASL in a cold-atom system}

\begin{figure}
	\includegraphics[width=1.0\linewidth]{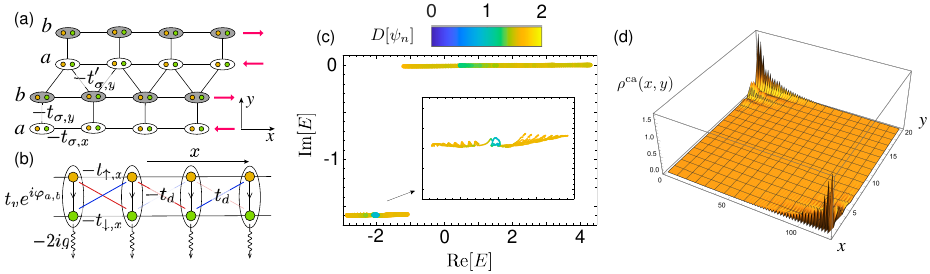}
	\caption{
(a) The sketch of the tight-binding model. of Eqs.~\ref{eq:TB1} and \ref{eq:TB2}. (b) Two-leg system for each sublattice [white and gray rows in (a)], with yellow and green sites representing up and down pseudospins. (a) and (b) are adopted from Ref. \citep{li2020topological}. (c) Energy spectrum with $g=0.8$. Inset displays the zoom-in view of eigenenergies marked by the arrow. (d) Summed distribution $\rho^{\rm ca}(x,y)=\sum_n\vert\psi_n(x,y)\vert^2$ corresponding to (c), where the summation runs over $n_D=30$ most localized eigenstates. Numerically we consider $N_x=60$ and $N_y=10$ unit cells along the two directions in (c) and (d). Other parameters are $\varphi=\pi/2$ and $\left\{t_{+,x},t_{-,x},t_{+,y},t_{-,y},t_{+,y}',t_{-,y}',\Delta_{-},t_d,t_v\right\}=\left\{0.38,0.43,-0.11,-0.41,0.44,0.14,2.07,-0.09,0.16\right\}$\citep{li2020topological}.
}
	\label{cold_atom}
\end{figure}

To demonstrate the experimental feasibility of ASL, we consider cold atoms loaded in a two-dimensional optical lattice with atom loss introduced from a resonant optical beam~\citep{li2020topological}.
In the tight-binding approximation, its Bloch Hamiltnoian is given by
 $H(\mathbf{k})=\mathbf{\Psi}^\dagger(\mathbf{k}) h_{\rm 2D}(\mathbf{k})\mathbf{\Psi}(\mathbf{k})$ with $\mathbf{\Psi}^\dagger(\mathbf{k})=(\hat{a}^\dagger_{\uparrow\mathbf{k}},\hat{b}^\dagger_{\uparrow\mathbf{k}},\hat{a}^\dagger_{\downarrow\mathbf{k}},\hat{b}^\dagger_{\downarrow\mathbf{k}})$,
with $\alpha^\dagger_{s \mathbf{k}}$ creating a particle of  pseudospin $s$ and crystal-momentum $\mathbf{k}$ on sublattice $\alpha$.
The Hamiltnoian matrix reads
\begin{equation}
	h_{\rm 2D}(\mathbf{k})=h_{\sigma}^+(\mathbf{k})\sigma_0+h_{\sigma}^-(\mathbf{k})\sigma_3+h_{\tau}^+(\mathbf{k})\tau_0+h_{\tau}^-(\mathbf{k})\tau_3, \label{eq:TB1}
\end{equation}
\begin{equation}
\begin{split}
	&h_{\sigma}^{\pm}(\mathbf{k})=-(2t_{\pm,x}\cos k_x-\Delta_{\pm}\pm ig)\tau_0-\left\{t_{\pm,y}+t_{\pm,y}'[\cos k_y+\cos(k_y-k_x)]\right\}\tau_1-t_{\pm,y}'[\sin k_y+\sin(k_y-k_x)]\tau_2,\\
	&h_\tau^-(\mathbf{k})=(t_v\cos\varphi)\sigma_1+(t_v\sin\varphi)\sigma_2, \\
	&h_\tau^+(\mathbf{k})=(2t_d\sin k_x)\sigma_2.\label{eq:TB2}
\end{split}
\end{equation}
Here $\tau_i$ and $\sigma_i$ $(i=1,2,3)$ are two sets of Pauli matrices acting on the sublattice and pseudospin (lattice orbitals) spaces, respectively, with $\tau_0$ and $\sigma_0$ their corresponding $2\times2$ identity matrices. The various coupling amplitudes $t_v$, $t_d$, $t_{\pm,\alpha}=(t_{\uparrow,\alpha}\pm t_{\downarrow,\alpha})/2$, and $t_{\pm,\alpha}'=(t_{\uparrow,\alpha}'\pm t_{\downarrow,\alpha}')/2$ arise from overlap integrals between lattice orbitals.
Similar to the HN-SSH model in the main text, this model supports both topological localization along $y$ direction and non-Hermitian non-reciprocal pumping along $x$ direction, induced by periodical driving of the lattice and/or the atom loss~\citep{li2020topological}.
Therefore ASL can be expected to arise from effective bulk coupling between the top and bottom edges.

\begin{figure}
	\includegraphics[width=0.75\linewidth]{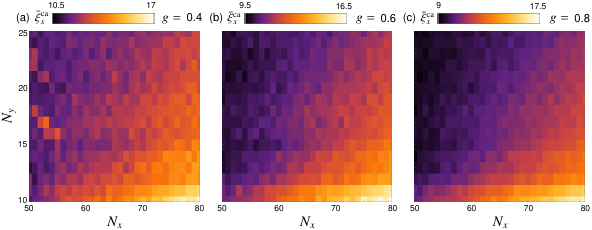}
	\caption{ASL in the cold atom system with loss. (a)-(c) The average localization length $\bar{\xi}_x^{\rm ca}=\sum_n^{n_{\rm D}}\xi_x^{\rm ca}/n_{\rm D}$ of $n_D=30$ most localized states versus $N_x$ and $N_y$, with $g=0.4,0.6$ and $0.8$ respectively. Other  parameters are $\varphi=\pi/2$ and $\left\{t_{+,x},t_{-,x},t_{+,y},t_{-,y},t_{+,y}',t_{-,y}',\Delta_{-},t_d,t_v\right\}=\left\{0.38,0.43,-0.11,-0.41,0.44,0.14,2.07,-0.09,0.16\right\}$\citep{li2020topological}.}
	\label{cold_atom2}
\end{figure}

The lattice strucutre of the model is sketched in Fig. \ref{cold_atom}(a) and (b). 
In Fig. \ref{cold_atom}(c) and (d), it is seen that this model also supports corner states with loop spectrum.
Since each of the top and bottom edges contains an internal pseudospin degree of freedom,
we define the localization length similar to the non-Hermitian BBH model in the main text (also see Sec. \ref{sec:length} in the Supplemental Materials),
\begin{equation*}
	\xi_x^{\rm ca}=\frac{2(N_x-1)}{\ln\left\{[\vert\psi_n^{a,\uparrow}(N_x,1)\vert\vert\psi_n^{a,\downarrow}(N_x,1)\vert]/[\vert\psi_n^{a,\uparrow}(1,1)\vert\vert\psi_n^{a,\downarrow}(1,1)\vert]\right\}},
\end{equation*}
to further characterize the localization along $x-$direction on the top and bottom edges. 
Numerically, we consider the average localization length of the $n_{\rm D}$ most localized states, 
$\bar{\xi}_x^{\rm ca}=\sum_n^{n_{\rm D}} \xi_x^{\rm ca}/n_{\rm D}$, and demonstrate the diagrams of $\bar{\xi}_x^{\rm ca}$ versus $N_x$ and $N_y$, the sizes of the system, in Fig. \ref{cold_atom2}. 
The ASL is verified by the observed size-dependence of $\bar{\xi}_x^{\rm ca}$ similar to that of the HN-SSH model [see Fig. 2(c) in the main text], which becomes more pronounced as the atom loss rate increases.

\red{
\section{ASL in systems with long-range coupling}
}

\red{In this section, we investigate ASL in our models with extra long-range coupling. We find that ASL still exist, yet its specific behavior can be altered by long-range coupling.}

\red{
\subsection{HN-SSH model with exponentially decaying long-range coupling}
}
\begin{figure}
	\includegraphics[width=1\linewidth]{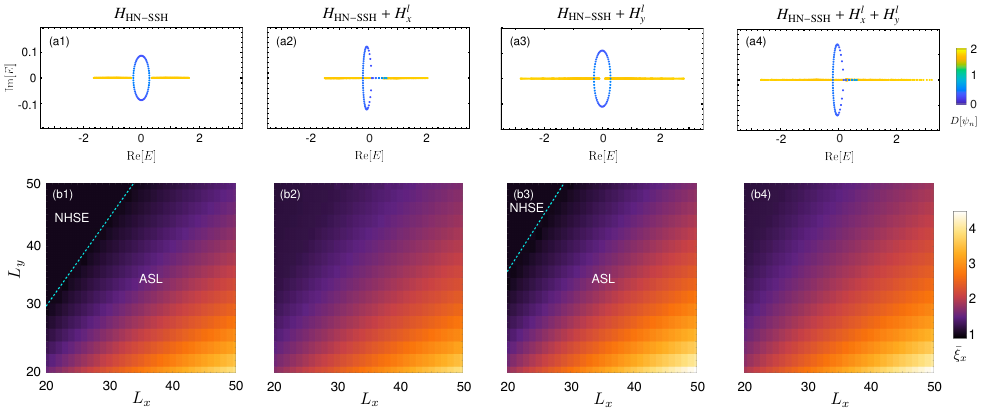}
	\caption{\red{(a) Energy spectra of HN-SSH model with exponentially decaying long-range coupling. The decay length is set to be $l_d=1.5$. The size of the system is set to be $L_x=L_y=30$. Eigenenergies are marked by different colors according to the fractal dimension $D[\psi]$. (a1) $H_{\rm HN-SSH}$, without long-range coupling. (a2) $H_{\rm HN-SSH}+H_x^l$, with long-range coupling added only along $x$ direction. (a3) $H_{\rm HN-SSH}+H_y^l$, with long-range coupling added only along $y$ direction. (a4) $H_{\rm HN-SSH}+H_x^l+H_y^l$, with long-range coupling added along both $x$ and $y$ direction. 
(b) Average localization length along $x$ direction of the corner states in (a), $\bar{\xi}_x$, versus the size of system along $x$ and $y$ directions, $L_x$ and $L_y$. (b1)-(b4) Ccorrespond to (a1)-(a4). Cyan dashed line in (b1) is the analytical result of $\xi_x^{\rm AS}=\xi_x^{\rm skin}$, which is the transition line between ASL and NHSE for the system without long-range coupling. A similar transition line is obtained numerically in (b3).
The NHSE region disappears in (b2) and (b4), and becomes smaller in (b3).
Parameters are $t_x^+=0.35,t_x^-=0.05,t_1=0.25,t_2=1$. }
}
	\label{HN-SSH_decay_E_xi}
\end{figure}

\red{
We start with the HN-SSH model in the main text, whose Hamiltonian in real space reads, 
\begin{eqnarray}
	H_{\rm HN-SSH}&=&\sum_{x=1}^{L_x-1}\sum_{y=1}^{L_y/2}\left(t_x^+ c_{x,y,A}^\dagger c_{x+1,y,A}+t_x^- c_{x+1,y,A}^\dagger c_{x,y,A}+t_x^- c_{x,y,B}^\dagger c_{x+1,y,B}+t_x^+ c_{x+1,y,B}^\dagger c_{x,y,B}\right)\notag\\
	&&+\sum_{x=1}^{L_x}\left[\sum_{y=1}^{L_y/2}t_1(c_{x,y,A}^\dagger c_{x,y,B}+c_{x,y,B}^\dagger c_{x,y,A})+\sum_{y=1}^{L_y/2-1}t_2(c_{x,y+1,A}^\dagger c_{x,y,B}+c_{x,y,B}^\dagger c_{x,y+1,A})\right], 
	\label{eq_HNSSH_realspace}
\end{eqnarray}
where $c_{x,y,\alpha}^{(\dagger)}$ ($\alpha=A,B$) is the annihilation (creation) operator of $\alpha$ site at position $(x,y)$, $t_x^+, t_x^-$ are the nonreciprocal hopping amplitudes along $x$ direction, and $t_1,t_2$ are the staggered hopping amplitudes along $y$ direction ($t_1<t_2$ is chosen to generate topological localization on top and bottom edges). 
In addition, we consider exponentially decaying long-range coupling along $x$ and $y$ directions, described by 
\begin{eqnarray}
	H_x^l=\sum_{x=1}^{L_x-1}\sum_{y=1}^{L_y/2}\sum_{l_x=2}^{L_x-x}\left(t_x^+ c_{x,y,A}^\dagger c_{x+l_x,y,A}+t_x^- c_{x+l_x,y,A}^\dagger c_{x,y,A}+t_x^- c_{x,y,B}^\dagger c_{x+l_x,y,B}+t_x^+ c_{x+l_x,y,B}^\dagger c_{x,y,B}\right)\exp[-(l_x-1)/l_{d,x}]
\end{eqnarray}
\begin{eqnarray}
	H_y^l&=&\sum_{x=1}^{L_x}\sum_{y=1}^{L_y/2}\sum_{l_y=2}^{L_y/2-y}t_1(c_{x,y,A}^\dagger c_{x,y+l_y-1,B}+c_{x,y+l_y-1,B}^\dagger c_{x,y,A})\exp[-(l_y-1)/l_{d,y}]\notag\\
	&&+\sum_{x=1}^{L_x}\sum_{y=1}^{L_y/2-1}\sum_{l_y=2}^{L_y/2+1-y}t_2(c_{x,y+l_y,A}^\dagger c_{x,y,B}+c_{x,y,B}^\dagger c_{x,y+l_y,A})\exp[-(l_y-1)/l_{d,y}], 
\end{eqnarray}
where $l_{d,x(y)}$ is the decay length along $x(y)$ direction. In the folowing, we set $l_{d,x}=l_{d,y}=l_d$ for simplicity. 
}

\red{
Compared to the case without long-range coupling [$H_{\rm HN-SSH}$, Fig. \ref{HN-SSH_decay_E_xi}(a1)], 
we find that long-range coupling along different directions affects the spectrum in different manners.
For instance, when long-range coupling is added only along the $x$ direction [$H_{\rm HN-SSH}+H_x^l$, Fig. \ref{HN-SSH_decay_E_xi}(a2)], the shape of the boundary spectrum changes, transforming from the original loop into a combination of loop and line. Conversely, when long-range coupling is added only along the $y$ direction [$H_{\rm HN-SSH}+H_y^l$, Fig. \ref{HN-SSH_decay_E_xi}(a3)], the shape of the boundary spectrum retains its loop-like structure, but its size changes. 
Finally, the case with long-range coupling  added along both $x$ and $y$ directions [$H_{\rm HN-SSH}+H_x^l+H_y^l$] is shown in Fig. \ref{HN-SSH_decay_E_xi}(a4), where the boundary spectrum transforms into the loop-line structure, with its size also slightly increased.
Nevertheless, the boundary spectrum always retains the loop-like feature that suggests size-dependent localization. 
}

\red{
In the main text, we consider the case without long-range coupling, where the corner states exhibit NHSE and ASL in different parameter regimes, with their demarcation given by $\xi_x^{\rm AS}=\xi_x^{\rm skin}$, as shown in Fig. \ref{HN-SSH_decay_E_xi}(b1). 
With long-range coupling added, the regime of ASL is seen to persists, while that of NHSE disappears in Fig. \ref{HN-SSH_decay_E_xi}(b2) and (b4), and becomes smaller in (b3).
These observations can still be understood using the effective 1D ladder model consisting of  the top and bottom edges.
That is, taking these two edges as two 1D chains,  long-range coupling along $x$ direction can enhance the size-dependent features along each of them \citep{wang2023scaling}, allowing ASL to persist even when the effective inter-chain coupling through the bulk is weak (i.e., with a large $L_y$).
On the other hand, long-range coupling along $y$ direction enhances the effective coupling between the two chains, thus only reducing but not eliminating the region of the NHSE.
}


\red{
\subsection{Non-Hermitian BBH model with exponentially decaying long-range coupling}
}

\begin{figure}
	\includegraphics[width=1\linewidth]{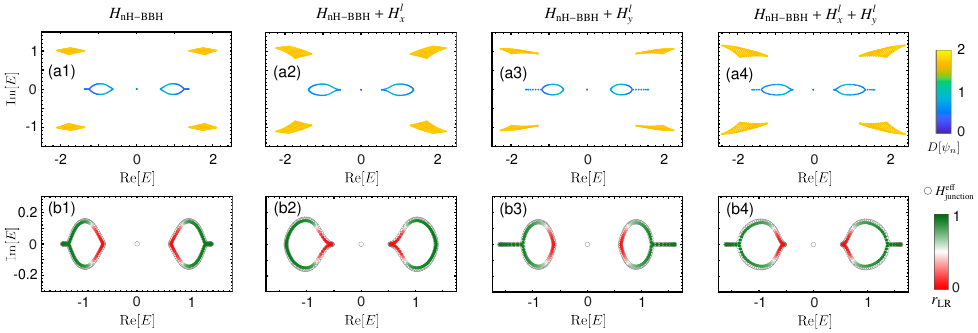}
	\caption{\red{(a) Energy spectrum of the non-Hermitian BBH model with exponentially decaying long-range coupling. Eigenenergies are marked by different colors according to the fractal dimension $D[\psi]$. (a1) $H_{\rm nH-BBH}$, without long-range coupling. (a2) $H_{\rm nH-BBH}+H_x^l$, with long-range coupling added only along $x$ direction. (a3) $H_{\rm nH-BBH}+H_y^l$, with long-range coupling added only along $y$ direction. (a4) $H_{\rm nH-BBH}+H_x^l+h_y^l$, with long-range coupling added along both $x$ and $y$ directions. (b) Comparison between the eigenenergies of the edge states (colored solid dots) in (a) and that of $H_{\rm junction}^{\rm eff}$ [Eq. \eqref{H_junction}], the effective 1D junction model formed by the edges of the 2D system. Colors indicate the edge distribution ratio $r_{\rm LR}$ of each edge state. $r_{\rm LR}\approx 1$ and $0$ correspond to eigenstates distributing mostly on the left/right and top/bottom edges, respectively. It is seen that the eigenenergies of $H_{\rm junction}^{\rm eff}$ match well with those edge states. Parameters are $t_x^+=1.5,t_y^+=2.5,t_x^-=t_y^-=0.5,t'=0.25,N_x=N_y=20$, and $l_d=0.5$. }}
	\label{BBH_E_decay}
\end{figure}

\red{
Similar to the HN-SSH model above, we now consider exponentially decaying long-range coupling in the non-Hermitian BBH model. 
The Hamiltonian of non-Hermitian BBH model in real space is, 
\begin{eqnarray}
	H_{\rm nH-BBH}&=&\sum_{y=1}^{N_y}\sum_{x=1}^{N_x}t'\left(c_{x,y,A}^\dagger c_{x,y,B}+c_{x,y,B}^\dagger c_{x,y,A}+c_{x,y,C}^\dagger c_{x,y,D}+c_{x,y,D}^\dagger c_{x,y,D}\right)\notag\\
	&&+\sum_{x=1}^{N_x}\sum_{y=1}^{N_y}t'\left(c_{x,y,A}^\dagger c_{x,y,C}+c_{x,y,C}^\dagger c_{x,y,A}+c_{x,y,B}^\dagger c_{x,y,D}+c_{x,y,D}^\dagger c_{x,y,B}\right)\notag\\
	&&+\sum_{y=1}^{N_y}\sum_{x=1}^{N_x-1}\left(t_x^+ c_{x,y,B}^\dagger c_{x+1,y,A}+t_x^- c_{x+1,y,A}^\dagger c_{x,y,B}+t_x^- c_{x,y,C}^\dagger c_{x+1,y,D}+t_x^+ c_{x+1,y,D}^\dagger c_{x,y,C}\right)\notag\\
	&&+\sum_{x=1}^{N_x}\sum_{y=1}^{N_y-1}\left(t_y^- c_{x,y,C}^\dagger c_{x,y+1,B}+t_y^+ c_{x,y+1,B}^\dagger c_{x,y,C}+t_y^+ c_{x,y,D}^\dagger c_{x,y+1,B}+t_y^- c_{x,y+1,B}^\dagger c_{x,y,D}\right), 
\end{eqnarray}
where $c_{x,y,\alpha}^{(\dagger)}$ ($\alpha=A,B,C,D$) is the annihilation (creation) operator of $\alpha$ site at position $(x,y)$, $t'$ is the intracell hopping amplitude, and $t_{x(y)}^+,t_{x(y)}^-$ are the nonreciprocal intercell hopping amplitudes along $x(y)$ direction. 
And the exponential decaying long-range coupling along $x$ and $y$ directions are described by 
\begin{eqnarray}
	H_{x}^l&=&\sum_{y=1}^{N_y}\sum_{x=1}^{N_x}\sum_{l_x=2}^{N_x+1-x}t'\left(c_{x,y,A}^\dagger c_{x+l_x-1,y,B}+c_{x+l_x-1,y,B}^\dagger c_{x,y,A}+c_{x,y,C}^\dagger c_{x+l_x-1,y,D}+c_{x+l_x-1,y,D}^\dagger c_{x,y,D}\right)\exp[-(l_x-1)/l_{d,x}]\notag\\
	&&+\sum_{y=1}^{N_y}\sum_{x=1}^{N_x-1}\sum_{l_x=2}^{N_x-x}\left(t_x^+ c_{x,y,B}^\dagger c_{x+l_x,y,A}+t_x^- c_{x+l_x,y,A}^\dagger c_{x,y,B}+t_x^- c_{x,y,C}^\dagger c_{x+l_x,y,D}+t_x^+ c_{x+l_x,y,D}^\dagger c_{x,y,C}\right)\exp[-(l_x-1)/l_{d,x}], 
\end{eqnarray}
\begin{eqnarray}
	H_{y}^l&=&\sum_{x=1}^{N_x}\sum_{y=1}^{N_y}\sum_{l_y=2}^{N_y+1-y}t'\left(c_{x,y,A}^\dagger c_{x,y+l_y-1,C}+c_{x,y+l_y-1,C}^\dagger c_{x,y,A}+c_{x,y,B}^\dagger c_{x,y+l_y-1,D}+c_{x,y+l_y-1,D}^\dagger c_{x,y,B}\right)\exp[-(l_y-1)/l_{d,y}]\notag\\
	&&+\sum_{x=1}^{N_x}\sum_{y=1}^{N_y-1}\sum_{l_y=2}^{N_y-y}\left(t_y^- c_{x,y,C}^\dagger c_{x,y+l_y,B}+t_y^+ c_{x,y+l_y,B}^\dagger c_{x,y,C}+t_y^+ c_{x,y,D}^\dagger c_{x,y+l_y,B}+t_y^- c_{x,y+l_y,B}^\dagger c_{x,y,D}\right)\exp[-(l_y-1)/l_{d,y}], 
\end{eqnarray}
where $l_{d,x(y)}$ is the decay length along $x(y)$ direction. In the following, we set $l_{d,x}=l_{d,y}=l_d$ for simplicity. 
}

\red{
In Fig. \ref{BBH_E_decay}, we compare the spectral features for scenarios with long-range coupling added along different directions, which is found to enhance the size-dependent feature along one direction, and suppress it along the other.
For example, comparing to the scenario without long-range coupling [$H_{\rm nH-BBH}$, Fig. \ref{BBH_E_decay}(a1) and (b1)], 
we find that long-range coupling along $x$ direction [$H_{\rm nH-BBH}+H_x^l$, Fig. \ref{BBH_E_decay}(a2) and (b2)] partially shrinks the loop-like spectrum into lines for boundary states at the top and bottom edges  [red dots with $r_{\rm LR}\approx0$ in Fig. \ref{BBH_E_decay}(b)],
while opens the line-like spectrum into loops for boundary states at the left and right edges  [green dots with $r_{\rm LR}\approx0$ in Fig. \ref{BBH_E_decay}(b)].
Conversely long-range coupling along $y$ direction only [$H_{\rm nH-BBH}+H_y^l$, Fig. \ref{BBH_E_decay}(a2) and (b2)] affect the spectrum oppositely for boundary states at different edges.
Finally, loop-like boundary spectrum persists in all these scenarios, including the one with long-range coupling is added along both directions [$H_{\rm nH-BBH}+h_x^l+H_y^l$, Fig. \ref{BBH_E_decay}(a4) and (b4)], indicating the existence ASL. 
}

\red{
Interestingly, we find that the effective 1D junction model is still applicable under long-range coupling [Fig. \ref{BBH_E_decay}(b)]. 
Physically, extra long-range coupling along one direction, say $x$, enhances the connection between left and right edges and thus the size-dependent feature for corresponding eigenstates (see our discussion in Sec. \ref{sec:junction}).
However, its suppression on size-dependent feature along top and bottom edges (namely, $x$ direction) is against previous knowledges in a single non-Hermitian chain under OBCs, where long-range coupling enhances the scale-free localization \citep{wang2023scaling}. A possible explanation may be that the long-range coupling along one direction ($x$) enhances the connection within a single edge (top and bottom), making it more independent from the other edges and thus behave more like under OBCs; yet it is not strong enough (decays too fast with increasing hopping range) to induce the scale-free localization. 
However, further in-depth analysis is required to provide a complete understanding of these phenomena.
}

\begin{figure}
	\includegraphics[width=0.6\linewidth]{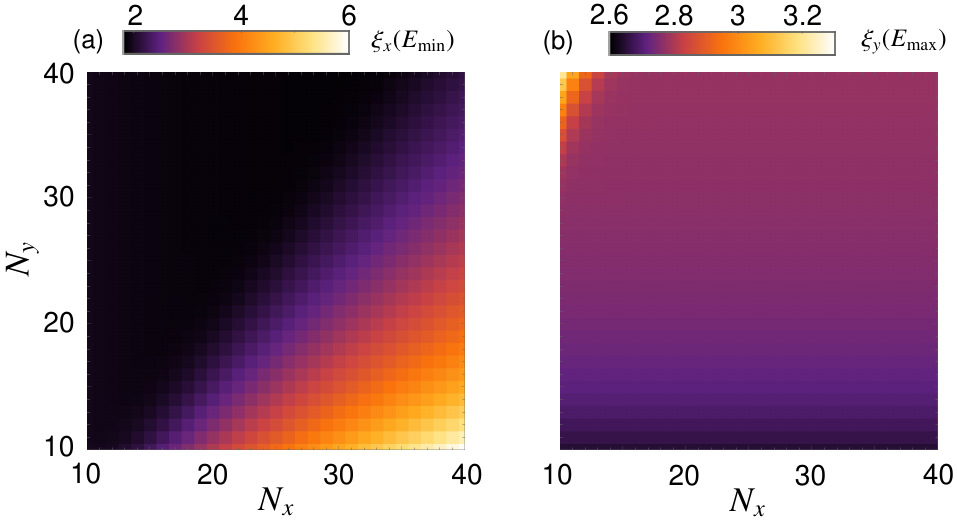}
	\caption{\red{Localization length along $x(y)$ direction of corner state with the minimal (maximal) absolute value of real energy of the system $H_{\rm nH-BBH}+H_x^l+H_y^l$, $\xi_x(E_{\rm min})[\xi_y(E_{\rm max})]$, versus the system's size $N_x$ and $N_y$. Parameters are $t_x^+=1.5,t_y^+=2.5,t_x^-=t_y^-=0.5,t'=0.25$, and $l_d=0.5$. }}
	\label{BBH_xi_decay}
\end{figure}

\red{
Finally, as shown in Fig. \ref{BBH_xi_decay}, we calculate the localization length along $x$ ($y$) direction of the corner state with the minimal (maximal) absolute value of real energy, i.e. $E_{\rm min}:\text{Re}[E_{\rm min}]=\text{Min}\vert\text{Re}[E]\vert$ ($E_{\rm max}:\text{Re}[E_{\rm max}]=\text{Max}\vert\text{Re}[E]\vert$), to illustrate the size-dependent behaviors of boundary states on top and bottom (left and right) edges. We can see that ASL still exists in the presence with long-range coupling along both directions.
In addition, we note other types of size-dependent effects may arise in regions beyond ASL [top-left region in Fig. \ref{BBH_xi_decay}(a) and bottom-right region in Fig. \ref{BBH_xi_decay}(b)], which awaits further exploration.}

\bibliography{ASLrefs}

\begin{thebibliography}{68}%
\makeatletter
\providecommand \@ifxundefined [1]{%
 \@ifx{#1\undefined}
}%
\providecommand \@ifnum [1]{%
 \ifnum #1\expandafter \@firstoftwo
 \else \expandafter \@secondoftwo
 \fi
}%
\providecommand \@ifx [1]{%
 \ifx #1\expandafter \@firstoftwo
 \else \expandafter \@secondoftwo
 \fi
}%
\providecommand \natexlab [1]{#1}%
\providecommand \enquote  [1]{``#1''}%
\providecommand \bibnamefont  [1]{#1}%
\providecommand \bibfnamefont [1]{#1}%
\providecommand \citenamefont [1]{#1}%
\providecommand \href@noop [0]{\@secondoftwo}%
\providecommand \href [0]{\begingroup \@sanitize@url \@href}%
\providecommand \@href[1]{\@@startlink{#1}\@@href}%
\providecommand \@@href[1]{\endgroup#1\@@endlink}%
\providecommand \@sanitize@url [0]{\catcode `\\12\catcode `\$12\catcode
  `\&12\catcode `\#12\catcode `\^12\catcode `\_12\catcode `\%12\relax}%
\providecommand \@@startlink[1]{}%
\providecommand \@@endlink[0]{}%
\providecommand \url  [0]{\begingroup\@sanitize@url \@url }%
\providecommand \@url [1]{\endgroup\@href {#1}{\urlprefix }}%
\providecommand \urlprefix  [0]{URL }%
\providecommand \Eprint [0]{\href }%
\providecommand \doibase [0]{https://doi.org/}%
\providecommand \selectlanguage [0]{\@gobble}%
\providecommand \bibinfo  [0]{\@secondoftwo}%
\providecommand \bibfield  [0]{\@secondoftwo}%
\providecommand \translation [1]{[#1]}%
\providecommand \BibitemOpen [0]{}%
\providecommand \bibitemStop [0]{}%
\providecommand \bibitemNoStop [0]{.\EOS\space}%
\providecommand \EOS [0]{\spacefactor3000\relax}%
\providecommand \BibitemShut  [1]{\csname bibitem#1\endcsname}%
\let\auto@bib@innerbib\@empty
\bibitem [{\citenamefont {Anderson}(1958)}]{anderson1958absence}%
  \BibitemOpen
  \bibfield  {author} {\bibinfo {author} {\bibfnamefont {P.~W.}\ \bibnamefont
  {Anderson}},\ }\href@noop {} {\bibfield  {journal} {\bibinfo  {journal}
  {Physical review}\ }\textbf {\bibinfo {volume} {109}},\ \bibinfo {pages}
  {1492} (\bibinfo {year} {1958})}\BibitemShut {NoStop}%
\bibitem [{\citenamefont {Hasan}\ and\ \citenamefont
  {Kane}(2010)}]{hasan2010colloquium}%
  \BibitemOpen
  \bibfield  {author} {\bibinfo {author} {\bibfnamefont {M.~Z.}\ \bibnamefont
  {Hasan}}\ and\ \bibinfo {author} {\bibfnamefont {C.~L.}\ \bibnamefont
  {Kane}},\ }\href@noop {} {\bibfield  {journal} {\bibinfo  {journal} {Reviews
  of modern physics}\ }\textbf {\bibinfo {volume} {82}},\ \bibinfo {pages}
  {3045} (\bibinfo {year} {2010})}\BibitemShut {NoStop}%
\bibitem [{\citenamefont {Qi}\ and\ \citenamefont
  {Zhang}(2011)}]{qi2011topological}%
  \BibitemOpen
  \bibfield  {author} {\bibinfo {author} {\bibfnamefont {X.-L.}\ \bibnamefont
  {Qi}}\ and\ \bibinfo {author} {\bibfnamefont {S.-C.}\ \bibnamefont {Zhang}},\
  }\href@noop {} {\bibfield  {journal} {\bibinfo  {journal} {Reviews of modern
  physics}\ }\textbf {\bibinfo {volume} {83}},\ \bibinfo {pages} {1057}
  (\bibinfo {year} {2011})}\BibitemShut {NoStop}%
\bibitem [{\citenamefont {Lee}(2016)}]{lee2016anomalous}%
  \BibitemOpen
  \bibfield  {author} {\bibinfo {author} {\bibfnamefont {T.~E.}\ \bibnamefont
  {Lee}},\ }\href@noop {} {\bibfield  {journal} {\bibinfo  {journal} {Physical
  review letters}\ }\textbf {\bibinfo {volume} {116}},\ \bibinfo {pages}
  {133903} (\bibinfo {year} {2016})}\BibitemShut {NoStop}%
\bibitem [{\citenamefont {Martinez~Alvarez}\ \emph {et~al.}(2018)\citenamefont
  {Martinez~Alvarez}, \citenamefont {Barrios~Vargas},\ and\ \citenamefont
  {Foa~Torres}}]{martinez2018non}%
  \BibitemOpen
  \bibfield  {author} {\bibinfo {author} {\bibfnamefont {V.~M.}\ \bibnamefont
  {Martinez~Alvarez}}, \bibinfo {author} {\bibfnamefont {J.~E.}\ \bibnamefont
  {Barrios~Vargas}},\ and\ \bibinfo {author} {\bibfnamefont {L.~E.~F.}\
  \bibnamefont {Foa~Torres}},\ }\href
  {https://doi.org/10.1103/PhysRevB.97.121401} {\bibfield  {journal} {\bibinfo
  {journal} {Phys. Rev. B}\ }\textbf {\bibinfo {volume} {97}},\ \bibinfo
  {pages} {121401} (\bibinfo {year} {2018})}\BibitemShut {NoStop}%
\bibitem [{\citenamefont {Yao}\ and\ \citenamefont {Wang}(2018)}]{yao2018edge}%
  \BibitemOpen
  \bibfield  {author} {\bibinfo {author} {\bibfnamefont {S.}~\bibnamefont
  {Yao}}\ and\ \bibinfo {author} {\bibfnamefont {Z.}~\bibnamefont {Wang}},\
  }\href@noop {} {\bibfield  {journal} {\bibinfo  {journal} {Physical review
  letters}\ }\textbf {\bibinfo {volume} {121}},\ \bibinfo {pages} {086803}
  (\bibinfo {year} {2018})}\BibitemShut {NoStop}%
\bibitem [{\citenamefont {Yokomizo}\ and\ \citenamefont
  {Murakami}(2019)}]{yokomizo2019non}%
  \BibitemOpen
  \bibfield  {author} {\bibinfo {author} {\bibfnamefont {K.}~\bibnamefont
  {Yokomizo}}\ and\ \bibinfo {author} {\bibfnamefont {S.}~\bibnamefont
  {Murakami}},\ }\href@noop {} {\bibfield  {journal} {\bibinfo  {journal}
  {Physical review letters}\ }\textbf {\bibinfo {volume} {123}},\ \bibinfo
  {pages} {066404} (\bibinfo {year} {2019})}\BibitemShut {NoStop}%
\bibitem [{\citenamefont {Borgnia}\ \emph {et~al.}(2020)\citenamefont
  {Borgnia}, \citenamefont {Kruchkov},\ and\ \citenamefont
  {Slager}}]{borgnia2020non}%
  \BibitemOpen
  \bibfield  {author} {\bibinfo {author} {\bibfnamefont {D.~S.}\ \bibnamefont
  {Borgnia}}, \bibinfo {author} {\bibfnamefont {A.~J.}\ \bibnamefont
  {Kruchkov}},\ and\ \bibinfo {author} {\bibfnamefont {R.-J.}\ \bibnamefont
  {Slager}},\ }\href@noop {} {\bibfield  {journal} {\bibinfo  {journal}
  {Physical review letters}\ }\textbf {\bibinfo {volume} {124}},\ \bibinfo
  {pages} {056802} (\bibinfo {year} {2020})}\BibitemShut {NoStop}%
\bibitem [{\citenamefont {Okuma}\ \emph {et~al.}(2020)\citenamefont {Okuma},
  \citenamefont {Kawabata}, \citenamefont {Shiozaki},\ and\ \citenamefont
  {Sato}}]{okuma2020topological}%
  \BibitemOpen
  \bibfield  {author} {\bibinfo {author} {\bibfnamefont {N.}~\bibnamefont
  {Okuma}}, \bibinfo {author} {\bibfnamefont {K.}~\bibnamefont {Kawabata}},
  \bibinfo {author} {\bibfnamefont {K.}~\bibnamefont {Shiozaki}},\ and\
  \bibinfo {author} {\bibfnamefont {M.}~\bibnamefont {Sato}},\ }\href@noop {}
  {\bibfield  {journal} {\bibinfo  {journal} {Physical review letters}\
  }\textbf {\bibinfo {volume} {124}},\ \bibinfo {pages} {086801} (\bibinfo
  {year} {2020})}\BibitemShut {NoStop}%
\bibitem [{\citenamefont {Zhang}\ \emph {et~al.}(2020)\citenamefont {Zhang},
  \citenamefont {Yang},\ and\ \citenamefont {Fang}}]{zhang2020correspondence}%
  \BibitemOpen
  \bibfield  {author} {\bibinfo {author} {\bibfnamefont {K.}~\bibnamefont
  {Zhang}}, \bibinfo {author} {\bibfnamefont {Z.}~\bibnamefont {Yang}},\ and\
  \bibinfo {author} {\bibfnamefont {C.}~\bibnamefont {Fang}},\ }\href@noop {}
  {\bibfield  {journal} {\bibinfo  {journal} {Physical Review Letters}\
  }\textbf {\bibinfo {volume} {125}},\ \bibinfo {pages} {126402} (\bibinfo
  {year} {2020})}\BibitemShut {NoStop}%
\bibitem [{\citenamefont {Lin}\ \emph {et~al.}(2023)\citenamefont {Lin},
  \citenamefont {Tai}, \citenamefont {Li},\ and\ \citenamefont
  {Lee}}]{lin2023topological}%
  \BibitemOpen
  \bibfield  {author} {\bibinfo {author} {\bibfnamefont {R.}~\bibnamefont
  {Lin}}, \bibinfo {author} {\bibfnamefont {T.}~\bibnamefont {Tai}}, \bibinfo
  {author} {\bibfnamefont {L.}~\bibnamefont {Li}},\ and\ \bibinfo {author}
  {\bibfnamefont {C.~H.}\ \bibnamefont {Lee}},\ }\href@noop {} {\bibfield
  {journal} {\bibinfo  {journal} {Frontiers of Physics}\ }\textbf {\bibinfo
  {volume} {18}},\ \bibinfo {pages} {53605} (\bibinfo {year}
  {2023})}\BibitemShut {NoStop}%
\bibitem [{\citenamefont {Torres}(2019)}]{torres2019perspective}%
  \BibitemOpen
  \bibfield  {author} {\bibinfo {author} {\bibfnamefont {L.~E.~F.}\
  \bibnamefont {Torres}},\ }\href@noop {} {\bibfield  {journal} {\bibinfo
  {journal} {Journal of Physics: Materials}\ }\textbf {\bibinfo {volume} {3}},\
  \bibinfo {pages} {014002} (\bibinfo {year} {2019})}\BibitemShut {NoStop}%
\bibitem [{\citenamefont {Peng}\ \emph {et~al.}(2022)\citenamefont {Peng},
  \citenamefont {Jie}, \citenamefont {Yu},\ and\ \citenamefont
  {Wang}}]{PhysRevB.106.L161402}%
  \BibitemOpen
  \bibfield  {author} {\bibinfo {author} {\bibfnamefont {Y.}~\bibnamefont
  {Peng}}, \bibinfo {author} {\bibfnamefont {J.}~\bibnamefont {Jie}}, \bibinfo
  {author} {\bibfnamefont {D.}~\bibnamefont {Yu}},\ and\ \bibinfo {author}
  {\bibfnamefont {Y.}~\bibnamefont {Wang}},\ }\href
  {https://doi.org/10.1103/PhysRevB.106.L161402} {\bibfield  {journal}
  {\bibinfo  {journal} {Phys. Rev. B}\ }\textbf {\bibinfo {volume} {106}},\
  \bibinfo {pages} {L161402} (\bibinfo {year} {2022})}\BibitemShut {NoStop}%
\bibitem [{\citenamefont {Rui}\ \emph {et~al.}(2022)\citenamefont {Rui},
  \citenamefont {Zheng}, \citenamefont {Wang},\ and\ \citenamefont
  {Wang}}]{PhysRevLett.128.226401}%
  \BibitemOpen
  \bibfield  {author} {\bibinfo {author} {\bibfnamefont {W.~B.}\ \bibnamefont
  {Rui}}, \bibinfo {author} {\bibfnamefont {Z.}~\bibnamefont {Zheng}}, \bibinfo
  {author} {\bibfnamefont {C.}~\bibnamefont {Wang}},\ and\ \bibinfo {author}
  {\bibfnamefont {Z.~D.}\ \bibnamefont {Wang}},\ }\href
  {https://doi.org/10.1103/PhysRevLett.128.226401} {\bibfield  {journal}
  {\bibinfo  {journal} {Phys. Rev. Lett.}\ }\textbf {\bibinfo {volume} {128}},\
  \bibinfo {pages} {226401} (\bibinfo {year} {2022})}\BibitemShut {NoStop}%
\bibitem [{\citenamefont {Rui}\ \emph {et~al.}(2023{\natexlab{a}})\citenamefont
  {Rui}, \citenamefont {Zhao},\ and\ \citenamefont
  {Wang}}]{PhysRevLett.131.176402}%
  \BibitemOpen
  \bibfield  {author} {\bibinfo {author} {\bibfnamefont {W.~B.}\ \bibnamefont
  {Rui}}, \bibinfo {author} {\bibfnamefont {Y.~X.}\ \bibnamefont {Zhao}},\ and\
  \bibinfo {author} {\bibfnamefont {Z.~D.}\ \bibnamefont {Wang}},\ }\href
  {https://doi.org/10.1103/PhysRevLett.131.176402} {\bibfield  {journal}
  {\bibinfo  {journal} {Phys. Rev. Lett.}\ }\textbf {\bibinfo {volume} {131}},\
  \bibinfo {pages} {176402} (\bibinfo {year} {2023}{\natexlab{a}})}\BibitemShut
  {NoStop}%
\bibitem [{\citenamefont {Rui}\ \emph {et~al.}(2023{\natexlab{b}})\citenamefont
  {Rui}, \citenamefont {Zhao},\ and\ \citenamefont
  {Wang}}]{PhysRevB.108.165105}%
  \BibitemOpen
  \bibfield  {author} {\bibinfo {author} {\bibfnamefont {W.~B.}\ \bibnamefont
  {Rui}}, \bibinfo {author} {\bibfnamefont {Y.~X.}\ \bibnamefont {Zhao}},\ and\
  \bibinfo {author} {\bibfnamefont {Z.~D.}\ \bibnamefont {Wang}},\ }\href
  {https://doi.org/10.1103/PhysRevB.108.165105} {\bibfield  {journal} {\bibinfo
   {journal} {Phys. Rev. B}\ }\textbf {\bibinfo {volume} {108}},\ \bibinfo
  {pages} {165105} (\bibinfo {year} {2023}{\natexlab{b}})}\BibitemShut
  {NoStop}%
\bibitem [{\citenamefont {Li}\ \emph {et~al.}(2020{\natexlab{a}})\citenamefont
  {Li}, \citenamefont {Lee}, \citenamefont {Mu},\ and\ \citenamefont
  {Gong}}]{li2020critical}%
  \BibitemOpen
  \bibfield  {author} {\bibinfo {author} {\bibfnamefont {L.}~\bibnamefont
  {Li}}, \bibinfo {author} {\bibfnamefont {C.~H.}\ \bibnamefont {Lee}},
  \bibinfo {author} {\bibfnamefont {S.}~\bibnamefont {Mu}},\ and\ \bibinfo
  {author} {\bibfnamefont {J.}~\bibnamefont {Gong}},\ }\href@noop {} {\bibfield
   {journal} {\bibinfo  {journal} {Nature communications}\ }\textbf {\bibinfo
  {volume} {11}},\ \bibinfo {pages} {5491} (\bibinfo {year}
  {2020}{\natexlab{a}})}\BibitemShut {NoStop}%
\bibitem [{\citenamefont {Li}\ \emph {et~al.}(2021)\citenamefont {Li},
  \citenamefont {Lee},\ and\ \citenamefont {Gong}}]{li2021impurity}%
  \BibitemOpen
  \bibfield  {author} {\bibinfo {author} {\bibfnamefont {L.}~\bibnamefont
  {Li}}, \bibinfo {author} {\bibfnamefont {C.~H.}\ \bibnamefont {Lee}},\ and\
  \bibinfo {author} {\bibfnamefont {J.}~\bibnamefont {Gong}},\ }\href@noop {}
  {\bibfield  {journal} {\bibinfo  {journal} {Communications Physics}\ }\textbf
  {\bibinfo {volume} {4}},\ \bibinfo {pages} {42} (\bibinfo {year}
  {2021})}\BibitemShut {NoStop}%
\bibitem [{\citenamefont {Guo}\ \emph {et~al.}(2023)\citenamefont {Guo},
  \citenamefont {Wang}, \citenamefont {Hu},\ and\ \citenamefont
  {Chen}}]{guo2023accumulation}%
  \BibitemOpen
  \bibfield  {author} {\bibinfo {author} {\bibfnamefont {C.-X.}\ \bibnamefont
  {Guo}}, \bibinfo {author} {\bibfnamefont {X.}~\bibnamefont {Wang}}, \bibinfo
  {author} {\bibfnamefont {H.}~\bibnamefont {Hu}},\ and\ \bibinfo {author}
  {\bibfnamefont {S.}~\bibnamefont {Chen}},\ }\href@noop {} {\bibfield
  {journal} {\bibinfo  {journal} {Physical Review B}\ }\textbf {\bibinfo
  {volume} {107}},\ \bibinfo {pages} {134121} (\bibinfo {year}
  {2023})}\BibitemShut {NoStop}%
\bibitem [{\citenamefont {Li}\ \emph {et~al.}(2023{\natexlab{a}})\citenamefont
  {Li}, \citenamefont {Wang}, \citenamefont {Song},\ and\ \citenamefont
  {Wang}}]{li2023scale}%
  \BibitemOpen
  \bibfield  {author} {\bibinfo {author} {\bibfnamefont {B.}~\bibnamefont
  {Li}}, \bibinfo {author} {\bibfnamefont {H.-R.}\ \bibnamefont {Wang}},
  \bibinfo {author} {\bibfnamefont {F.}~\bibnamefont {Song}},\ and\ \bibinfo
  {author} {\bibfnamefont {Z.}~\bibnamefont {Wang}},\ }\href@noop {} {\bibfield
   {journal} {\bibinfo  {journal} {Physical Review B}\ }\textbf {\bibinfo
  {volume} {108}},\ \bibinfo {pages} {L161409} (\bibinfo {year}
  {2023}{\natexlab{a}})}\BibitemShut {NoStop}%
\bibitem [{\citenamefont {Ke}\ \emph {et~al.}(2023)\citenamefont {Ke},
  \citenamefont {Huang}, \citenamefont {Liu}, \citenamefont {Kivshar},\ and\
  \citenamefont {Lee}}]{PhysRevLett.131.103604}%
  \BibitemOpen
  \bibfield  {author} {\bibinfo {author} {\bibfnamefont {Y.}~\bibnamefont
  {Ke}}, \bibinfo {author} {\bibfnamefont {J.}~\bibnamefont {Huang}}, \bibinfo
  {author} {\bibfnamefont {W.}~\bibnamefont {Liu}}, \bibinfo {author}
  {\bibfnamefont {Y.}~\bibnamefont {Kivshar}},\ and\ \bibinfo {author}
  {\bibfnamefont {C.}~\bibnamefont {Lee}},\ }\href
  {https://doi.org/10.1103/PhysRevLett.131.103604} {\bibfield  {journal}
  {\bibinfo  {journal} {Phys. Rev. Lett.}\ }\textbf {\bibinfo {volume} {131}},\
  \bibinfo {pages} {103604} (\bibinfo {year} {2023})}\BibitemShut {NoStop}%
\bibitem [{\citenamefont {Zheng}\ \emph {et~al.}(2024)\citenamefont {Zheng},
  \citenamefont {Qiao}, \citenamefont {Wang}, \citenamefont {Cao},\ and\
  \citenamefont {Chen}}]{PhysRevLett.132.086502}%
  \BibitemOpen
  \bibfield  {author} {\bibinfo {author} {\bibfnamefont {M.}~\bibnamefont
  {Zheng}}, \bibinfo {author} {\bibfnamefont {Y.}~\bibnamefont {Qiao}},
  \bibinfo {author} {\bibfnamefont {Y.}~\bibnamefont {Wang}}, \bibinfo {author}
  {\bibfnamefont {J.}~\bibnamefont {Cao}},\ and\ \bibinfo {author}
  {\bibfnamefont {S.}~\bibnamefont {Chen}},\ }\href
  {https://doi.org/10.1103/PhysRevLett.132.086502} {\bibfield  {journal}
  {\bibinfo  {journal} {Phys. Rev. Lett.}\ }\textbf {\bibinfo {volume} {132}},\
  \bibinfo {pages} {086502} (\bibinfo {year} {2024})}\BibitemShut {NoStop}%
\bibitem [{\citenamefont {Guo}\ \emph {et~al.}(2024)\citenamefont {Guo},
  \citenamefont {Su}, \citenamefont {Wang}, \citenamefont {Li}, \citenamefont
  {Wang}, \citenamefont {Ruan}, \citenamefont {Du}, \citenamefont {Zheng},
  \citenamefont {Chen},\ and\ \citenamefont {Hu}}]{guo2024scale}%
  \BibitemOpen
  \bibfield  {author} {\bibinfo {author} {\bibfnamefont {C.-X.}\ \bibnamefont
  {Guo}}, \bibinfo {author} {\bibfnamefont {L.}~\bibnamefont {Su}}, \bibinfo
  {author} {\bibfnamefont {Y.}~\bibnamefont {Wang}}, \bibinfo {author}
  {\bibfnamefont {L.}~\bibnamefont {Li}}, \bibinfo {author} {\bibfnamefont
  {J.}~\bibnamefont {Wang}}, \bibinfo {author} {\bibfnamefont {X.}~\bibnamefont
  {Ruan}}, \bibinfo {author} {\bibfnamefont {Y.}~\bibnamefont {Du}}, \bibinfo
  {author} {\bibfnamefont {D.}~\bibnamefont {Zheng}}, \bibinfo {author}
  {\bibfnamefont {S.}~\bibnamefont {Chen}},\ and\ \bibinfo {author}
  {\bibfnamefont {H.}~\bibnamefont {Hu}},\ }\href@noop {} {\bibfield  {journal}
  {\bibinfo  {journal} {Nature Communications}\ }\textbf {\bibinfo {volume}
  {15}},\ \bibinfo {pages} {9120} (\bibinfo {year} {2024})}\BibitemShut
  {NoStop}%
\bibitem [{\citenamefont {Yokomizo}\ and\ \citenamefont
  {Murakami}(2021)}]{yokomizo2021scaling}%
  \BibitemOpen
  \bibfield  {author} {\bibinfo {author} {\bibfnamefont {K.}~\bibnamefont
  {Yokomizo}}\ and\ \bibinfo {author} {\bibfnamefont {S.}~\bibnamefont
  {Murakami}},\ }\href@noop {} {\bibfield  {journal} {\bibinfo  {journal}
  {Physical Review B}\ }\textbf {\bibinfo {volume} {104}},\ \bibinfo {pages}
  {165117} (\bibinfo {year} {2021})}\BibitemShut {NoStop}%
\bibitem [{\citenamefont {Qin}\ \emph {et~al.}(2023)\citenamefont {Qin},
  \citenamefont {Ma}, \citenamefont {Shen},\ and\ \citenamefont
  {Lee}}]{qin2023universal}%
  \BibitemOpen
  \bibfield  {author} {\bibinfo {author} {\bibfnamefont {F.}~\bibnamefont
  {Qin}}, \bibinfo {author} {\bibfnamefont {Y.}~\bibnamefont {Ma}}, \bibinfo
  {author} {\bibfnamefont {R.}~\bibnamefont {Shen}},\ and\ \bibinfo {author}
  {\bibfnamefont {C.~H.}\ \bibnamefont {Lee}},\ }\href@noop {} {\bibfield
  {journal} {\bibinfo  {journal} {Physical Review B}\ }\textbf {\bibinfo
  {volume} {107}},\ \bibinfo {pages} {155430} (\bibinfo {year}
  {2023})}\BibitemShut {NoStop}%
\bibitem [{\citenamefont {Lee}(2022)}]{lee2022exceptional}%
  \BibitemOpen
  \bibfield  {author} {\bibinfo {author} {\bibfnamefont {C.~H.}\ \bibnamefont
  {Lee}},\ }\href@noop {} {\bibfield  {journal} {\bibinfo  {journal} {Physical
  Review Letters}\ }\textbf {\bibinfo {volume} {128}},\ \bibinfo {pages}
  {010402} (\bibinfo {year} {2022})}\BibitemShut {NoStop}%
\bibitem [{\citenamefont {Zhu}\ and\ \citenamefont {Li}(2024)}]{zhu2024brief}%
  \BibitemOpen
  \bibfield  {author} {\bibinfo {author} {\bibfnamefont {W.}~\bibnamefont
  {Zhu}}\ and\ \bibinfo {author} {\bibfnamefont {L.}~\bibnamefont {Li}},\
  }\href@noop {} {\bibfield  {journal} {\bibinfo  {journal} {Journal of
  Physics: Condensed Matter}\ } (\bibinfo {year} {2024})}\BibitemShut {NoStop}%
\bibitem [{\citenamefont {Lee}\ \emph {et~al.}(2019)\citenamefont {Lee},
  \citenamefont {Li},\ and\ \citenamefont {Gong}}]{lee2019hybrid}%
  \BibitemOpen
  \bibfield  {author} {\bibinfo {author} {\bibfnamefont {C.~H.}\ \bibnamefont
  {Lee}}, \bibinfo {author} {\bibfnamefont {L.}~\bibnamefont {Li}},\ and\
  \bibinfo {author} {\bibfnamefont {J.}~\bibnamefont {Gong}},\ }\href@noop {}
  {\bibfield  {journal} {\bibinfo  {journal} {Physical review letters}\
  }\textbf {\bibinfo {volume} {123}},\ \bibinfo {pages} {016805} (\bibinfo
  {year} {2019})}\BibitemShut {NoStop}%
\bibitem [{\citenamefont {Li}\ \emph {et~al.}(2022)\citenamefont {Li},
  \citenamefont {Liang}, \citenamefont {Wang}, \citenamefont {Lu},\ and\
  \citenamefont {Liu}}]{li2022gain}%
  \BibitemOpen
  \bibfield  {author} {\bibinfo {author} {\bibfnamefont {Y.}~\bibnamefont
  {Li}}, \bibinfo {author} {\bibfnamefont {C.}~\bibnamefont {Liang}}, \bibinfo
  {author} {\bibfnamefont {C.}~\bibnamefont {Wang}}, \bibinfo {author}
  {\bibfnamefont {C.}~\bibnamefont {Lu}},\ and\ \bibinfo {author}
  {\bibfnamefont {Y.-C.}\ \bibnamefont {Liu}},\ }\href@noop {} {\bibfield
  {journal} {\bibinfo  {journal} {Physical Review Letters}\ }\textbf {\bibinfo
  {volume} {128}},\ \bibinfo {pages} {223903} (\bibinfo {year}
  {2022})}\BibitemShut {NoStop}%
\bibitem [{\citenamefont {Zhu}\ and\ \citenamefont
  {Gong}(2022)}]{zhu2022hybrid}%
  \BibitemOpen
  \bibfield  {author} {\bibinfo {author} {\bibfnamefont {W.}~\bibnamefont
  {Zhu}}\ and\ \bibinfo {author} {\bibfnamefont {J.}~\bibnamefont {Gong}},\
  }\href@noop {} {\bibfield  {journal} {\bibinfo  {journal} {Physical Review
  B}\ }\textbf {\bibinfo {volume} {106}},\ \bibinfo {pages} {035425} (\bibinfo
  {year} {2022})}\BibitemShut {NoStop}%
\bibitem [{\citenamefont {Li}\ \emph {et~al.}(2020{\natexlab{b}})\citenamefont
  {Li}, \citenamefont {Lee},\ and\ \citenamefont {Gong}}]{li2020topological}%
  \BibitemOpen
  \bibfield  {author} {\bibinfo {author} {\bibfnamefont {L.}~\bibnamefont
  {Li}}, \bibinfo {author} {\bibfnamefont {C.~H.}\ \bibnamefont {Lee}},\ and\
  \bibinfo {author} {\bibfnamefont {J.}~\bibnamefont {Gong}},\ }\href@noop {}
  {\bibfield  {journal} {\bibinfo  {journal} {Physical review letters}\
  }\textbf {\bibinfo {volume} {124}},\ \bibinfo {pages} {250402} (\bibinfo
  {year} {2020}{\natexlab{b}})}\BibitemShut {NoStop}%
\bibitem [{\citenamefont {Ou}\ \emph {et~al.}(2023)\citenamefont {Ou},
  \citenamefont {Wang},\ and\ \citenamefont {Li}}]{ou2023non}%
  \BibitemOpen
  \bibfield  {author} {\bibinfo {author} {\bibfnamefont {Z.}~\bibnamefont
  {Ou}}, \bibinfo {author} {\bibfnamefont {Y.}~\bibnamefont {Wang}},\ and\
  \bibinfo {author} {\bibfnamefont {L.}~\bibnamefont {Li}},\ }\href@noop {}
  {\bibfield  {journal} {\bibinfo  {journal} {Physical Review B}\ }\textbf
  {\bibinfo {volume} {107}},\ \bibinfo {pages} {L161404} (\bibinfo {year}
  {2023})}\BibitemShut {NoStop}%
\bibitem [{\citenamefont {Kawabata}\ \emph {et~al.}(2020)\citenamefont
  {Kawabata}, \citenamefont {Sato},\ and\ \citenamefont
  {Shiozaki}}]{kawabata2020higher}%
  \BibitemOpen
  \bibfield  {author} {\bibinfo {author} {\bibfnamefont {K.}~\bibnamefont
  {Kawabata}}, \bibinfo {author} {\bibfnamefont {M.}~\bibnamefont {Sato}},\
  and\ \bibinfo {author} {\bibfnamefont {K.}~\bibnamefont {Shiozaki}},\
  }\href@noop {} {\bibfield  {journal} {\bibinfo  {journal} {Physical Review
  B}\ }\textbf {\bibinfo {volume} {102}},\ \bibinfo {pages} {205118} (\bibinfo
  {year} {2020})}\BibitemShut {NoStop}%
\bibitem [{\citenamefont {Okugawa}\ \emph {et~al.}(2020)\citenamefont
  {Okugawa}, \citenamefont {Takahashi},\ and\ \citenamefont
  {Yokomizo}}]{okugawa2020second}%
  \BibitemOpen
  \bibfield  {author} {\bibinfo {author} {\bibfnamefont {R.}~\bibnamefont
  {Okugawa}}, \bibinfo {author} {\bibfnamefont {R.}~\bibnamefont {Takahashi}},\
  and\ \bibinfo {author} {\bibfnamefont {K.}~\bibnamefont {Yokomizo}},\
  }\href@noop {} {\bibfield  {journal} {\bibinfo  {journal} {Physical Review
  B}\ }\textbf {\bibinfo {volume} {102}},\ \bibinfo {pages} {241202} (\bibinfo
  {year} {2020})}\BibitemShut {NoStop}%
\bibitem [{\citenamefont {Fu}\ \emph {et~al.}(2021)\citenamefont {Fu},
  \citenamefont {Hu},\ and\ \citenamefont {Wan}}]{fu2021non}%
  \BibitemOpen
  \bibfield  {author} {\bibinfo {author} {\bibfnamefont {Y.}~\bibnamefont
  {Fu}}, \bibinfo {author} {\bibfnamefont {J.}~\bibnamefont {Hu}},\ and\
  \bibinfo {author} {\bibfnamefont {S.}~\bibnamefont {Wan}},\ }\href@noop {}
  {\bibfield  {journal} {\bibinfo  {journal} {Physical Review B}\ }\textbf
  {\bibinfo {volume} {103}},\ \bibinfo {pages} {045420} (\bibinfo {year}
  {2021})}\BibitemShut {NoStop}%
\bibitem [{\citenamefont {Yamamoto}\ \emph {et~al.}(2019)\citenamefont
  {Yamamoto}, \citenamefont {Nakagawa}, \citenamefont {Adachi}, \citenamefont
  {Takasan}, \citenamefont {Ueda},\ and\ \citenamefont
  {Kawakami}}]{PhysRevLett.123.123601}%
  \BibitemOpen
  \bibfield  {author} {\bibinfo {author} {\bibfnamefont {K.}~\bibnamefont
  {Yamamoto}}, \bibinfo {author} {\bibfnamefont {M.}~\bibnamefont {Nakagawa}},
  \bibinfo {author} {\bibfnamefont {K.}~\bibnamefont {Adachi}}, \bibinfo
  {author} {\bibfnamefont {K.}~\bibnamefont {Takasan}}, \bibinfo {author}
  {\bibfnamefont {M.}~\bibnamefont {Ueda}},\ and\ \bibinfo {author}
  {\bibfnamefont {N.}~\bibnamefont {Kawakami}},\ }\href
  {https://doi.org/10.1103/PhysRevLett.123.123601} {\bibfield  {journal}
  {\bibinfo  {journal} {Phys. Rev. Lett.}\ }\textbf {\bibinfo {volume} {123}},\
  \bibinfo {pages} {123601} (\bibinfo {year} {2019})}\BibitemShut {NoStop}%
\bibitem [{\citenamefont {Liu}\ \emph {et~al.}(2020)\citenamefont {Liu},
  \citenamefont {He}, \citenamefont {Yoshida}, \citenamefont {Xiang},\ and\
  \citenamefont {Nori}}]{PhysRevB.102.235151}%
  \BibitemOpen
  \bibfield  {author} {\bibinfo {author} {\bibfnamefont {T.}~\bibnamefont
  {Liu}}, \bibinfo {author} {\bibfnamefont {J.~J.}\ \bibnamefont {He}},
  \bibinfo {author} {\bibfnamefont {T.}~\bibnamefont {Yoshida}}, \bibinfo
  {author} {\bibfnamefont {Z.-L.}\ \bibnamefont {Xiang}},\ and\ \bibinfo
  {author} {\bibfnamefont {F.}~\bibnamefont {Nori}},\ }\href
  {https://doi.org/10.1103/PhysRevB.102.235151} {\bibfield  {journal} {\bibinfo
   {journal} {Phys. Rev. B}\ }\textbf {\bibinfo {volume} {102}},\ \bibinfo
  {pages} {235151} (\bibinfo {year} {2020})}\BibitemShut {NoStop}%
\bibitem [{\citenamefont {Zhang}\ \emph
  {et~al.}(2022{\natexlab{a}})\citenamefont {Zhang}, \citenamefont {Denner},
  \citenamefont {Bzdu{\v{s}}ek}, \citenamefont {Sentef},\ and\ \citenamefont
  {Neupert}}]{zhang2022symmetry}%
  \BibitemOpen
  \bibfield  {author} {\bibinfo {author} {\bibfnamefont {S.-B.}\ \bibnamefont
  {Zhang}}, \bibinfo {author} {\bibfnamefont {M.~M.}\ \bibnamefont {Denner}},
  \bibinfo {author} {\bibfnamefont {T.}~\bibnamefont {Bzdu{\v{s}}ek}}, \bibinfo
  {author} {\bibfnamefont {M.~A.}\ \bibnamefont {Sentef}},\ and\ \bibinfo
  {author} {\bibfnamefont {T.}~\bibnamefont {Neupert}},\ }\href@noop {}
  {\bibfield  {journal} {\bibinfo  {journal} {Physical review B}\ }\textbf
  {\bibinfo {volume} {106}},\ \bibinfo {pages} {L121102} (\bibinfo {year}
  {2022}{\natexlab{a}})}\BibitemShut {NoStop}%
\bibitem [{\citenamefont {Faugno}\ and\ \citenamefont
  {Ozawa}(2022)}]{faugno2022interaction}%
  \BibitemOpen
  \bibfield  {author} {\bibinfo {author} {\bibfnamefont {W.~N.}\ \bibnamefont
  {Faugno}}\ and\ \bibinfo {author} {\bibfnamefont {T.}~\bibnamefont {Ozawa}},\
  }\href@noop {} {\bibfield  {journal} {\bibinfo  {journal} {Physical review
  letters}\ }\textbf {\bibinfo {volume} {129}},\ \bibinfo {pages} {180401}
  (\bibinfo {year} {2022})}\BibitemShut {NoStop}%
\bibitem [{\citenamefont {Qin}\ and\ \citenamefont
  {Li}(2024)}]{qin2024occupation}%
  \BibitemOpen
  \bibfield  {author} {\bibinfo {author} {\bibfnamefont {Y.}~\bibnamefont
  {Qin}}\ and\ \bibinfo {author} {\bibfnamefont {L.}~\bibnamefont {Li}},\
  }\href@noop {} {\bibfield  {journal} {\bibinfo  {journal} {Physical Review
  Letters}\ }\textbf {\bibinfo {volume} {132}},\ \bibinfo {pages} {096501}
  (\bibinfo {year} {2024})}\BibitemShut {NoStop}%
\bibitem [{\citenamefont {Yoshida}\ \emph {et~al.}(2024)\citenamefont
  {Yoshida}, \citenamefont {Zhang}, \citenamefont {Neupert},\ and\
  \citenamefont {Kawakami}}]{yoshida2024non}%
  \BibitemOpen
  \bibfield  {author} {\bibinfo {author} {\bibfnamefont {T.}~\bibnamefont
  {Yoshida}}, \bibinfo {author} {\bibfnamefont {S.-B.}\ \bibnamefont {Zhang}},
  \bibinfo {author} {\bibfnamefont {T.}~\bibnamefont {Neupert}},\ and\ \bibinfo
  {author} {\bibfnamefont {N.}~\bibnamefont {Kawakami}},\ }\href@noop {}
  {\bibfield  {journal} {\bibinfo  {journal} {Physical Review Letters}\
  }\textbf {\bibinfo {volume} {133}},\ \bibinfo {pages} {076502} (\bibinfo
  {year} {2024})}\BibitemShut {NoStop}%
\bibitem [{\citenamefont {Shimomura}\ and\ \citenamefont
  {Sato}(2024)}]{PhysRevLett.133.136502}%
  \BibitemOpen
  \bibfield  {author} {\bibinfo {author} {\bibfnamefont {K.}~\bibnamefont
  {Shimomura}}\ and\ \bibinfo {author} {\bibfnamefont {M.}~\bibnamefont
  {Sato}},\ }\href {https://doi.org/10.1103/PhysRevLett.133.136502} {\bibfield
  {journal} {\bibinfo  {journal} {Phys. Rev. Lett.}\ }\textbf {\bibinfo
  {volume} {133}},\ \bibinfo {pages} {136502} (\bibinfo {year}
  {2024})}\BibitemShut {NoStop}%
\bibitem [{\citenamefont {Hatano}\ and\ \citenamefont
  {Nelson}(1996)}]{hatano1996localization}%
  \BibitemOpen
  \bibfield  {author} {\bibinfo {author} {\bibfnamefont {N.}~\bibnamefont
  {Hatano}}\ and\ \bibinfo {author} {\bibfnamefont {D.~R.}\ \bibnamefont
  {Nelson}},\ }\href@noop {} {\bibfield  {journal} {\bibinfo  {journal}
  {Physical review letters}\ }\textbf {\bibinfo {volume} {77}},\ \bibinfo
  {pages} {570} (\bibinfo {year} {1996})}\BibitemShut {NoStop}%
\bibitem [{\citenamefont {Hatano}\ and\ \citenamefont
  {Nelson}(1998)}]{hatano1998non}%
  \BibitemOpen
  \bibfield  {author} {\bibinfo {author} {\bibfnamefont {N.}~\bibnamefont
  {Hatano}}\ and\ \bibinfo {author} {\bibfnamefont {D.~R.}\ \bibnamefont
  {Nelson}},\ }\href@noop {} {\bibfield  {journal} {\bibinfo  {journal}
  {Physical Review B}\ }\textbf {\bibinfo {volume} {58}},\ \bibinfo {pages}
  {8384} (\bibinfo {year} {1998})}\BibitemShut {NoStop}%
\bibitem [{\citenamefont {Su}\ \emph {et~al.}(1979)\citenamefont {Su},
  \citenamefont {Schrieffer},\ and\ \citenamefont {Heeger}}]{su1979solitons}%
  \BibitemOpen
  \bibfield  {author} {\bibinfo {author} {\bibfnamefont {W.-P.}\ \bibnamefont
  {Su}}, \bibinfo {author} {\bibfnamefont {J.~R.}\ \bibnamefont {Schrieffer}},\
  and\ \bibinfo {author} {\bibfnamefont {A.~J.}\ \bibnamefont {Heeger}},\
  }\href@noop {} {\bibfield  {journal} {\bibinfo  {journal} {Physical review
  letters}\ }\textbf {\bibinfo {volume} {42}},\ \bibinfo {pages} {1698}
  (\bibinfo {year} {1979})}\BibitemShut {NoStop}%
\bibitem [{\citenamefont {Wegner}(1980)}]{wegner1980inverse}%
  \BibitemOpen
  \bibfield  {author} {\bibinfo {author} {\bibfnamefont {F.}~\bibnamefont
  {Wegner}},\ }\href@noop {} {\bibfield  {journal} {\bibinfo  {journal}
  {Zeitschrift f{\"u}r Physik B Condensed Matter}\ }\textbf {\bibinfo {volume}
  {36}},\ \bibinfo {pages} {209} (\bibinfo {year} {1980})}\BibitemShut
  {NoStop}%
\bibitem [{\citenamefont {Ganeshan}\ \emph {et~al.}(2015)\citenamefont
  {Ganeshan}, \citenamefont {Pixley},\ and\ \citenamefont
  {Das~Sarma}}]{ganeshan2015nearest}%
  \BibitemOpen
  \bibfield  {author} {\bibinfo {author} {\bibfnamefont {S.}~\bibnamefont
  {Ganeshan}}, \bibinfo {author} {\bibfnamefont {J.~H.}\ \bibnamefont
  {Pixley}},\ and\ \bibinfo {author} {\bibfnamefont {S.}~\bibnamefont
  {Das~Sarma}},\ }\href {https://doi.org/10.1103/PhysRevLett.114.146601}
  {\bibfield  {journal} {\bibinfo  {journal} {Phys. Rev. Lett.}\ }\textbf
  {\bibinfo {volume} {114}},\ \bibinfo {pages} {146601} (\bibinfo {year}
  {2015})}\BibitemShut {NoStop}%
\bibitem [{sup()}]{suppmat}%
  \BibitemOpen
  \href@noop {} {\bibinfo  {journal} {Supplemental Materials}\ }\BibitemShut
  {NoStop}%
\bibitem [{\citenamefont {Kawabata}\ and\ \citenamefont
  {Ryu}(2021)}]{kawabata2021nonunitary}%
  \BibitemOpen
\bibfield  {journal} {  }\bibfield  {author} {\bibinfo {author} {\bibfnamefont
  {K.}~\bibnamefont {Kawabata}}\ and\ \bibinfo {author} {\bibfnamefont
  {S.}~\bibnamefont {Ryu}},\ }\href@noop {} {\bibfield  {journal} {\bibinfo
  {journal} {Physical review letters}\ }\textbf {\bibinfo {volume} {126}},\
  \bibinfo {pages} {166801} (\bibinfo {year} {2021})}\BibitemShut {NoStop}%
\bibitem [{\citenamefont {Davies}\ \emph {et~al.}(2024)\citenamefont {Davies},
  \citenamefont {Barandun}, \citenamefont {Hiltunen}, \citenamefont {Craster},\
  and\ \citenamefont {Ammari}}]{davies2024two}%
  \BibitemOpen
  \bibfield  {author} {\bibinfo {author} {\bibfnamefont {B.}~\bibnamefont
  {Davies}}, \bibinfo {author} {\bibfnamefont {S.}~\bibnamefont {Barandun}},
  \bibinfo {author} {\bibfnamefont {E.~O.}\ \bibnamefont {Hiltunen}}, \bibinfo
  {author} {\bibfnamefont {R.~V.}\ \bibnamefont {Craster}},\ and\ \bibinfo
  {author} {\bibfnamefont {H.}~\bibnamefont {Ammari}},\ }\href@noop {}
  {\bibfield  {journal} {\bibinfo  {journal} {arXiv preprint arXiv:2403.12546}\
  } (\bibinfo {year} {2024})}\BibitemShut {NoStop}%
\bibitem [{\citenamefont {Zhang}\ \emph {et~al.}(2024)\citenamefont {Zhang},
  \citenamefont {Shu},\ and\ \citenamefont {Sun}}]{zhang2024algebraic}%
  \BibitemOpen
  \bibfield  {author} {\bibinfo {author} {\bibfnamefont {K.}~\bibnamefont
  {Zhang}}, \bibinfo {author} {\bibfnamefont {C.}~\bibnamefont {Shu}},\ and\
  \bibinfo {author} {\bibfnamefont {K.}~\bibnamefont {Sun}},\ }\href@noop {}
  {\bibfield  {journal} {\bibinfo  {journal} {arXiv preprint arXiv:2406.06682}\
  } (\bibinfo {year} {2024})}\BibitemShut {NoStop}%
\bibitem [{\citenamefont {Benalcazar}\ \emph
  {et~al.}(2017{\natexlab{a}})\citenamefont {Benalcazar}, \citenamefont
  {Bernevig},\ and\ \citenamefont {Hughes}}]{benalcazar2017quantized}%
  \BibitemOpen
  \bibfield  {author} {\bibinfo {author} {\bibfnamefont {W.~A.}\ \bibnamefont
  {Benalcazar}}, \bibinfo {author} {\bibfnamefont {B.~A.}\ \bibnamefont
  {Bernevig}},\ and\ \bibinfo {author} {\bibfnamefont {T.~L.}\ \bibnamefont
  {Hughes}},\ }\href@noop {} {\bibfield  {journal} {\bibinfo  {journal}
  {Science}\ }\textbf {\bibinfo {volume} {357}},\ \bibinfo {pages} {61}
  (\bibinfo {year} {2017}{\natexlab{a}})}\BibitemShut {NoStop}%
\bibitem [{\citenamefont {Benalcazar}\ \emph
  {et~al.}(2017{\natexlab{b}})\citenamefont {Benalcazar}, \citenamefont
  {Bernevig},\ and\ \citenamefont {Hughes}}]{benalcazar2017electric}%
  \BibitemOpen
  \bibfield  {author} {\bibinfo {author} {\bibfnamefont {W.~A.}\ \bibnamefont
  {Benalcazar}}, \bibinfo {author} {\bibfnamefont {B.~A.}\ \bibnamefont
  {Bernevig}},\ and\ \bibinfo {author} {\bibfnamefont {T.~L.}\ \bibnamefont
  {Hughes}},\ }\href@noop {} {\bibfield  {journal} {\bibinfo  {journal}
  {Physical Review B}\ }\textbf {\bibinfo {volume} {96}},\ \bibinfo {pages}
  {245115} (\bibinfo {year} {2017}{\natexlab{b}})}\BibitemShut {NoStop}%
\bibitem [{Note1()}]{Note1}%
  \BibitemOpen
  \bibinfo {note} {The junction system of the edges effectively form a 1D
  system with closed boundary conditions. In contrast, in Refs.~\protect \citep
  {budich2020non,guo2021exact}, the extreme sensitivity is a property of OBC
  systems with NHSE, where extra weak couplings effectively close the
  boundaries and greatly affect the GBZ description of the OBC
  eigensolutions.}\BibitemShut {Stop}%
\bibitem [{\citenamefont {Li}\ \emph {et~al.}(2023{\natexlab{b}})\citenamefont
  {Li}, \citenamefont {Trauzettel}, \citenamefont {Neupert},\ and\
  \citenamefont {Zhang}}]{li2023enhancement}%
  \BibitemOpen
  \bibfield  {author} {\bibinfo {author} {\bibfnamefont {C.-A.}\ \bibnamefont
  {Li}}, \bibinfo {author} {\bibfnamefont {B.}~\bibnamefont {Trauzettel}},
  \bibinfo {author} {\bibfnamefont {T.}~\bibnamefont {Neupert}},\ and\ \bibinfo
  {author} {\bibfnamefont {S.-B.}\ \bibnamefont {Zhang}},\ }\href@noop {}
  {\bibfield  {journal} {\bibinfo  {journal} {Physical Review Letters}\
  }\textbf {\bibinfo {volume} {131}},\ \bibinfo {pages} {116601} (\bibinfo
  {year} {2023}{\natexlab{b}})}\BibitemShut {NoStop}%
\bibitem [{\citenamefont {Zou}\ \emph {et~al.}(2021)\citenamefont {Zou},
  \citenamefont {Chen}, \citenamefont {He}, \citenamefont {Bao}, \citenamefont
  {Lee}, \citenamefont {Sun},\ and\ \citenamefont
  {Zhang}}]{zou2021observation}%
  \BibitemOpen
  \bibfield  {author} {\bibinfo {author} {\bibfnamefont {D.}~\bibnamefont
  {Zou}}, \bibinfo {author} {\bibfnamefont {T.}~\bibnamefont {Chen}}, \bibinfo
  {author} {\bibfnamefont {W.}~\bibnamefont {He}}, \bibinfo {author}
  {\bibfnamefont {J.}~\bibnamefont {Bao}}, \bibinfo {author} {\bibfnamefont
  {C.~H.}\ \bibnamefont {Lee}}, \bibinfo {author} {\bibfnamefont
  {H.}~\bibnamefont {Sun}},\ and\ \bibinfo {author} {\bibfnamefont
  {X.}~\bibnamefont {Zhang}},\ }\href@noop {} {\bibfield  {journal} {\bibinfo
  {journal} {Nature Communications}\ }\textbf {\bibinfo {volume} {12}},\
  \bibinfo {pages} {7201} (\bibinfo {year} {2021})}\BibitemShut {NoStop}%
\bibitem [{\citenamefont {Palacios}\ \emph {et~al.}(2021)\citenamefont
  {Palacios}, \citenamefont {Tchoumakov}, \citenamefont {Guix}, \citenamefont
  {Pagonabarraga}, \citenamefont {S{\'a}nchez},\ and\ \citenamefont
  {G.~Grushin}}]{palacios2021guided}%
  \BibitemOpen
  \bibfield  {author} {\bibinfo {author} {\bibfnamefont {L.~S.}\ \bibnamefont
  {Palacios}}, \bibinfo {author} {\bibfnamefont {S.}~\bibnamefont
  {Tchoumakov}}, \bibinfo {author} {\bibfnamefont {M.}~\bibnamefont {Guix}},
  \bibinfo {author} {\bibfnamefont {I.}~\bibnamefont {Pagonabarraga}}, \bibinfo
  {author} {\bibfnamefont {S.}~\bibnamefont {S{\'a}nchez}},\ and\ \bibinfo
  {author} {\bibfnamefont {A.}~\bibnamefont {G.~Grushin}},\ }\href@noop {}
  {\bibfield  {journal} {\bibinfo  {journal} {Nature Communications}\ }\textbf
  {\bibinfo {volume} {12}},\ \bibinfo {pages} {4691} (\bibinfo {year}
  {2021})}\BibitemShut {NoStop}%
\bibitem [{\citenamefont {Zhang}\ \emph {et~al.}(2021)\citenamefont {Zhang},
  \citenamefont {Tian}, \citenamefont {Jiang}, \citenamefont {Lu},\ and\
  \citenamefont {Chen}}]{zhang2021observation}%
  \BibitemOpen
  \bibfield  {author} {\bibinfo {author} {\bibfnamefont {X.}~\bibnamefont
  {Zhang}}, \bibinfo {author} {\bibfnamefont {Y.}~\bibnamefont {Tian}},
  \bibinfo {author} {\bibfnamefont {J.-H.}\ \bibnamefont {Jiang}}, \bibinfo
  {author} {\bibfnamefont {M.-H.}\ \bibnamefont {Lu}},\ and\ \bibinfo {author}
  {\bibfnamefont {Y.-F.}\ \bibnamefont {Chen}},\ }\href@noop {} {\bibfield
  {journal} {\bibinfo  {journal} {Nature communications}\ }\textbf {\bibinfo
  {volume} {12}},\ \bibinfo {pages} {5377} (\bibinfo {year}
  {2021})}\BibitemShut {NoStop}%
\bibitem [{\citenamefont {Liang}\ \emph {et~al.}(2022)\citenamefont {Liang},
  \citenamefont {Xie}, \citenamefont {Dong}, \citenamefont {Li}, \citenamefont
  {Li}, \citenamefont {Gadway}, \citenamefont {Yi},\ and\ \citenamefont
  {Yan}}]{liang2022dynamic}%
  \BibitemOpen
  \bibfield  {author} {\bibinfo {author} {\bibfnamefont {Q.}~\bibnamefont
  {Liang}}, \bibinfo {author} {\bibfnamefont {D.}~\bibnamefont {Xie}}, \bibinfo
  {author} {\bibfnamefont {Z.}~\bibnamefont {Dong}}, \bibinfo {author}
  {\bibfnamefont {H.}~\bibnamefont {Li}}, \bibinfo {author} {\bibfnamefont
  {H.}~\bibnamefont {Li}}, \bibinfo {author} {\bibfnamefont {B.}~\bibnamefont
  {Gadway}}, \bibinfo {author} {\bibfnamefont {W.}~\bibnamefont {Yi}},\ and\
  \bibinfo {author} {\bibfnamefont {B.}~\bibnamefont {Yan}},\ }\href
  {https://doi.org/10.1103/PhysRevLett.129.070401} {\bibfield  {journal}
  {\bibinfo  {journal} {Phys. Rev. Lett.}\ }\textbf {\bibinfo {volume} {129}},\
  \bibinfo {pages} {070401} (\bibinfo {year} {2022})}\BibitemShut {NoStop}%
\bibitem [{\citenamefont {Zhao}\ \emph {et~al.}(2025)\citenamefont {Zhao},
  \citenamefont {Wang}, \citenamefont {He}, \citenamefont {Poon}, \citenamefont
  {Pak}, \citenamefont {Liu}, \citenamefont {Ren}, \citenamefont {Liu},\ and\
  \citenamefont {Jo}}]{zhao2025two}%
  \BibitemOpen
  \bibfield  {author} {\bibinfo {author} {\bibfnamefont {E.}~\bibnamefont
  {Zhao}}, \bibinfo {author} {\bibfnamefont {Z.}~\bibnamefont {Wang}}, \bibinfo
  {author} {\bibfnamefont {C.}~\bibnamefont {He}}, \bibinfo {author}
  {\bibfnamefont {T.~F.~J.}\ \bibnamefont {Poon}}, \bibinfo {author}
  {\bibfnamefont {K.~K.}\ \bibnamefont {Pak}}, \bibinfo {author} {\bibfnamefont
  {Y.-J.}\ \bibnamefont {Liu}}, \bibinfo {author} {\bibfnamefont
  {P.}~\bibnamefont {Ren}}, \bibinfo {author} {\bibfnamefont {X.-J.}\
  \bibnamefont {Liu}},\ and\ \bibinfo {author} {\bibfnamefont {G.-B.}\
  \bibnamefont {Jo}},\ }\href@noop {} {\bibfield  {journal} {\bibinfo
  {journal} {Nature}\ ,\ \bibinfo {pages} {1}} (\bibinfo {year}
  {2025})}\BibitemShut {NoStop}%
\bibitem [{\citenamefont {Zhang}\ \emph
  {et~al.}(2022{\natexlab{b}})\citenamefont {Zhang}, \citenamefont {Yang},\
  and\ \citenamefont {Fang}}]{zhang2022universal}%
  \BibitemOpen
  \bibfield  {author} {\bibinfo {author} {\bibfnamefont {K.}~\bibnamefont
  {Zhang}}, \bibinfo {author} {\bibfnamefont {Z.}~\bibnamefont {Yang}},\ and\
  \bibinfo {author} {\bibfnamefont {C.}~\bibnamefont {Fang}},\ }\href@noop {}
  {\bibfield  {journal} {\bibinfo  {journal} {Nature communications}\ }\textbf
  {\bibinfo {volume} {13}},\ \bibinfo {pages} {2496} (\bibinfo {year}
  {2022}{\natexlab{b}})}\BibitemShut {NoStop}%
\bibitem [{\citenamefont {Lei}\ \emph {et~al.}(2024)\citenamefont {Lei},
  \citenamefont {Lee},\ and\ \citenamefont {Li}}]{lei2024activating}%
  \BibitemOpen
  \bibfield  {author} {\bibinfo {author} {\bibfnamefont {Z.}~\bibnamefont
  {Lei}}, \bibinfo {author} {\bibfnamefont {C.~H.}\ \bibnamefont {Lee}},\ and\
  \bibinfo {author} {\bibfnamefont {L.}~\bibnamefont {Li}},\ }\href@noop {}
  {\bibfield  {journal} {\bibinfo  {journal} {Communications Physics}\ }\textbf
  {\bibinfo {volume} {7}},\ \bibinfo {pages} {100} (\bibinfo {year}
  {2024})}\BibitemShut {NoStop}%
\bibitem [{\citenamefont {Wang}\ \emph {et~al.}(2024)\citenamefont {Wang},
  \citenamefont {Song},\ and\ \citenamefont {Wang}}]{wang2024amoeba}%
  \BibitemOpen
  \bibfield  {author} {\bibinfo {author} {\bibfnamefont {H.-Y.}\ \bibnamefont
  {Wang}}, \bibinfo {author} {\bibfnamefont {F.}~\bibnamefont {Song}},\ and\
  \bibinfo {author} {\bibfnamefont {Z.}~\bibnamefont {Wang}},\ }\href@noop {}
  {\bibfield  {journal} {\bibinfo  {journal} {Physical Review X}\ }\textbf
  {\bibinfo {volume} {14}},\ \bibinfo {pages} {021011} (\bibinfo {year}
  {2024})}\BibitemShut {NoStop}%
\bibitem [{\citenamefont {Hu}(2024)}]{hu2024topological}%
  \BibitemOpen
  \bibfield  {author} {\bibinfo {author} {\bibfnamefont {H.}~\bibnamefont
  {Hu}},\ }\href@noop {} {\bibfield  {journal} {\bibinfo  {journal} {Science
  Bulletin (arXiv:2306.12022)}\ } (\bibinfo {year} {2024})}\BibitemShut
  {NoStop}%
\bibitem [{\citenamefont {Xiong}\ \emph {et~al.}(2024)\citenamefont {Xiong},
  \citenamefont {Xing},\ and\ \citenamefont {Hu}}]{xiong2024non}%
  \BibitemOpen
  \bibfield  {author} {\bibinfo {author} {\bibfnamefont {Y.}~\bibnamefont
  {Xiong}}, \bibinfo {author} {\bibfnamefont {Z.-Y.}\ \bibnamefont {Xing}},\
  and\ \bibinfo {author} {\bibfnamefont {H.}~\bibnamefont {Hu}},\ }\href@noop
  {} {\bibfield  {journal} {\bibinfo  {journal} {arXiv preprint
  arXiv:2407.01296}\ } (\bibinfo {year} {2024})}\BibitemShut {NoStop}%
\bibitem [{\citenamefont {Guo}\ \emph {et~al.}(2021)\citenamefont {Guo},
  \citenamefont {Liu}, \citenamefont {Zhao}, \citenamefont {Liu},\ and\
  \citenamefont {Chen}}]{guo2021exact}%
  \BibitemOpen
  \bibfield  {author} {\bibinfo {author} {\bibfnamefont {C.-X.}\ \bibnamefont
  {Guo}}, \bibinfo {author} {\bibfnamefont {C.-H.}\ \bibnamefont {Liu}},
  \bibinfo {author} {\bibfnamefont {X.-M.}\ \bibnamefont {Zhao}}, \bibinfo
  {author} {\bibfnamefont {Y.}~\bibnamefont {Liu}},\ and\ \bibinfo {author}
  {\bibfnamefont {S.}~\bibnamefont {Chen}},\ }\href@noop {} {\bibfield
  {journal} {\bibinfo  {journal} {Physical Review Letters}\ }\textbf {\bibinfo
  {volume} {127}},\ \bibinfo {pages} {116801} (\bibinfo {year}
  {2021})}\BibitemShut {NoStop}%
\bibitem [{\citenamefont {Wang}\ \emph {et~al.}(2023)\citenamefont {Wang},
  \citenamefont {Jen},\ and\ \citenamefont {You}}]{wang2023scaling}%
  \BibitemOpen
  \bibfield  {author} {\bibinfo {author} {\bibfnamefont {Y.-C.}\ \bibnamefont
  {Wang}}, \bibinfo {author} {\bibfnamefont {H.}~\bibnamefont {Jen}},\ and\
  \bibinfo {author} {\bibfnamefont {J.-S.}\ \bibnamefont {You}},\ }\href@noop
  {} {\bibfield  {journal} {\bibinfo  {journal} {Physical Review B}\ }\textbf
  {\bibinfo {volume} {108}},\ \bibinfo {pages} {085418} (\bibinfo {year}
  {2023})}\BibitemShut {NoStop}%
\bibitem [{\citenamefont {Budich}\ and\ \citenamefont
  {Bergholtz}(2020)}]{budich2020non}%
  \BibitemOpen
  \bibfield  {author} {\bibinfo {author} {\bibfnamefont {J.~C.}\ \bibnamefont
  {Budich}}\ and\ \bibinfo {author} {\bibfnamefont {E.~J.}\ \bibnamefont
  {Bergholtz}},\ }\href@noop {} {\bibfield  {journal} {\bibinfo  {journal}
  {Physical Review Letters}\ }\textbf {\bibinfo {volume} {125}},\ \bibinfo
  {pages} {180403} (\bibinfo {year} {2020})}\BibitemShut {NoStop}%
\end{thebibliography}%
\bibliographystyle{apsrev4-2}

\end{document}